\documentclass[aip,jcp,
  amsmath,amssymb,endnotes,
  a4paper,reprint]{revtex4-2}

\usepackage{graphicx}
\usepackage{dcolumn}

\usepackage{MnSymbol}
\usepackage{siunitx}

\newcommand{\Aeps}{A_\varepsilon}
\newcommand{\dd}{\mathrm{d}}
\newcommand{\ee}{\mathrm{e}}
\newcommand{\mbm}{\mathbf{m}}
\newcommand{\mbe}{\mathbf{e}}
\newcommand{\mbX}{\mathbf{X}}

\newcommand{\mbal}{\boldsymbol{\alpha}}
\newcommand{\aiso}{\alpha_\mathrm{el}}

\newcommand{\kB}{k_\mathrm{B}}
\newcommand{\NA}{N_\mathrm{A}}

\newcommand{\HtO}{H${}_2{}^{16}$O}

\begin{document}

\title{Comprehensive Quantum Calculation of the First Dielectric Virial Coefficient of Water}

\author{Giovanni Garberoglio}
\email{garberoglio@ectstar.eu}
\affiliation{European Centre for Theoretical Studies in Nuclear Physics
and Related Areas (ECT*), Fondazione Bruno Kessler, Trento I-38123, Italy.}

\author{Chiara Lissoni}
\thanks{These authors contributed equally.}
\author{Luca Spagnoli}
\thanks{These authors contributed equally.}
\affiliation{Physics Department, University of Trento, Trento I-38123, Italy.}

\author{Allan H. Harvey}
\email{allan.harvey@nist.gov}
\affiliation{Applied Chemicals and Materials Division, National
  Institute of Standards and Technology, Boulder, CO 80305, USA.}

\begin{abstract}
We present a complete calculation, fully accounting for quantum effects and for molecular
flexibility, of the first dielectric virial coefficient of water and its isotopologues. The
contribution of the electronic polarizability is computed from a state-of-the-art intramolecular
potential and polarizability surface from the literature, and its small temperature dependence is
quantified. The dipolar polarizability is calculated in a similar manner with an accurate
literature dipole-moment surface; it differs from the classical result both due to the different
molecular geometries sampled at different temperatures and due to the quantization of rotation. We
calculate the dipolar contribution independently from spectroscopic information in the HITRAN2020
database and find that the two methods yield consistent results. The resulting first dielectric
virial coefficient provides a complete description of the dielectric constant at low density that
can be used in humidity metrology and as a boundary condition for new formulations for the static
dielectric constant of water and heavy water.
\end{abstract}

\date{16 November 2023}

\maketitle


\section{Introduction}

Water is crucial in many scientific and industrial contexts.  Measurement
of the water content of a gas ({\em i.e.}, humidity) is needed in
studies of the atmosphere related to weather and climate processes, but
obtaining accurate, fast, and reproducible measurements is challenging.
There are also industrial contexts where knowledge of water content is
important; an example is natural gas transportation where water can freeze
out as ice or form hydrates, both of which are undesirable and potentially
dangerous.

One technology that has been proposed for humidity metrology is
measurement of the static dielectric constant.  Because the molecules in
dry air (and, for the most part, in natural gas) are nonpolar, the presence
of highly polar water molecules can have a significant effect on the
dielectric constant, even at fairly low concentrations.  Apparatus for
measuring this effect has been developed by Cuccaro \textit{et
  al.}\cite{Cuccaro_2012} for water in air and nitrogen and by Gavioso
\textit{et al.}\cite{Gavioso_2014} for water in methane and natural gas.
To apply such measurements in metrology, it is necessary to have an
accurate expression for the contribution of water molecules to the static
dielectric constant.

For a low-density gas with a dipole moment, the well-known classical
relationship of the static dielectric constant $\varepsilon$ to the static
isotropic electronic polarizability $\alpha$ and the squared magnitude of the
molecular dipole moment $\mu^2$ of the gas constituent is given by the
Debye--Langevin modification of the Clausius--Mossotti expression (CMDL):
\begin{equation}\label{Eq:CM}
\frac{\varepsilon - 1}{\varepsilon + 2} = \frac{4 \pi}{3} \NA \rho \left ( \alpha + \frac{\mu^2}{3
  k_\mathrm{B} T} \right ),
\end{equation}
where $\NA$ is the Avogadro constant, $\rho$ is the molar density,
and $k_\mathrm{B}$ is the Boltzmann constant.  The first dielectric virial coefficient
$A_\varepsilon$ is defined as the low-density limit of the proportionality constant between the
density and the Clausius--Mossotti quotient:
\begin{equation}\label{Eq:Aeps}
A_\varepsilon \equiv \lim_{\rho \to 0} \frac{1}{\rho}\frac{\varepsilon - 1}{\varepsilon +2}.
\end{equation}
For low-density gas mixtures, the left-hand side of Eq.~(\ref{Eq:CM}) is
simply the sum of $\rho A_\varepsilon$ of each pure component.

One might think that Eq. (\ref{Eq:CM}) provides a simple route to the
calculation of $A_\varepsilon$ for water, and therefore to humidity
metrology.  The isotropic electronic polarizability of the water molecule in the
ground rovibrational state has been calculated by \textit{ab initio}
quantum mechanics,\cite{DPS_2018} and the result is in good agreement with
extrapolation of gas-phase refractivity measurements to the static
limit.\cite{Schoedel_2006,Egan_2022} The dipole moment of the H$_2$O
molecule in the rovibrational ground state has been measured to a relative
uncertainty of $5 \times 10^{-5}$.\cite{Shostak_1991}

There are, however, several ways in which the classical Eq. (\ref{Eq:CM})
is oversimplified and slightly inaccurate.  The purpose of this paper is to
provide a rigorous accounting of all effects on $A_\varepsilon$, all of
which involve quantum mechanics in one form or another.

First, the quantization of rotation means that the classical expression is
inexact.  A first-order correction for this quantum effect was first
derived for rigid linear molecules by MacRury and Steele,~\cite{MacRury74} and
was generalized to rigid nonlinear molecules by Gray \textit{et al.}\cite{GGJ2}

Second, molecules are not rigid objects, and hence the electronic
polarizability has a small temperature dependence. 
Only at $0$~K does it assume the ground-state value; at higher temperatures
other rovibrational states are occupied, each of which has a slightly
different electronic polarizability. 

Third, there is a similar effect for the dipole moment. The excited
rovibrational states populated at finite temperatures have somewhat
different dipole moments than the ground-state value. Additionally, the
proper quantum mechanical derivation of the expression for the first
dielectric virial coefficient shows that the generalizaton of
Eq.~(\ref{Eq:CM}) does not simply involve the average value of the dipole
moment.

In this paper, we will discuss in detail how Eq.~(\ref{Eq:CM}) is modified
by quantum mechanical effects involving the rovibrational states of the
water molecule. Additionally, we will provide the first fully quantum
calculation of the first dielectric coefficient $A_\varepsilon$ of
Eq.~(\ref{Eq:Aeps}), discussing in detail the various quantum effects that
one might expect for H$_2$O and its isotopologues due to their small
moments of inertia and molecular flexibility.

\section{The first dielectric virial coefficient of a quantum polar molecule}

The statistical derivation of the CMDL equation shows that the dielectric constant of a gas depends
on the derivative of the polarization density $P(F)$ in an external electric field $F$
evaluated at zero field,~\cite{Hill58,GGJ2,Garberoglio_2021}
\begin{equation}
  \frac{\varepsilon-1}{\varepsilon+2} = \frac{4 \pi}{3} \left. \frac{\dd
    P(F)}{\dd F}\right|_{F=0}
  \simeq A_\varepsilon \rho = 
  \frac{4\pi}{3} \left. \frac{\dd p(F)}{\dd F}\right|_{F=0} ~ \rho \NA, \label{eq:Aeps}
\end{equation}
where in the right-hand side we have expanded $P(F)$ to first order as a function of the molar
density $\rho$,~\cite{Garberoglio_2021} that is $P(F) = \rho \NA p(F)$ where $p(F)$ is the dipole
moment of an isolated molecule in the external field $F$. The quantity
$A_\varepsilon$ is the first dielectric virial coefficient.

The dipole moment of a molecule in thermodynamic equilibrium at temperature $T$ in an electric
field is given by the expectation value
\begin{eqnarray}
  p(F) &=& \frac{ \mathrm{Tr}\left[
    (\mbm \cdot \mbe + \mbe \cdot \mbal \cdot \mbe F) ~  \exp\left(-\beta H(F) \right)
      \right]}{\mathrm{Tr}\left[\ee^{-\beta H(F)}\right]} \label{eq:p} \\
    H(F) &=& H_0 - F ~ \mbm \cdot \mbe - \frac{F^2}{2} ~ \mbe \cdot
  \mbal \cdot \mbe, \label{eq:H}
\end{eqnarray}
where $H_0$ is the Hamiltonian of the free molecule, $\beta = (\kB T)^{-1}$, $\mbe$ is a unit vector
in the direction of the externally applied electric field of magnitude $F$, $\mbal$ is the molecular
electronic polarizability tensor, and $\mbm$ is the molecular dipole moment.  Notice that, in
general, both the dipole moment $\mbm$ and the electronic polarizability tensor $\mbal$ depend on
the specific molecular orientation and configuration which, in the case of water, is determined by 6
degrees of freedom. Usually, these are given by three Euler angles defining the relation between the
molecule-fixed and the the laboratory-fixed coordinate system,~\cite{Zare} the lengths $r_1$ and
$r_2$ of the two O--H bonds, and the angle $\theta$ between their directions.  The denominator in
Eq.~(\ref{eq:p}) is the partition function of a single molecule, and will be denoted by
$Q_1(\beta,F)$.

\subsection{Classical and semiclassical limit for rigid rotors}

The evaluation of the derivative at zero external field appearing in equations (\ref{eq:Aeps}),
(\ref{eq:p}), and (\ref{eq:H}) can be performed explicitly. Its derivation is reported in the
Supplementary Material and it shows that, in general, the first dielectric virial coefficient
has two contributions,
\begin{equation}
  A_\varepsilon = \frac{4 \pi}{3} \NA \left( \alpha_\mathrm{el} + \alpha_\mathrm{dip} \right),  
  \label{eq:Aeps_comp}
\end{equation}
where the first depends on the electronic polarizability surface $\mbal$, while the second depends on the
dipole-moment surface $\mbm$.
In the classical limit, one recovers Eq.~(\ref{Eq:CM}) 
\begin{equation}
  A_\varepsilon^{\mathrm{(cl)}} = \frac{4 \pi}{3} \NA \left\langle
  \aiso + \frac{\beta |\mbm|^2}{3} 
  \right\rangle,
  \label{eq:Acl}
\end{equation}
where the average $\langle \cdots \rangle$ is over the internal configurations of the molecule; this
is the result quoted in Ref.~\onlinecite{MacRury74} and shows how to interpret the quantities
$\alpha$ and $\mu^2$ in Eq.~(\ref{Eq:CM}), which are the values of the electronic polarizability
and the squared magnitude of the dipole moment at the assumed rigid configuration of water.
In a simple classical rigid model of water ({\em e.g.}, when the bond lengths
$r_1$ and $r_2$ and the angle $\theta$ are fixed at their ground-state
values~\cite{Czako:09}), the contribution to
$A_\varepsilon$ from the dipole moment is $A_\varepsilon^{(\mathrm{dip,cl})}
\sim 70.2$~cm${}^3$/mol at $T=300$~K, whereas that from the
electronic polarizability tensor is on the order of
$A_\varepsilon^{(\mathrm{el,cl})} \sim 3.7$~cm${}^3$/mol.

Semiclassical corrections to the dipolar contribution in Eq.~(\ref{eq:Acl}) have been derived in
Ref.~\onlinecite{GGJ2}, and are given by
\begin{multline}
  \frac{\Delta
    A_\varepsilon^{\mathrm{(dip,semi)}}}{A_\varepsilon^{\mathrm{(dip,cl)}}} = \\
  -\frac{\beta \hbar^2}{12 |\mbm|^2}
  \left(
  \frac{m_y^2 + m_z^2}{I_x} +
  \frac{m_z^2 + m_x^2}{I_y} +
  \frac{m_x^2 + m_y^2}{I_z}    
  \right),
  \label{eq:Asemi}  
\end{multline}
where $I_i$ are the principal moments of inertia of the molecule and $m_i$ are
the components of the dipole moment along the principal axis. For water at $300$~K, the
semiclassical correction of Eq.~(\ref{eq:Asemi}) is $A_\varepsilon^{\mathrm{(semi)}} =
-2.0$~cm${}^3$/mol, which in relative terms is roughly 3\% of the dipolar contribution.

\subsection{Quantum statistics}

In the general case, one has to be careful in performing the derivative with respect to $F$ in
Eq.~(\ref{eq:Aeps}), due to the presence of non-commuting operators.
Denoting by $|i\rangle$ a complete set of molecular rovibrational quantum states and by $E_i$ the
corresponding energies, the fully quantum expressions for the two terms in Eq.~(\ref{eq:Aeps_comp})
are (see the Supplementary Material for the derivation)
\begin{eqnarray}
\alpha_\mathrm{el}(T) &=& \frac{1}{Q_1(\beta,0)} \sum_{i}
\frac{\langle i | \mathrm{Tr}(\mbal) | i \rangle}{3}
~ \ee^{-\beta E_i},
\label{eq:alpha_el} \\
\alpha_\mathrm{dip}(T) &=&
\frac{1}{Q_1(\beta,0)} 
  \sum_{i \neq j}
  |\langle i | \mbm \cdot \mbe | j \rangle|^2
  \frac{\ee^{-\beta E_i} - \ee^{-\beta E_j}}{E_j-E_i}.
  \label{eq:alpha_dip}
\end{eqnarray}
The quantum mechanical formula for $\alpha_\mathrm{dip}$ was first derived by Illinger and
Smyth.~\cite{Illinger60.1}
In general, a molecular eigenstate $|i\rangle$ is characterized by a set of rotational and
vibrational quantum numbers and hence one can split the sum in Eq.~(\ref{eq:alpha_dip}) into two
contributions: the first corresponds to those $i \leftrightarrow j$ transitions where the
vibrational state changes, resulting in the so-called {\em vibrational polarizability}. The sum over
the remaining transitions, which have the same vibrational quantum numbers, is called the {\em
  rotational polarizability}.~\cite{Bishop90}

In the $T \to 0$ limit, the electronic polarizability (\ref{eq:alpha_el})
becomes just an average over the ground rovibrational state $|0\rangle$,
that is
\begin{equation}
  \alpha_\mathrm{el}^0 = \frac{1}{3}
  \langle 0 | \mathrm{Tr}(\mbal) |0 \rangle,
  \label{eq:alpha_el_0}
\end{equation}
whereas the dipolar polarizability becomes~\cite{Bishop90}
\begin{equation}
  \alpha_\mathrm{dip}^0 = 2 \sum_{j \neq 0}
  \frac{|\langle 0 | \mbm \cdot \mbe | j \rangle|^2}{E_j-E_0}
  = \frac{2}{3} \sum_{j \neq 0}
  \frac{|\langle 0 | \mbm | j \rangle|^2}{E_j-E_0},
\end{equation}
where we have used rotational invariance to substitute $\mbe \cdot \mbal
\cdot \mbe = \mathrm{Tr}(\mbal)/3$ and 
$|\langle i | \mbm \cdot \mbe | j\rangle|^2 =
|\langle i | \mbm | j\rangle|^2 / 3$.

In the high-temperature (classical) limit, $T \to \infty$, one can write
\begin{equation}
  \ee^{-\beta E_i} - \ee^{-\beta E_j}
  \simeq \ee^{-\beta E_i} \beta (E_j - E_i),
\end{equation}
and Eq.~(\ref{eq:alpha_dip}) becomes 
\begin{equation}
  \alpha_\mathrm{dip}^\mathrm{class}(T) = \frac{\beta}{3}
  \frac{\sum_i \langle i | |\mbm|^2 | i \rangle \ee^{-\beta E_i}}{Q_1(\beta,0)},
\end{equation}
which can be seen as a generalization of the classical expression
(\ref{eq:Acl}) where the square of the permanent dipole is replaced by
its quantum thermal average, which depends only on the diagonal matrix
elements of the corresponding operator. In general, however, the dipolar
contribution to the first dielectric virial coefficient depends on all the
matrix elements of the dipole moment, as evidenced by
Eq.~(\ref{eq:alpha_dip}).

\section{Dipolar polarizability from spectroscopy}

The dipolar polarizability (\ref{eq:alpha_dip}) depends on the squared matrix
elements of the dipole operator $\mbm$, which is the same operator that
determines the Einstein coefficient associated with an electromagnetic
transition between states of energy $E_i$ and $E_j$.~\cite{Hilborn82}
These coefficients, or equivalently the line intensities, of several
thousands of transitions are available in the HITRAN2020
database~\cite{HITRAN2020} for many water isotopologues, as well as other
molecules. This paves the way to an experimental determination of the
dipolar polarizability from spectroscopic data.

To this end, let us rewrite Eq.~(\ref{eq:alpha_dip}) according to the following considerations: first
of all, we notice that the quantity to be summed is invariant under the exchange $i \leftrightarrow
j$. Hence, the dipolar polarizability is given by twice the sum performed over those states for
which $E_j > E_i$. Second, there might be degeneracies among the energy levels, so let us denote
by $d_i$ the number of states having energy $E_i$; in the case of water $d_i = (2J+1) g_i$ where $J$
is the angular momentum of state $i$ and $g_i$ is its nuclear-spin degeneracy. A general state
$|i\rangle$ can then be labeled by its energy $E_i$ and an integer number $\xi$ running between $1$
and $d_i$.  Finally, let us define the average dipole matrix element squared between levels with
energy $E_i$ and $E_j$ as
\begin{equation}
  M_{ij} = \frac{1}{d_i d_j}
  \sum_{\xi=1}^{d_i} \sum_{\eta=1}^{d_j}
  |\langle E_i, \xi | \mbm \cdot \mbe | E_j, \eta \rangle|^2,
\end{equation}
and the transition frequency $\omega_{ij} = (E_j - E_i)/\hbar$.
With these definitions, we can write the dipolar polarizability as
\begin{equation}
  \alpha_\mathrm{dip}(T) = \frac{2}{Q_1(\beta)} \sum_{E_j > E_i}
  \frac{d_i d_j M_{ij}}{\hbar \omega_{ij}} \ee^{-\beta E_i} \left(1 - \ee^{-\beta \hbar \omega_{ij}}\right).
\end{equation}

On the other hand, the spectral line intensities $S_{ij}$ for the
transitions between molecular levels of energy $E_i$ and $E_j$ reported in the HITRAN2020
database are given by (in our notation)~\cite{hitran_lte,hitran_a} 
\begin{equation}
  S_{ij} = \frac{2 \pi}{\hbar c} I_A \omega_{ij} d_i d_j M_{ij} \frac{\ee^{-\beta_\mathrm{R} E_i}
    (1-\ee^{-\beta_\mathrm{R} \hbar \omega_{ij}})}{Q_1(\beta_\mathrm{R})},
  \label{eq:HS}
\end{equation}
where $I_A$ is the isotopic abundance of the species under consideration,
$\beta_\mathrm{R} = (\kB T_\mathrm{R})^{-1}$, and $T_\mathrm{R} = 296$~K is a reference temperature. 
Consequently, one can write
\begin{eqnarray}
  \alpha_\mathrm{dip}(T) &=& \frac{c}{\pi ~ I_A} \sum_{E_j > E_i}
  \frac{S_{ij}}{\omega_{ij}^2}
  \frac{Q_1(\beta_\mathrm{R})}  {\ee^{-\beta_\mathrm{R} E_i} (1-\ee^{-\beta_\mathrm{R} \hbar
      \omega_{ij}})}
  \times \nonumber \\
  & &
  \frac{\ee^{-\beta E_i} (1-\ee^{-\beta \hbar \omega_{ij}})}{Q_1(\beta)},
  \label{eq:H_alpha}
\end{eqnarray}
which enables the calculation of the dipolar polarizability based on the quantities $S_{ij}$,
$\omega_{ij}$, $E_i$, $Q_1(\beta)$, that are all reported the HITRAN2020 database. Since the
HITRAN2020 database also reports quantum numbers for the upper and lower states of the transitions,
the calculation of the vibrational and rotational contributions to the dipolar polarizability
using Eq.~(\ref{eq:H_alpha}) is straightforward.

The HITRAN2020 database also reports the uncertainty associated with the line intensities $S_{ij}$;
we have used this information to propagate it to the uncertainty of $A_{\varepsilon}^\mathrm{(dip)}$
using Eq.~(\ref{eq:H_alpha}). In those cases where HITRAN2020 reported a range for the uncertainty,
we made the conservative choice of taking the highest value and considering it to be a standard
uncertainty.

\section{{\em Ab initio} calculation of the first dielectric virial coefficient}
\label{sec:theory}

The quantities $\alpha_\mathrm{el}(T)$ and $\alpha_\mathrm{dip}(T)$ can be computed from the
knowledge of the intramolecular potential, the trace of the polarizability $\alpha(\mbX) =
\mathrm{Tr}(\mbal)$, and the dipole moment $\mbm(\mbX)$, where $\mbX$ denotes a set of
intramolecular coordinates ({\em e.g.}, the two bond lengths and the HOH angle). All of these
quantities are available from {\em ab initio} electronic structure calculations.  In particular, we
used the recent PES15K~\cite{PES15K} potential-energy surface to compute the intramolecular
potential, the dipole-moment surface $\mbm(\mbX)$ CKAPTEN from Ref.~\onlinecite{DMS_2018}, and the
isotropic electronic polarizability $\alpha(\mbX)$ from Ref.~\onlinecite{DPS_2018}.  In
Ref.~\onlinecite{DPS_2018}, the authors provide a fitting function based on extensive calculations of the polarizability
for many molecular configurations (the DPS-H${}_2$O-3K database), solving the electronic structure
at the CCSDT/daTZ level of theory. In addition, they perfomed a few evaluations of the
polarizability surface at a more accurate level of theory (CCSDT/daQZ + $\Delta
\alpha_\mathrm{basis}$ + $\Delta \alpha_\mathrm{core}$). The values obtained in this case are much
more accurate for configurations close to the equilibrium configuration of water. Therefore, we
slightly changed their parametrization of the polarizability surface (from their Table
II~\cite{DPS_2018}) in order to obtain more accurate values around the equilibrium geometry. In
practice, we changed the first line of their Table II (corresponding to $(ijk) = (000)$) to the
values reported in their Table III, that is $\alpha_{xx} = 9.8744$, $\alpha_{yy} = 9.2233$, and
$\alpha_{zz} = 9.5190$ (in atomic units). This corresponds to a rigid shift of the
isotropic part of the CKAPTEN polarizability surface by $\Delta \alpha_\mathrm{iso} = -0.039133$
atomic units, which in relative terms is approximately 0.4\%.

There are two main ways to compute the first dielectric virial coefficient for water. The first is
to diagonalize the Schr{\"o}dinger equation for the nuclear motion of the water molecule, 
compute the temperature-dependent electronic polarizability of Eq.~(\ref{eq:alpha_el}) and the
dipolar polarizability of Eq.~(\ref{eq:alpha_dip}), and obtain the first dielectic virial coefficient
from Eq.~(\ref{eq:Aeps_comp}).
Although there are efficient approaches to solve the quantum-mechanical three-body problem and
obtain the rovibrational eigenstates of water molecules, a complete diagonalization of the
intramolecular Hamiltonian becomes progressively more difficult for molecules with a larger number of
atoms. We will show in the following how the two temperature-dependent contributions to
$A_\varepsilon(T)$ can be obtained from a path-integral approach, which has a much more favorable
computational scaling for polyatomic molecules.

\subsection{The discrete variable representation approach}
\label{sec:DVR}

In the case of water or other triatomic molecules, it is convenient to use the form of the
three-body Hamiltonian in the molecule-fixed frame derived by Sutcliffe and Tennyson using Jacobi
coordinates.~\cite{Sutcliffe87} In this approach, one obtains, for each value of the total molecular
angular momentum $J$, a vibrational Hamiltonian that depends on three coordinates: the moduli of the
two Jacobi vectors, $R_1$ and $R_2$, and the value of the angle between them, $\Theta$. The full
rovibrational spectrum of water can be obtained by diagonalizing the vibrational Hamiltonian for
progressively larger values of the molecular angular momentum $J$.  The Hamiltonian is conveniently
written using the so-called Discrete Variable Representation (DVR),~\cite{Light2k, Szalay93, Baye15}
which has been used in many investigations of water properties.~\cite{DVR3D,Czako_DVR}

In practice, the rovibrational wavefunction is written as
\begin{multline}
  \Psi_{J \nu m I}(q,\Omega,I) = \\
    \sqrt{\frac{2J+1}{8 \pi^2}} \sum_{k=-J}^J
    \psi_{J \nu}(q,k) D^J_{mk}(\Omega) \chi({\cal I}_{J \nu}, I),
\label{eq:water_psi}
\end{multline}
where $q = (R_1, R_2, \Theta)$ denotes the molecular vibrational coordinates, $\psi_{J \nu}(q,k)$
are the eigenfunctions of the rovibrational Hamiltonian in the molecule-fixed frame for a given
value of $J$,
$D^J_{mk}(\Omega)$ are Wigner rotation matrices from the molecule-fixed to the laboratory-fixed
frames, and $\chi({\cal I}_{J \nu}, I)$ denotes the wavefunction of nuclear spins, with total spin
${\cal I}_{J \nu}$ and projection $I$ along the laboratory $Z$ axis. In the case of water molecules
with two identical hydrogen isotopes, the overall wavefunction in Eq.~(\ref{eq:water_psi}) must have
a specific symmetry upon exchange of the two hydrogens, reflecting their fermionic (H or T,
antisymmetric wavefunction) or bosonic (D, symmetric wavefunction) nature.  The need of a
well-defined exchange symmetry for the total wavefunction implies that the nuclear-spin state of the
two hydrogens depends on exchange symmetry of the rovibrational state $\psi_{J \nu}$.

In the case of two hydrogen atoms (with nuclear spin $1/2$), molecular configurations have either
${\cal I}_{J \nu} = 0$ (para-water) or ${\cal I}_{J \nu} = 1$ (ortho-water), with multiplicities
$g_\mathrm{para} = 1$ and $g_\mathrm{ortho} = 3$, respectively.  In the case of two deuterium atoms
(with nuclear spin $1$), the ortho spin isomer has total spin ${\cal I}$ which is either 0 or 2
(that is, degeneracy $g_\mathrm{ortho} = 6$) or total spin $1$, with degeneracy $g_\mathrm{para} =
3$.  For the sake of completeness, we recall that nuclear spin degeneracies can also come from the
oxygen spin, although in this case they do not depend on the rovibrational state. This is
particularly relevant for the ${}^{17}$O oxygen isotope, which has ${\cal I} = 5/2$. The other
oxygen isotopes have ${\cal I}=0$.

Notice that the energy levels on the molecular Hamiltonian depend only on
the quantum numbers $J$ (the total angular momentum), $k$ (the projection of the angular momentum in
the molecule-fixed $z$ axis), and $\nu$ (that labels
the vibrational states obtained for given values of $J$ and $k$). Rotational
invariance implies a $2J+1$ degeneracy on the label $m$ in Eq.~(\ref{eq:water_psi}).

Evaluating the symmetry upon exchange of the wavefunctions $\psi_{J \nu k}(q)$, and hence the
degeneracy, is a non-trivial procedure,~\cite{Edit10} but this information is already included in the
HITRAN2020 database. Given the accuracy of our calculated energy levels, we assigned degeneracies by
looking at that of the closest energy level, for a given $J$, among the states present in
HITRAN2020.  Using the DVR approach, one can compute the electronic and dipolar contribution to the
polarizability directly from the quantum mechanical expressions of Eqs.~(\ref{eq:alpha_el}) and
(\ref{eq:alpha_dip}), respectively.  In the first case, one obtains
\begin{equation}
  \alpha_\mathrm{el}(T) = \sum_{J \nu}
  \frac{g_{J\nu} (2J+1) \ee^{-\beta E_{J \nu}}}{3 ~ Q_1(\beta,0)}
  \left\langle
  \psi_{J \nu} | \mathrm{Tr}(\mbal) | \psi_{J \nu}
  \right\rangle,
  \label{eq:ael_dvr}
\end{equation}
where $g_{J\nu}$ is the degeneracy of the given rovibrational state
and, in the DVR representation, one has
\begin{equation}
\left\langle
  \psi_{J \nu} | \mathrm{Tr}(\mbal) | \psi_{J \nu}
  \right\rangle = \sum_{q,k} \left| \psi_{J \nu k}(q) \right|^2 \mathrm{Tr}(\mbal(q)),
\end{equation}
since the trace of the polarizability tensor $\mbal(q)$ is a scalar
quantity and a diagonal operator in the DVR representation.
Notice that in the case of a rigid model of water, the electronic polarizability (\ref{eq:ael_dvr})
does not depend on temperature.

The quantum mechanical expression for $\alpha_\mathrm{dip}(T)$ is more complicated,
because one needs to evaluate the matrix element of the components of the dipole moment $\mbm$ in
the laboratory-fixed frame using wavefunctions defined in the molecule-fixed
frame.~\cite{LeSueur92,Zare,DVR3D} Its derivation is discussed in Appendix~\ref{sec:alpha_dip}, for
both rigid and flexible molecular models.

\subsection{The path-integral approach}

A first-principles evaluation of the first dielectric virial coefficient from
Eqs.~(\ref{eq:Aeps})--(\ref{eq:H}) can also be performed using the
path-integral formulation of quantum statistical mechanics.~\cite{FH} The main advantage of this
method is that it works directly in the Cartesian coordinates of all the atoms, and hence one does
not need the analytically complicated procedure of separating the center-of-mass, rotational, and
vibrational motion as needed for solving the Schr{\"o}dinger equation.

\subsubsection{Quantum rigid rotors}
\label{sec:quantum_rigid}

In the case of a rigid rotor, $\aiso$ is a constant and hence easily evaluated. Taking the
derivative with respect to $F$ of the first term in Eq.~(\ref{eq:Aeps}) requires some care, because
the dipole moment direction in the laboratory reference frame, $\mbm \cdot \mbe$ (which will be
denoted by $m_Z$ in the following), does not commute with $H_0$ which, in this case, is the
Hamiltonian of a quantum asymmetric rigid rotor.  It is convenient to specialize the trace as an
integration over all possible orientations of the molecule -- which will be denoted by $\Omega_1$ --
and at the same time use Trotter's identity to write
\begin{equation}
  \ee^{
    -\beta H_0 + \beta \mbm \cdot \mbe F
    }  \sim 
    \left( \ee^{-\beta H_0/P} \ee^{\beta (\mbm \cdot \mbe) F /P} \right)^P,
\label{eq:Trotter}
\end{equation}
which becomes an equality for a sufficiently large $P$. Inserting
$P-1$ completeness relations of the form 
$$
1 = \int |\Omega_k\rangle\langle \Omega_k| ~ \dd \Omega_k,
$$
between the $P$ products in Eq.~(\ref{eq:Trotter}), one ends up with the
expression
\begin{widetext}
\begin{eqnarray}
  A_\varepsilon &=& \frac{4\pi}{3}\aiso +
  \frac{4\pi}{3 Q_1} \frac{\partial}{\partial F} \int
  m_Z(\Omega_1) \prod_{k=1}^P \langle \Omega_k | \ee^{-\beta H_0/P} |
  \Omega_{k+1} \rangle ~ \ee^{\beta m_Z(\Omega_{k+1}) F/P}
  \prod_{k=1}^P \dd \Omega_k \label{eq:qAeps0} \\
  &=& \frac{4\pi}{3}\aiso + \frac{4\pi}{3 Q_1} \int
  m_Z(\Omega_1) \prod_{k=1}^P \langle \Omega_k | \ee^{-\beta H_0/P} |
  \Omega_{k+1} \rangle ~ \left[ \frac{\beta}{P} \sum_{k=1}^P m_Z(\Omega_{k})
    \right] ~ 
  \prod_{k=1}^P \dd \Omega_k \label{eq:qAeps1} \\
  &=& \frac{4\pi}{3}\aiso + \frac{4 \pi \beta}{3 Q_1} \int
  \prod_{k=1}^P \langle \Omega_k | \ee^{-\beta H_0/P} | \Omega_{k+1} \rangle
  ~ \left[ \frac{1}{P} \sum_{k=1}^P m_Z(\Omega_{k})
    \right]^2 ~ 
  \prod_{k=1}^P  \dd \Omega_k, \label{eq:qAeps2} \\
  &=& \frac{4\pi}{3}\aiso + \frac{4 \pi \beta}{9} \left\langle |\overline{\mbm}|^2 \right\rangle,
\end{eqnarray}  
\end{widetext}
where we have defined $\overline{\mbm} = \sum_k \mbm(\Omega_k) / P$. In
passing from (\ref{eq:qAeps1}) to (\ref{eq:qAeps2}), we have used 
the fact that singling out $m_Z(\Omega_1)$ in (\ref{eq:qAeps0}) is an
arbitrary choice that we have averaged upon. Notice also that we have
defined $\Omega_{P+1} = \Omega_1$.
The evaluation of the matrix elements $\langle \Omega_k | \ee^{-\beta H_0/P}
| \Omega_{k+1} \rangle$ for a rigid-rotor molecule is discussed in
Refs.~\onlinecite{Noya11_1,Noya11_2}.

\subsubsection{Quantum flexible molecules}

In dealing with flexible models of water, it is convenient to denote by
$\mbX$ the set of all the coordinates of the three atoms. In this case,
$\Aeps$ is still given by Eq.~(\ref{eq:Aeps}) where the Hamiltonian $H_0$
includes the kinetic energy of the three atoms, $K$, and the intramolecular
potential $V_\mathrm{int}(\mbX)$. Note that both the dipole moment $\mbm$
and the electronic polarizability tensor $\mbal$ depend on $\mbX$.  The path-integral
evaluation of quantities related to flexible molecules has been described
in detail in Ref.~\onlinecite{Garberoglio18}. The main result is that one
can map the quantum partition function to a classical partition function
where each atom is represented by a ring polymer with $P$ beads. This
approach provides an explicit expression of the interaction between
consecutive beads (that turns out to be an harmonic potential) and the
interaction among the ring polymers, which depends on
$V_\mathrm{int}(\mbX)$. In the case of H${}_2$O or D${}_2$O, one has to
consider the indistinguishability between the hydrogen or deuterium atoms
within a molecule. Although this can be done in the path-integral approach,
it can be shown that exchange effects are important only for temperatures
$T \lesssim 50$~K~\cite{Garberoglio18} and therefore they will be neglected in
this paper.

With a derivation analogous to that performed in the case of rigid rotors,
the path-integral expression for $\Aeps$ in the case of a flexible model
turns out to be
\begin{eqnarray}
  \Aeps &=& \frac{4 \pi}{3} \int 
  \left(
  \overline{\aiso} + \frac{\beta}{3} \left| \overline{\mbm} \right|^2
  \right) \times \nonumber \\
  & &
  {\cal P}(\mbX_1, \ldots, \mbX_P) \prod_{k=1}^P \dd \mbX_k      
    \label{eq:Aeps_flexi} \\
    \overline{\aiso} &=& \frac{1}{P} \sum_{k=1}^P \aiso(\mbX_k) \\
    \overline{\mbm} &=& \frac{1}{P} \sum_{k=1}^P \mbm(\mbX_k) \\
    {\cal P}(\mbX_1, \ldots, \mbX_P) &=&
    \frac{1}{Q_1}
    \prod_{k=1}^P \langle \mbX_k | \ee^{-\beta  T} | \mbX_{k+1} \rangle
    \times \nonumber \\
    & & 
    \exp\left(-\frac{\beta}{P} \sum_{k=1}^P V_\mathrm{int}(\mbX_k) \right),
    \label{eq:molprob}
\end{eqnarray}
where ${\cal P}(\cdots)$ is the probability of finding a specific
molecular configuration in the path-integral representation.

The first term in Eq.~(\ref{eq:Aeps_flexi}) is the path-integral
representation of the electronic contribution to the
electronic polarizability of
Eq.~(\ref{eq:alpha_el}),~\cite{Bishop86,Bishop90} whereas the second term
corresponds to the temperature-dependent dipolar polarizability,
Eq.~(\ref{eq:alpha_dip}).

\section{Numerical implementation}

\subsection{Quantum rigid rotors}

\subsubsection{Path-integral Monte Carlo}

In the case of quantum rigid rotors, our path-integral Monte Carlo (PIMC) simulation followed the
procedure outlined in Refs.~\onlinecite{Noya11_1, Noya11_2}.
We considered the underlying rigid model of water by using the ground-state geometric
parameters in Ref.~\onlinecite{Czako:09} ($r_1 = r_2 = 0.97565$~\AA\ and $\theta =
104.43^\circ$ for H${}_2$O; $r_1 = r_2 = 0.97077$~\AA\ and $\theta =
104.408^\circ$ for D${}_2$O), and those developed in
Ref.~\onlinecite{HDO-geo} for HD${}^{16}$O
($r_1=0.97126$~\AA, $r_2=0.96947$~\AA, $\theta = 104.01^\circ$), where
$r_1$ and $r_2$ are the two bondlengths and $\theta$ is the bending angle
of the water molecule.

We found convergence of the values of the dipolar polarizability using $P =
\mathrm{nint}(5 + 700~\mathrm{K}/T)$, where $T$ is the temperature and
$\mathrm{nint}(x)$ denotes the nearest integer to the number $x$.  We
performed $100000$ Monte Carlo moves (corresponding to an attempted rotation
of a molecule in one of the imaginary-time slices) for equilibration, and we
then averaged the values of the dipole moment on $256$ independent runs each one sampling $1000$
configurations, separated by $50$ Monte Carlo moves. 

\subsubsection{Hamiltonian diagonalization}
\label{sec:rigid_hamiltonian}

In the case of a rigid water model, the coordinates $q$ in Eq.~(\ref{eq:water_psi}) are fixed, and
the wavefunctions are given by
\begin{multline}
  \Psi_{J \nu m I}(\Omega,I) = \\
    \sqrt{\frac{2J+1}{8 \pi^2}} \sum_{k=-J}^J
    \psi_{J \nu k} D^J_{mk}(\Omega) \chi({\cal I}_{J \nu}, I).
\label{eq:water_psi_rigid}
\end{multline}
The quantities $\psi_{J \nu k}$ can
be obtained by diagonalization of the rigid-rotor Hamiltonian in the molecule-fixed
frame, that is
\begin{equation}
  H_\mathrm{R} = \frac{\hbar^2}{2} \left(
  \frac{J_a^2}{I_a} + \frac{J_b^2}{I_b} + \frac{J_c^2}{I_c}
  \right),
  \label{eq:HR}
\end{equation}
where $J_i$ are the angular momentum operators in the molecular frame, and $I_a \leq I_b \leq I_c$
are the principal inertia moments of the molecule.~\cite{Zare} The index $\nu$ of the rigid-rotor
eigenfunction labels the rotational states of $H_\mathrm{R}$ for a given total angular momentum $J$
and therefore is an integer between $-J$ and $J$ (inclusive).

In the case of rigid asymmetric rotors, such as water, the eigenfunctions for a given total angular
momentum $J$ are usually labeled with the notation $J_{K_a K_c}$ where $K_a$ and $K_c$ denote the
absolute value of the projection $K$ of the angular momentum in the molecular frame in the limits
$I_b \to I_a < I_c$ (oblate symmetrical top) and $I_a < I_b \to I_c$ (prolate symmetrical top). In
both of these limits, $K$ is a good quantum number and states with the same value of $|K|$ are
degenerate. The nuclear-spin degeneracy is related to the value of $K_a + K_c$: for H${}_2{}^{16}$O
one has that $g_{J\nu} = 1$ if $K_a + K_c$ is even (para-H${}_2{}^{16}$O), and $g_{J\nu}=3$ if $K_a + K_c$
is odd (ortho-H${}_2{}^{16}$O), while for D${}_2{}^{16}$O one has that $g_{J\nu} = 6$ if $K_a + K_c$
is even (ortho-D${}_2{}^{16}$O), and $g_{J \nu}=3$ if $K_a + K_c$ is odd (para-D${}_2{}^{16}$O).

We performed rigid-model calculations by numerical diagonalization of the Hamiltonian of
Eq.~(\ref{eq:HR}) up to $J=40$.

\subsection{Quantum flexible molecules}

\subsubsection{Path-integral Monte Carlo}
We checked convergence of various components of the first dielectric virial
coefficient with the number of beads $P$, and we found that it was reached
using $P = 70 + \mathrm{nint}(10000~\mathrm{K}/T)$ for every possible
isotopologue.
In order to perform PIMC calculations, we developed code based on the hybrid Monte Carlo
method~\cite{hmc2} to sample molecular configurations according to the probability of
Eq.~(\ref{eq:molprob}). We used $200000$ steps for equilibration and then
evaluated average values using at least 512 independent runs of $600000$ steps,
sampling observables every $1000$ steps.

\subsubsection{Discrete Variable Representation}
We developed a DVR code in house. $N_\mathrm{r} = 28$ and $N_\theta = 28$ basis set points
for the radial and angular coordinates, respectively, were sufficient to
obtain rovibrational energies within one part in $10^6$ from the reference
values calculated in Ref.~\onlinecite{PES15K} for $J=0$.
Limitations in memory and CPU time available prevented us from computing
rovibrational states at angular momenta higher than $J=19$. Comparison
between the partition functions obtained with our approach and the
reference ones available in the HITRAN2020 database~\cite{HITRAN2020} showed
that this limitation results in a systematic uncertainty of at most 0.6\%
for H${}_2$O at the highest temperature at which we have
used this approach, $T=500$~K.
As a further check of our implementation, we computed the average values of the
O--H bond-length and H-O-H angle for H${}_2$O and D${}_2$O in the $J=0$
subspace, and they were found to agree with the results of
Ref.~\onlinecite{Czako:09} to within one part in $10^5$.

\section{Results and discussion for H${}_2{}^{16}$O}

In this section, we will discuss in detail our results relative to the most common water
isotopologue, H${}_2{}^{16}$O. The results for two other isotopologues, HD${}^{16}$O and
D${}_2{}^{16}$O, are reported in the Supplementary Material.

\subsection{Electronic polarizability}

\begin{figure}
  \center\includegraphics[width=0.9\linewidth]{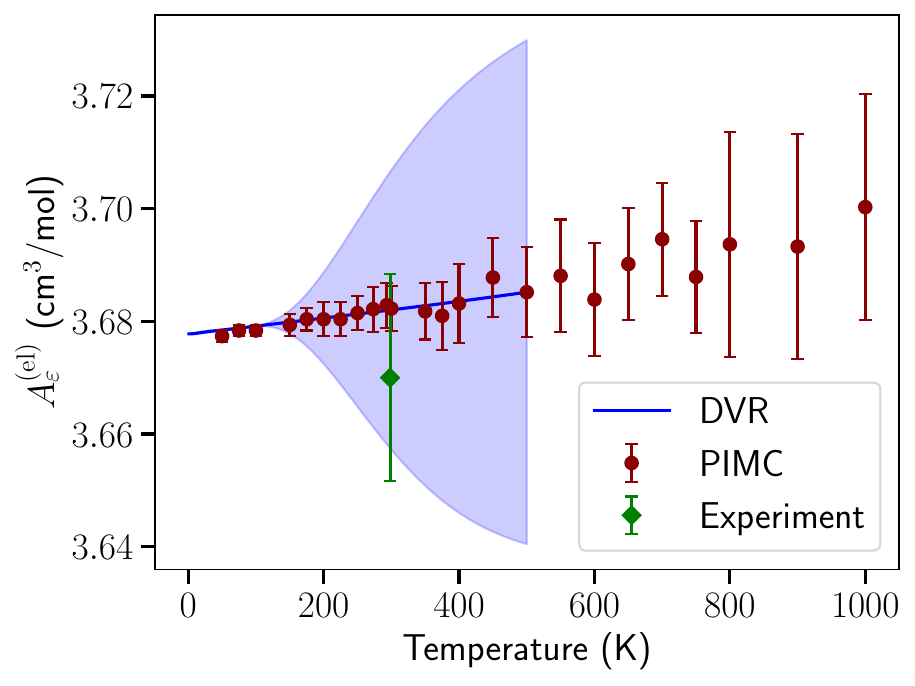}    
  \caption{The electronic polarizability contribution to $A_\varepsilon$,
    $A_{\varepsilon}^\mathrm{(el)}$, of H${}_2{}^{16}$O as a function of temperature, obtained from
    computer simulations.
    The blue line is the result of our DVR calculations, together with an estimated uncertainty
    coming from the limited number of angular momenta that have been considered, reported as a blue
    area.
    Filled circles are the results of the path-integral simulations and the diamond is the
    experimental result of Ref.~\onlinecite{Egan_2022}, adjusted to zero frequency using the dipole
    oscillator strength sums of Zeiss and Meath.~\cite{Zeiss77} All uncertainties are reported at
    the $k=2$ coverage value and do not include the propagation of the unknown uncertainty of the
    water electronic polarizability surface.}
  \label{fig:Aeps_pol_H2O}
\end{figure}

Let us begin our discussion by considering the electronic polarizability contribution to the first
virial coefficient, that is $\alpha_\mathrm{el}(T)$ defined in Eq.~(\ref{eq:alpha_el}).  As a first
approximation, {\em e.g.}, by using a classical rigid model for the water molecule, it can be
calculated as the value of the electronic polarizability surface at the average geometric parameters
(distances and angles) of the molecular ground state;~\cite{Czako:09} in this case one obtains the
value $A_{\varepsilon}^\mathrm{(el)} = 3.659$~cm${}^3$/mol.  However, one should in fact average the
value of the electronic polarizability surface over the distribution of configurations sampled by
the ground state of the water molecule. This procedure, performed using the DVR ground-state
wavefunction, provides $A_{\varepsilon}^\mathrm{(el)} = 3.678$~cm${}^3$/mol and shows that a simple
classical model underestimates the actual value by $-0.5\%$ in the $T \to 0$ limit.  Comparing the
ground-state averaged value of the electronic polarizability with its value at the geometry where
the intramolecular potential has its minimum ($r_1 = r_2 = 0.9579$~\AA\ and $\theta =
104.512^\circ$, corresponding to $A_{\varepsilon}^\mathrm{(el,min)} = 3.566$~cm${}^3$/mol),
one immediately obtains an estimate of the so-called vibrational contribution to the isotropic
electronic polarizability, which is due to the zero-point (ZP) motion of the water molecule. In our
case we obtain $\alpha_\mathrm{el}^\mathrm{ZP} = 0.2986$~a.u., in very good agreement with analogous
calculations in the literature.~\cite{Avila05,DPS_2018}

The actual values for the electronic polarizability contribution to the first
virial coefficient of water are reported in Table~\ref{tab:Aeps_pol_H2O} and
Fig.~\ref{fig:Aeps_pol_H2O}. We notice that our two calculation methods,
PIMC and DVR, agree within mutual uncertainties at all the
temperatures investigated in the present work. The PIMC approach is relatively
noisy, but the DVR calculations clearly show that the electronic
polarizability is a slightly increasing function of the temperature,
exceeding its ground-state value by $\sim 0.04\%$ at $T=100$~K.
Unfortunately, we are not aware of any characterization of the uncertainty
associated with the electronic polarizability surface~\cite{DPS_2018} and hence we
cannot provide a precise quantitative assessment of the uncertainty of
$A_{\varepsilon}^\mathrm{(el)}$.

\begin{table}[h]
  \caption{\label{tab:Aeps_pol_H2O}
    The values of the electronic polarizability contribution to $A_\varepsilon$
    for H${}_2{}^{16}$O. All of the uncertainties are reported at $k=2$ coverage and do
    not include the propagation of the unknown
    uncertainty of the water electronic polarizability surface.}
  \begin{ruledtabular}
  \begin{tabular}{dcc}
    \multicolumn{1}{c}{Temperature} & $A_{\varepsilon}^\mathrm{(el)}$ (path-integral) &
    $A_{\varepsilon}^\mathrm{(el)}$ (DVR) \\
      \multicolumn{1}{c}{(K)} & (cm${}^3$/mol) & (cm${}^3$/mol) \\
      \hline
1   &$      –       $&$ 3.67777 \pm 0.00001 $ \\
10  &$      –       $&$ 3.67786 \pm 0.00007 $ \\
25  &$      –       $&$ 3.67814 \pm 0.00016 $ \\
50  &$  3.677   \pm 0.001   $&$ 3.67845 \pm 0.00014 $ \\
75  &$  3.678   \pm 0.001   $&$ 3.67880 \pm 0.00013 $ \\
100 &$  3.678   \pm 0.001   $&$ 3.67915 \pm 0.00002 $ \\
125 &$  3.678   \pm 0.001   $&$ 3.6795  \pm 0.0010  $ \\
150 &$  3.679   \pm 0.002   $&$ 3.680   \pm 0.002   $ \\
175 &$  3.680   \pm 0.002   $&$ 3.680   \pm 0.005   $ \\
200 &$  3.680   \pm 0.003   $&$ 3.681   \pm 0.008   $ \\
225 &$  3.680   \pm 0.003   $&$ 3.681   \pm 0.012   $ \\
250 &$  3.681   \pm 0.003   $&$ 3.681   \pm 0.016   $ \\
273.16  &$  3.682   \pm 0.004   $&$ 3.68    \pm 0.02    $ \\
293.15  &$  3.683   \pm 0.004   $&$ 3.68    \pm 0.02    $ \\
300 &$  3.682   \pm 0.004   $&$ 3.68    \pm 0.02    $ \\
325 &$  3.682   \pm 0.004   $&$ 3.68    \pm 0.02    $ \\
350 &$  3.682   \pm 0.005   $&$ 3.68    \pm 0.03    $ \\
375 &$  3.681   \pm 0.006   $&$ 3.68    \pm 0.03    $ \\
400 &$  3.683   \pm 0.007   $&$ 3.68    \pm 0.04    $ \\
450 &$  3.688   \pm 0.007   $&$ 3.68    \pm 0.04    $ \\
500 &$  3.685   \pm 0.007   $&$ 3.69    \pm 0.04    $ \\
550 &$  3.688   \pm 0.008   $&$     –       $ \\
600 &$  3.68    \pm 0.01    $&$     –       $ \\
650 &$  3.69    \pm 0.01    $&$     –       $ \\
700 &$  3.69    \pm 0.01    $&$     –       $ \\
750 &$  3.69    \pm 0.01    $&$     –       $ \\
800 &$  3.69    \pm 0.02    $&$     –       $ \\
900 &$  3.69    \pm 0.02    $&$     –       $ \\
1000    &$  3.70    \pm 0.02    $&$     –       $ \\
1250    &$  3.70    \pm 0.03    $&$     –       $ \\
1500    &$  3.72    \pm 0.04    $&$     –       $ \\
1750    &$  3.71    \pm 0.04    $&$     –       $ \\
2000    &$  3.74    \pm 0.06    $&$     –       $
  \end{tabular}
  \end{ruledtabular}
\end{table}  

These results for $A_{\varepsilon}^\mathrm{(el)}$ can be compared to optical measurements of the
refractivity of water vapor. The two most precise studies of this quantity were performed by
Sch{\"o}del {\em et al.}~\cite{Schoedel_2006} and by Egan.~\cite{Egan_2022} In order to compare
with our static values of $A_{\varepsilon}^\mathrm{(el)}$, the results must be adjusted to zero
frequency; this can be done with the dipole oscillator strength sums of Zeiss and
Meath.~\cite{Zeiss77}  The resulting values (at 293.15~K in both cases) are approximately
3.67~cm${}^3$/mol, with expanded uncertainties on the order of 0.5\%.  This is in reasonable
agreement with the values calculated here, although the comparison suggests that the polarizability
surface of Lao {\em et al.}~\cite{DPS_2018} may yield polarizabilities that are slightly too large
(another possibility is inaccuracy in the dipole oscillator strengths of Ref.~\onlinecite{Zeiss77}).

We also developed a correlation for $A_{\varepsilon}^\mathrm{(el)}$ by fitting the numerical data
using a function of the form
\begin{equation}
  A_{\varepsilon}^\mathrm{(el)}(T) = a + \frac{b T}{1 + \exp\left[-(T-c)/T_0 \right]}.
  \label{eq:Ael_corr}
\end{equation}
The values of the fitted parameters in Eq.~(\ref{eq:Ael_corr}) for the water isotopologues reported in
this study are reported in Table~\ref{tab:Ael_corr}. The correlation reproduces the values reported
in Table~\ref{tab:Aeps_pol_H2O} within the assigned uncertainties in the temperature range between 1
and 2000~K.

{\squeezetable
\begin{table}
  \caption{Values of the parameters in Eq.~(\ref{eq:Ael_corr}) for the water isotopologues
    studied in this paper. The parameter $T_0$ has been set to $1$~K.}
  \begin{ruledtabular}
    \begin{tabular}{c|ddd}
    Isotopologue & \multicolumn{1}{c}{$a$} &
    \multicolumn{1}{c}{$b/10^{-5}$} &
    \multicolumn{1}{c}{$c$} \\
    & \multicolumn{1}{c}{(cm${}^3$/(mol K))} &
    \multicolumn{1}{c}{(cm${}^3$/(mol K))} &
    \multicolumn{1}{c}{(K)} \\
    \hline
    H${}_2{}^{16}$O & 3.67777 & 1.38466 & 8.84684 \\
    HD${}^{16}$O    & 3.66227 & 1.3733  & 9.63151 \\
    D${}_2{}^{16}$O & 3.6466  & 1.39401 & 5.3719  \\
  \end{tabular}      
  \end{ruledtabular}
  \label{tab:Ael_corr}  
\end{table}
}

\subsection{Dipolar polarizability}


\begin{table*}
  \caption{\label{tab:Aeps_dip_H2O} The values of the dipolar contribution to $A_\varepsilon$ for
    H${}_2{}^{16}$O using various models, and its total value (last column) from path-integral
    simulations.  All of the uncertainties are reported at $k=2$ coverage and do not include the
    propagation of the unknown uncertainty of the water dipole-moment surface and potential-energy
    surface. A breakdown of the rotational and vibrational contributions to
    $A_{\varepsilon}^\mathrm{(dip)}$ for H${}_2$O is reported in Table~I of the Supplementary Material.}
  \begin{ruledtabular}
  \begin{tabular}{ddccc|c}
    \multicolumn{1}{c}{Temperature} &
    \multicolumn{1}{c}{$A_{\varepsilon}^\mathrm{(dip)}  $ (semiclassical)} &
    \multicolumn{1}{c}{$A_{\varepsilon}^\mathrm{(dip)}$ (HITRAN2020)} &
    $A_{\varepsilon}^\mathrm{(dip)}$ (rigid) &
    $A_{\varepsilon}^\mathrm{(dip)}$ (flexible) &
    $A_{\varepsilon}$ (flexible) \\    
    \multicolumn{1}{c}{(K)} &
    \multicolumn{1}{c}{(cm${}^3$/mol)} &
    \multicolumn{1}{c}{(cm${}^3$/mol)} &
    (cm${}^3$/mol) &
    (cm${}^3$/mol) &    
    (cm${}^3$/mol) \\
    \hline
50  &   349.494 &   350 $\pm$   12  &   356.2   $\pm$   0.4 &   350 $\pm$   6   &   353 $\pm$   6   \\
75  &   248.944 &   248 $\pm$   9   &   251.18  $\pm$   0.12    &   248 $\pm$   2   &   252 $\pm$   2   \\
100 &   192.688 &   191 $\pm$   7   &   193.62  $\pm$   0.07    &   191.1 $\pm$   1.3   &   194.8 $\pm$ 1.3 \\
125 &   157.021 &   155 $\pm$   5   &   157.54  $\pm$   0.05    &   156.1 $\pm$   1.0   &   159.7 $\pm$   1.0   \\
150 &   132.445 &   131 $\pm$   5   &   132.75  $\pm$   0.03    &   131.5   $\pm$   0.7 &   135.2   $\pm$   0.7 \\
175 &   114.501 &   114 $\pm$   4   &   114.70  $\pm$   0.02    &   113.3   $\pm$   0.6 &   117.0   $\pm$   0.6 \\
200 &   100.829 &   100 $\pm$   4   &   100.985 $\pm$   0.015   &   100.1   $\pm$   0.6 &   103.8   $\pm$   0.6 \\
225 &   90.069  &   89  $\pm$   3   &   90.173  $\pm$   0.013   &   89.4    $\pm$   0.6 &   93.1    $\pm$   0.6 \\
250 &   81.381  &   81  $\pm$   3   &   81.482  $\pm$   0.010   &   81.0    $\pm$   0.6 &   84.7    $\pm$   0.6 \\
273.16  &   74.704  &   74  $\pm$   3   &   74.779  $\pm$   0.009   &   74.3    $\pm$   0.6 &   78.0    $\pm$   0.6 \\
293.15  &   69.763  &   69  $\pm$   2   &   69.833  $\pm$   0.008   &   69.1    $\pm$   0.6 &   72.8    $\pm$   0.6 \\
300 &   68.216  &   68  $\pm$   2   &   68.290  $\pm$   0.007   &   67.6    $\pm$   0.6 &   71.3    $\pm$   0.6 \\
325 &   63.110  &   63  $\pm$   2   &   63.162  $\pm$   0.006   &   62.8    $\pm$   0.6 &   66.5    $\pm$   0.6 \\
350 &   58.715  &   58  $\pm$   2   &   58.761  $\pm$   0.005   &   58.5    $\pm$   0.6 &   62.2    $\pm$   0.6 \\
375 &   54.892  &   55  $\pm$   2   &   54.931  $\pm$   0.005   &   54.5    $\pm$   0.6 &   58.2    $\pm$   0.6 \\
400 &   51.536  &   51  $\pm$   2   &   51.575  $\pm$   0.004   &   51.2    $\pm$   0.6 &   54.9    $\pm$   0.6 \\
450 &   45.920  &   46  $\pm$   2   &   45.951  $\pm$   0.003   &   45.8    $\pm$   0.5 &   49.5    $\pm$   0.5 \\
500 &   41.408  &   41.3    $\pm$   1.4 &   41.431  $\pm$   0.003   &   41.4    $\pm$   0.7 &   45.1    $\pm$   0.7 \\
550 &   37.703  &   37.6    $\pm$   1.3 &   37.726  $\pm$   0.002   &   37.3    $\pm$   0.6 &   41.0    $\pm$   0.6 \\
600 &   34.606  &   34.5    $\pm$   1.2 &   34.6220 $\pm$   0.0018  &   34.5    $\pm$   0.7 &   38.2    $\pm$   0.7 \\
650 &   31.980  &   31.9    $\pm$   1.1 &   31.9963 $\pm$   0.0016  &   32.1    $\pm$   0.8 &   35.8    $\pm$   0.8 \\
700 &   29.724  &   29.7    $\pm$   1.0 &   29.7340 $\pm$   0.0014  &   29.6    $\pm$   0.7 &   33.3    $\pm$   0.7 \\
750 &   27.765  &   27.7    $\pm$   1.0 &   27.7743 $\pm$   0.0011  &   27.7    $\pm$   0.6 &   31.4    $\pm$   0.6 \\
800 &   26.048  &   26.0    $\pm$   0.9 &   26.0568 $\pm$   0.0010  &   26.0    $\pm$   0.7 &   29.7    $\pm$   0.7 \\
900 &   23.182  &   23.1    $\pm$   0.8 &   23.1874 $\pm$   0.0008  &   23.1    $\pm$   0.7 &   26.9    $\pm$   0.7 \\
1000    &   20.884  &   20.8    $\pm$   0.8 &   20.8891 $\pm$   0.0006  &   20.6    $\pm$   0.6 &   24.3    $\pm$   0.6 \\
1250    &   16.736  &   16.5    $\pm$   0.7 &   16.7398 $\pm$   0.0004  &   16.5    $\pm$   0.7 &   20.2    $\pm$   0.7 \\
1500    &   13.962  &   13.5    $\pm$   0.7 &   13.9650 $\pm$   0.0003  &   13.9    $\pm$   0.6 &   17.6    $\pm$   0.6 \\
1750    &   11.977  &   11.1    $\pm$   0.6 &   11.9789 $\pm$   0.0002  &   11.8    $\pm$   0.5 &   15.6    $\pm$   0.5 \\
2000    &   10.487  &   9.2 $\pm$   0.6 &   10.48780    $\pm$   0.00016 &   10.4    $\pm$   0.6 &   14.1    $\pm$   0.6     
  \end{tabular}    
  \end{ruledtabular}  
\end{table*}      

In order to discuss the dipolar contribution to the first dielectric virial
coefficient, which is reported in Table~\ref{tab:Aeps_dip_H2O} for various
water models, let us begin by considering a classical rigid model
(cf. Eq.~(\ref{eq:Acl})). The most striking difference with respect to
the electronic polarizability contribution is the explicit appearance of a
dependence on the inverse of the temperature. As a consequence, the values
of $A_{\varepsilon}^\mathrm{(dip)}$ cover a significantly larger range than
those of $A_{\varepsilon}^\mathrm{(el)}$. Therefore, we will plot the product of
$A_{\varepsilon}^\mathrm{(dip)}$ and the temperature $T$.

Using the same water geometry as before,~\cite{Czako:09} the dipole moment
evaluated using the latest surface~\cite{DPS_2018} would provide a value
$\mu = 1.860$~D ($1$~D $\approx 3.33564 \times 10^{-30}$~C~m); the square root of
the average value of the dipole moment squared on the ground state of water
is $\sqrt{\langle |\mbm|^2 \rangle} = 1.857$~D using the same model.
However, the smallness of the moments of inertia of water makes quantum
rotational effects sizable; these can be investigated either with the
semiclassical correction of Eq.~(\ref{eq:Asemi}) or with the more accurate
path-integral simulation for rigid rotors described in
Sec.~\ref{sec:quantum_rigid}.

\begin{figure}[h]
  \center\includegraphics[width=0.9\linewidth]{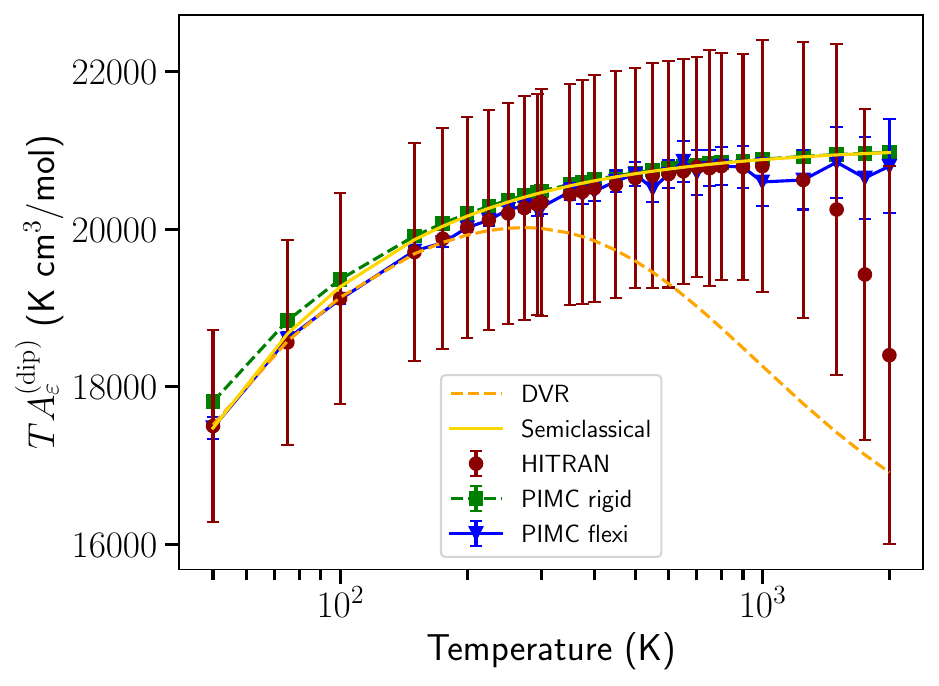}  
  \caption{The dipolar contribution to $A_\varepsilon$, $A_{\varepsilon}^\mathrm{(dip)}$, multiplied
    by the temperature $T$ of H${}_2{}^{16}$O as a function of temperature.  Red circles: values and
    uncertainties obtained from HITRAN2020 data. Green squares: values obtained using a rigid
    quantum mechanical model. Blue triangles: values obtained using a flexible quantum mechanical
    model solved using PIMC. Dashed orange line: values obtained using a flexible quantum mechanical
    model, solved using DVR (the discrepancies for $T>300$~K are due to insufficient
    convergence in the calculation of angular momentum states). Solid gold line: semiclassical
    values from Eq.~(\ref{eq:Asemi}).  All theoretical uncertainties are reported at
    the $k=2$ coverage value and do not include the propagation of the unknown uncertainty of the
    water dipole-moment surface. The lines joining the symbols are a guide to the eye.}
  \label{fig:Aeps_dip_H2O}
\end{figure}

We report in Fig.~\ref{fig:Aeps_dip_H2O} and Table~\ref{tab:Aeps_dip_H2O} the values of the dipolar
contribution to $A_\varepsilon$ for several water models. The semiclassical approach and quantum
rigid approach are in very good agreement for temperatures $T \gtrsim 100$~K.  Additionally, quantum
rotational effects are already appreciable at room temperature, where they contribute to a reduction
of the dielectric virial coefficient by $\sim 3 \%$ with respect to a classical value. Quantum
rotational effects become progressively more important at lower temperatures, resulting in a
reduction of $\sim 10 \%$ at $100$~K and $\sim 20 \%$ at $50$~K.  The results of rigid-model PIMC
calculations are in perfect agreement with the results obtained by diagonalization of the
Hamiltonian of Eq.~(\ref{eq:HR}).

Figure~\ref{fig:Aeps_dip_H2O} also shows the effect of molecular flexibility in
determining $A_{\varepsilon}^\mathrm{(dip)}$, as well as presenting the
experimental values derived from the HITRAN2020 database. In general, the
addition of flexibility results in a reduction of the dipolar contribution
to the first dielectric virial coefficient, which is particularly evident
at $T \leq 150$~K. The path-integral calculations are in very good
agreement with values derived from spectroscopy, falling well within the estimated
experimental uncertainty. We emphasize that we are not aware of any
uncertainty estimates for the electronic-polarizability or dipole-moment
surfaces, so we cannot provide a rigorous uncertainty analysis
on $A_\varepsilon$ at present and only its statistical contribution is
reported. For the sake of a meaningful comparison, we collected
enough statistics in the Monte Carlo calculations to make the statistical
uncertainty smaller than the experimental one. The average values of our
simulations and those computed from HITRAN2020 are in very good agreement at all
the temperatures studied.

The computed and experimental values of $A_{\varepsilon}^\mathrm{(dip)}$ begin to differ at high
temperature; this is evident for $T \geq 1500$~K in the case of H${}_2{}^{16}$O and at even smaller
temperatures for other isotopologues (as shown in Figures 2 and 5 of the Supplementary
Material). We think that this discrepancy is due to the limited coverage of high-energy
rovibrational states in the HITRAN2020 database, which limits the accuracy of the deduced values of
$A_{\varepsilon}^\mathrm{(dip)}$ at high temperatures.

Figure~\ref{fig:Aeps_dip_H2O} also reports the calculation of $A_{\varepsilon}^\mathrm{(dip)}$
performed using the DVR approach. The agreement with the path-integral calculation and with HITRAN2020
data is very good at temperatures $T \lesssim 300$~K. For higher temperatures, the DVR approach
suffers from the limited number of angular momentum states $J$ that we have been able to compute,
and therefore the DVR values diverge from those obtained using PIMC.

The last column of Table~\ref{tab:Aeps_dip_H2O} reports our theoretical estimate for the first
dielectric virial coefficient of water, which has been obtained by summing the value of the dipolar
contribution obtained using a flexible model of water (next-to-last column of the same table) and
the electronic contribution reported in Table~\ref{tab:Aeps_pol_H2O}.

Finally, we have also fitted the values of $A_{\varepsilon}^\mathrm{(dip)}(T)$ computed using PIMC to
a correlation of the form
\begin{equation}
  A_{\varepsilon}^\mathrm{(dip)}(T) = \frac{a (1 + d/T) / T}{1 + \exp\left[-(T-b)/c \right]}.
  \label{eq:Adip_corr}
\end{equation}
Values of the parameters $a$, $b$, $c$, and $d$ in Eq.~(\ref{eq:Adip_corr}) are reported in
Table~\ref{tab:Adip_corr}.

\begin{table*}
  \caption{Values of the parameters in Eqs.~(\ref{eq:Adip_corr}) and
    (\ref{eq:Adip_corr2}) for the water isotopologues studied in this paper.}
  \begin{ruledtabular}
  \begin{tabular}{c|dddd}
    Isotopologue & \multicolumn{1}{c}{$a$} &
    \multicolumn{1}{c}{$b$} &
    \multicolumn{1}{c}{$c$} &
    \multicolumn{1}{c}{$d$} \\    
    & \multicolumn{1}{c}{(K cm${}^3$/mol)} &
    \multicolumn{1}{c}{(K)} &
    \multicolumn{1}{c}{(K)} &
    \multicolumn{1}{c}{(K${}^2$ cm${}^3$/mol)} \\   
    \hline
    H${}_2{}^{16}$O & 20945.9 &   -693.079 &  184.074 & -7.46202 \\
    HD${}^{16}$O    & 21950.5 & -11979.3   & 4072.31  & -6.30806 \\
    D${}_2{}^{16}$O & 23949.4 & -17378.8   & 9154.42  & -4.5188  \\
  \end{tabular}
  \end{ruledtabular}  
  \label{tab:Adip_corr}  
\end{table*}

\subsection{Vibrational polarizability}

\begin{figure}[h]
  \center\includegraphics[width=0.9\linewidth]{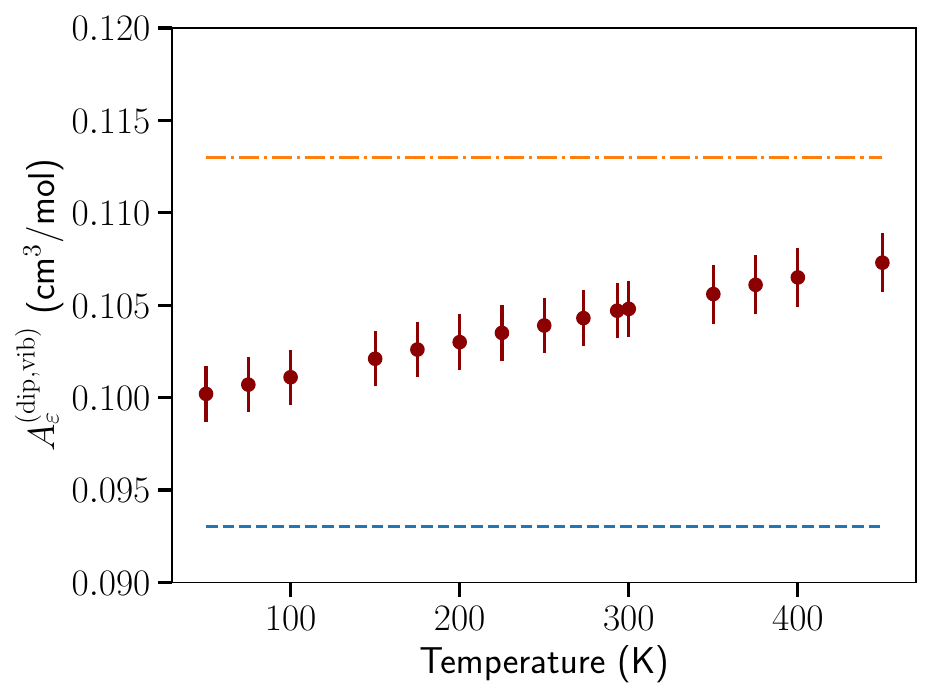}      
\caption{The vibrational polarizability contribution to $A_\varepsilon$,
  $A^\textrm{(dip,vib)}_{\varepsilon}$, of H$_2{}^{16}$O as a function of temperature,
  computed from spectroscopic information in the HITRAN2020 database.~\cite{HITRAN2020} The blue
  dashed line is the value 0.093~cm$^3$/mol calculated from a more crude use of spectroscopic data
  by Bishop and Cheung.\cite{Bishop_1982} The orange dot-dashed line is the value 0.113~cm$^3$/mol
  from the \textit{ab initio} calculations of Ruud \textit{et al.}\cite{Ruud_2000}}
\label{fig:vibpol}
\end{figure}

Because the HITRAN2020 database allows us to distinguish between vibrational transitions and those
in which only the rotational quantum numbers change, we can separate the calculated
$A_{\varepsilon}^\mathrm{(dip)}$ into vibrational and rotational parts. The detailed results for
H$_2{}^{16}$O, HD$^{16}$O, and D$_2{}^{16}$O are given in the Supplementary Material.  The
rotational contribution $\left( A^\mathrm{(dip,rot)}_{\varepsilon}\right)$ is larger than
the vibrational contribution $\left( A^\mathrm{(dip,vib)}_{\varepsilon}\right)$ by roughly
a factor of 600 at 300~K; our computed $A^\mathrm{(dip,vib)}_{\varepsilon}$ have a small
temperature dependence and for H$_2{}^{16}$O have values slightly above 0.10~cm$^3$/mol at typical
temperatures of experimental interest (see Fig.~\ref{fig:vibpol}).

The vibrational contribution to $A_{\varepsilon}^\mathrm{(dip)}$ can be related to the molecule's
vibrational polarizability, $\alpha_\mathrm{vib}$, by
\begin{equation}
A^\mathrm{(dip,vib)}_{\varepsilon} = \frac{4\pi}{3} N_{\mathrm{A}} \alpha_\mathrm{vib} .
\end{equation}
Our calculations produce $\alpha_\mathrm{vib}$ whose magnitude is roughly 3\% of the magnitude of
the electronic polarizability; this is often considered an additional contribution when compiling
the static polarizabilities of molecules.\cite{Hohm_2013} 

The vibrational polarizability has been a subject of some study, allowing comparison to previous
estimates.  Bishop and Cheung\cite{Bishop_1982} estimated $\alpha_\mathrm{vib}$ for H$_2$O and HDO
based on the positions and integrated intensities of the three primary vibrational bands of the
ground state. This can be thought of as an approximation to Eq. (25) in which each vibrational band
is lumped into one line.  Ruud \textit{et al.}\cite{Ruud_2000} estimated $\alpha_\mathrm{vib}$ from
\textit{ab initio} calculations of a perturbation expansion of each normal vibration.  The values of
$A^\mathrm{(dip,vib)}_{\varepsilon}$ derived from Refs. \citenum{Bishop_1982} and
\citenum{Ruud_2000} for H$_2$O are 0.093~cm$^3$/mol and 0.113~cm$^3$/mol, respectively. These
literature estimates are in reasonable agreement with our results, as shown in Fig.~\ref{fig:vibpol}.

Bishop and Cheung made a similar estimate for the HDO molecule, producing
$A^\mathrm{(dip,vib)}_{\varepsilon}$ = 0.107 cm$^3$/mol, which underestimates the
vibrational polarizability (see Table IV in the Supplementary Material) by an amount somewhat less
than that shown in Fig.~\ref{fig:vibpol} for H$_2$O.

\subsection{Dipole moment}

\begin{figure}
  \center\includegraphics[width=0.9\linewidth]{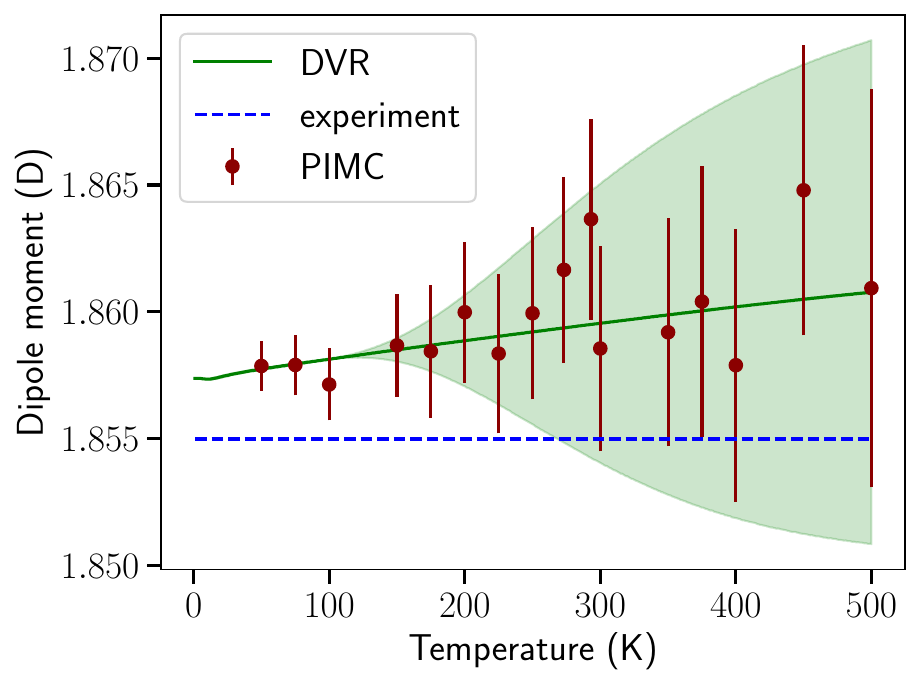}    
\caption{The average dipole moment of \HtO\ from our DVR (solid line) and PIMC (points)
  calculations. The shaded area denotes an estimate of the uncertainty of the DVR calculation. The
  PIMC error bars represent the expanded $(k=2)$ statistical uncertainty of the Monte Carlo
  calculation.  The dashed line reports the experimental ground-state value of $1.85498$~D from
  Ref.~\onlinecite{Shostak_1991}.}
  \label{fig:mu_H2O}
\end{figure}

We also computed the temperature-dependent value of the average dipole moment, defined as
\begin{equation}
  \mu(T) = \sqrt{\sum_{J\nu} \frac{g_{J\nu} (2J+1) \ee^{-\beta E_{J\nu}}}{Q_1(\beta,0)}
  \langle \psi_{J\nu} | |\mbm|^2 |\psi_{J\nu}\rangle},
  \label{eq:dip}
\end{equation}
where, as above, $J$ and $\nu$ are the total angular momentum of the water molecule, $\nu$ are
vibrational quantum numbers and $g_{J\nu}$ is the degeneracy of these states.  The results are
reported in Fig.~\ref{fig:mu_H2O}, where we also plot the experimental ground-state value of
$\mu$.~\cite{Shostak_1991} Similarly to what has been observed above, the PIMC calculations are
rather noisy, but agree with the DVR results up to the highest temperature that we have
investigated. We also report an estimate of the uncertainty of our DVR calculations, which is mainly
due to the finite number of angular momentum states $J$ considered in this work.  A slight increase
of the dipole moment with temperature is apparent in both cases (on the order of $0.2\%$ from $0$~K
to $500$~K), although the trend is probably clearer from the DVR results.

The value of the dipole moment at 0~K is that in the ground rovibrational state.  The DVR
calculations yield a dipole moment of 1.8574~D for H$_2$$^{16}$O, which is about 0.13\% larger than
the highly accurate ground-state value of 1.85498(9)~D measured by Shostak \textit{et al.}~\cite{Shostak_1991} A
similar calculation for D$_2{}^{16}$O yields a ground-state dipole moment of 1.8565~D, which exceeds
by approximately 0.11\% the experimental value of 1.8545(4)~D measured by Dyke and Muenter.~\cite{Dyke_1973}
These two comparisons with high-accuracy experimental results suggest that the dipole-moment surface
we used,~\cite{DMS_2018} which was designed more for spectroscopic applications than for accurate values of the
dipole moment itself, is slightly biased toward high values.

\subsection{Rescaled dipolar polarizability}

The slightly too large dipole moments discussed in the previous section will result in a slightly
overestimated dipolar polarizability.  This can, however, be simply corrected for because all
contributions to the dipolar polarizability are proportional to the square of the dipole moment.  We
can therefore produce an improved estimate of the dipolar polarizability by multiplying
Eq.~(\ref{eq:Adip_corr}) by the square of the ratio of the true ground-state dipole moment to that
obtained here; for that purpose we use the H$_2$O value since it is the most accurately measured,
producing a factor of $(1.85498/1.8574)^2 \approx 0.9974$.  The rescaled dipolar polarizability is
therefore given by
  \begin{equation}
  A_{\varepsilon}^\mathrm{(dip,r)}(T) = 0.9974 \frac{a(1+d/T)/T}{1 + \exp\left[-(T-b)/c \right]},
  \label{eq:Adip_corr2}
\end{equation}
with the parameters given in Table~\ref{tab:Adip_corr}.
At all temperatures, the magnitude of this correction is smaller than the standard statistical
uncertainty of the calculation of $A_{\varepsilon}^\mathrm{(dip)}$.

\begin{table}
  \caption{\label{tab:dip_H2O}
    The values of the average dipole moment $\mu(T)$ of H${}_2{}^{16}$O (Debye) from Eq.~(\ref{eq:dip})
    The PIMC uncertainties represent the expanded $(k=2)$ statistical uncertainty of the Monte Carlo
  calculation. Uncertainties do not include the propagation of the unknown uncertainty of the water
  dipole-moment surface.}
    \begin{ruledtabular}
  \begin{tabular}{dcc}
    \multicolumn{1}{c}{Temperature} &
    \multicolumn{1}{c}{$\mu(T)$ (PIMC)} &
    \multicolumn{1}{c}{$\mu(T)$ (DVR)} \\
    \multicolumn{1}{c}{(K)} &
    \multicolumn{1}{c}{(D)} &
    \multicolumn{1}{c}{(D)} \\
    \hline
1   &$      –       $&$ 1.85737 \pm 0.00001 $\\
10  &$      –       $&$ 1.85734 \pm 0.00002 $\\
25  &$      –       $&$ 1.85749 \pm 0.00004 $\\
50  &$  1.858   \pm 0.001   $&$ 1.85774 \pm 0.00003 $\\
75  &$  1.858   \pm 0.001   $&$ 1.85794 \pm 0.00003 $\\
100 &$  1.857   \pm 0.001   $&$ 1.85813 \pm 0.00010 $\\
125	&$	1.859	\pm	0.001	$&$	1.8583	\pm	0.0003	$\\
150	&$	1.859	\pm	0.002	$&$	1.8585	\pm	0.0005	$\\
175	&$	1.858	\pm	0.003	$&$	1.8587	\pm	0.0010	$\\
200	&$	1.860	\pm	0.003	$&$	1.859	\pm	0.002	$\\
225	&$	1.858	\pm	0.003	$&$	1.859	\pm	0.003	$\\
250	&$	1.860	\pm	0.003	$&$	1.859	\pm	0.004	$\\
273.16	&$	1.862	\pm	0.004	$&$	1.859	\pm	0.005	$\\
293.15	&$	1.864	\pm	0.004	$&$	1.860	\pm	0.005	$\\
300	&$	1.859	\pm	0.004	$&$	1.860	\pm	0.006	$\\
325	&$	1.860	\pm	0.004	$&$	1.860	\pm	0.006	$\\
350	&$	1.859	\pm	0.004	$&$	1.860	\pm	0.007	$\\
375	&$	1.860	\pm	0.005	$&$	1.860	\pm	0.008	$\\
400	&$	1.858	\pm	0.005	$&$	1.860	\pm	0.008	$\\
450	&$	1.865	\pm	0.006	$&$	1.860	\pm	0.009	$\\
500	&$	1.861	\pm	0.008	$&$	1.861	\pm	0.010	$\\
600 &$  1.858   \pm 0.009   $&$     –       $\\
650 &$  1.855   \pm 0.009   $&$     –       $\\
700 &$  1.873   \pm 0.011   $&$     –       $\\
750 &$  1.861   \pm 0.012   $&$     –       $\\
800 &$  1.858   \pm 0.013   $&$     –       $\\
900 &$  1.864   \pm 0.013   $&$     –       $\\
1000    &$  1.868   \pm 0.014   $&$     –       $\\
1250    &$  1.863   \pm 0.017   $&$     –       $\\
1500    &$  1.870   \pm 0.020   $&$     –       $\\
1750    &$  1.868   \pm 0.021   $&$     –       $\\
2000    &$  1.870   \pm 0.025   $&$     –       $
  \end{tabular}    
  \end{ruledtabular}  
\end{table}

\section{Conclusions}

This work has presented the first complete theoretical calculation of
water's first dielectric virial coefficient, $A_\varepsilon$, taking into
account the flexibility of the water molecule and state-of-the-art
descriptions of the variation of the electronic polarizability and the
dipole moment with molecular geometry.  The path-integral method, and in
some cases the DVR approach to the three-body Hamiltonian, are used to
perform the calculations with full accounting for quantum effects.

The contribution of the electronic polarizability to $A_\varepsilon$ is not
constant as is typically assumed, but increases slightly with temperature
due to the different polarizabilities of states other than the
rovibrational ground state.  Our results are consistent with the best
experimental values for this quantity, which are obtained from measurements
of the refractive index.\cite{Schoedel_2006,Egan_2022}

The contribution of the dipolar polarizability also differs somewhat from
the classical $\frac{\mu^2}{3k_\mathrm{B}T}$ functional form, both because
values of the dipole moment other than the ground-state value are sampled
at finite temperature and because of the quantization of rotation.  The
latter effect reduces the dipolar contribution to $A_\varepsilon$ by
roughly 3\% at room temperature.  The calculated dipolar contribution to
$A_\varepsilon$ also agrees well with estimates using line positions and
intensities in the HITRAN2020 database.

In addition to the dominant isotopologue H$_2$$^{16}$O, we performed calculations for D$_2$$^{16}$O
and HD$^{16}$O.  The data are presented in the Supplementary Material, but we note here that there
is nothing surprising in the results.  The electronic polarizability (and therefore
$A_{\varepsilon}^\mathrm{(el)}$) is smaller by amounts on the order of 0.5\% for HD$^{16}$O and 1\%
for D$_2$$^{16}$O. This probably reflects the shorter average length of O--D bonds compared to O--H
bonds.  The dipole moment is slightly reduced by D substitution, in agreement with the experimental
result for D$_2$$^{16}$O.~\cite{Dyke_1973}  The dipolar contribution $A_{\varepsilon}^\mathrm{(dip)}$ is not affected
(within the uncertainty of our calculations) by D substitution at high temperatures. However, below
about 500~K, D substitution noticeably increases $A_{\varepsilon}^\mathrm{(dip)}$. This is because the
substitution increases the moment of inertia, reducing the magnitude of the (negative) correction to
$A_{\varepsilon}^\mathrm{(dip)}$ due to the quantization of rotation [expressed semiclassically by
  Eq.~(\ref{eq:Asemi})].

Our results for $A_\varepsilon$ provide accurate, temperature-dependent data that can be used to
describe the effect of water on the static dielectric constant of gases for humidity metrology.
They can also serve as a low-density boundary condition for future comprehensive formulations for
the static dielectric constant of H$_2$O and D$_2$O. We note that the current international standard
formulation for the dielectric constant of H$_2$O\cite{Dielec_1997,Harvey2023} does not account for
the quantum effects studied in this work; it also uses a dipole moment that is roughly 1\% smaller
than the best experimental value,~\cite{Shostak_1991} which in our terminology produces a value of
$A_{\varepsilon}^\mathrm{(dip)}$ that is roughly 2\% too small.
Our results for $A_{\varepsilon}^\mathrm{(el)}$ could also be used to improve the standard formulation
for the refractive index of water,~\cite{Harvey2023,Harvey23_refractive} although this would require
a dispersion correction from our static values to optical frequencies.

Our recommended formula for $A_\varepsilon(T)$ is given by
\begin{equation}
A_\varepsilon(T) = A_{\varepsilon}^\mathrm{(el)}(T) + A^\mathrm{(dip,r)}_{\varepsilon}(T),
\end{equation}
where the electronic polarizability contribution $A_{\varepsilon}^\mathrm{(el)}$ is given by Eq.~(\ref{eq:Ael_corr})
and the (rescaled) dipolar contribution $A^\textrm{(dip,r)}_{\varepsilon}$ is given by 
Eq.~(\ref{eq:Adip_corr2}).

Extension beyond the low-density limit would require the second dielectric
virial coefficient, $B_\varepsilon$. $B_\varepsilon$ can be computed in a
straightforward way for noble gases,\cite{Garberoglio_2021} but the
calculation is much more difficult for a molecule like water.  The largest
effect would likely come from the correlation of molecular dipoles due to
the pair potential; this would be relatively straightforward for a rigid,
nonpolarizable model and Yang \textit{et al.}\cite{Yang_2017} performed
such a calculation for a simple water model.  A complete calculation of
$B_\varepsilon$ would require a multidimensional surface for the
nonadditive electronic polarizability and for changes in multipole moments
as the molecules mutually polarize each other.  Incorporating the
flexibility of the molecules would greatly increase the complexity, and is
likely impractical at present.  A classical calculation of $B_\varepsilon$
with a rigid, polarizable water model was performed by Stone \textit{et
  al.};\cite{Stone_2000} thus far the result has not been confirmed by
independent calculations and unfortunately there seem to be no reliable
experimental determinations of $B_\varepsilon$.  Additional rigorous
calculations of $B_\varepsilon$ for water would therefore be desirable.

\section{Supplementary Material}

The Supplementary Material includes the following: derivation of Eqs.~(\ref{eq:alpha_el}) and
(\ref{eq:alpha_dip}), a table with the breakdown of the vibrational and rotational contributions to
$A_{\varepsilon}^\mathrm{(dip)}$ for H${}_2$O.  Tables and figures reporting $\Aeps^{\mathrm{(el)}}$
and dipole moments of HDO and D${}_2$O. Tables and figures reporting the computed
$\Aeps^{\mathrm{(dip)}}$ for HDO and D${}_2$O, their division into vibrational and rotational
contributions, and their derivation from HITRAN2020 data

\section{Acknowledgments}
We thank Joseph T. Hodges of NIST for helpful discussions about HITRAN and spectroscopic data,
Patrick Egan of NIST for discussions on water's refractivity, and
the University of Trento for a generous allocation of computing resources on their HPC Cluster.

\section*{Author Declarations}
\subsection*{Conflict of Interest}
The authors have no conflicts to disclose.

\section*{Data Availability}
The data that support the findings of this study are available within the
article and its Supplementary Material.

\appendix

\section{Wavefunction expression for $\alpha_\mathrm{dip}(T)$}
\label{sec:alpha_dip}

\subsection{Flexible models}

The derivation of the expression of $\alpha_\mathrm{dip}(T)$ starts from the general formula of
Eq.~(\ref{eq:alpha_dip}) and the representation of the molecular quantum states of
Eq.~(\ref{eq:water_psi}).  In fact, the indices $i$ and $j$ in Eq.~(\ref{eq:alpha_dip}) stand for all
the quantum numbers needed to describe a given molecular state, that is $i \equiv J \nu m
I$. Correspondingly, we will indicate the quantum numbers corresponding to the index $j$ with primed
quantities, {\em i.e.}, $j \equiv J' \nu' m' I'$.  Since energies depend only on the quantum
numbers $J$ and $\nu$, and since the matrix element of the dipole moment operator does not act on
the nuclear spins, one can perform the sum over $I$ and $I'$ obtaining
\begin{equation}
  \sum_{I,I'} \left|
  \chi^*({\cal I}_{J\nu},I) \chi( {\cal I}_{J' \nu'}, I')
  \right|^2 = g_{J\nu} \delta_{g_{J\nu}, g_{J'\nu'}},
\end{equation}
that is, the sum over the nuclear spin states allows only ortho-ortho or para-para transitions, and
provides the corresponding degeneracy factor.

Additionally, particular care must be taken to evaluate the matrix element
of $\mbm \cdot \mbe$, where we will assume, without loss of generality, that we are
considering $\mbe$ aligned along the $Z$ axis in the laboratory frame.
However, the DVR procedure writes the wavefunction with coordinates $q$ that are
defined in the so-called molecular frame, where the orientation of the
molecule is fixed. In order to evaluate the matrix elements of $\mbm_Z$, one
has to recall that it transforms as the $0$-th component of a vector operator, which
means that it is given by~\cite{LeSueur92,Zare,DVR3D}
\begin{equation}
  \mbm_0^\mathrm{L} = \sum_{K=1}^1 \mbm_K^\mathrm{M}(q) D^{1*}_{0K}(\Omega),
\end{equation}
where the superscript `L' denotes the spherical components of the operator in the laboratory-fixed frame,
and the superscript `M' denotes spherical components in the molecule-fixed frame.
Using the identities
\begin{eqnarray}
  \int D^{J_1}_{m_1 k_1}(\Omega)  D^{J_2}_{m_2 k_2}(\Omega)  D^{J_3}_{m_3 k_3}(\Omega) \dd \Omega
  &=& \nonumber \\
  8 \pi^2
  \left(
  \begin{array}{ccc}
    J_1 & J_2 & J_3 \\
    m_1 & m_2 & m_3
  \end{array}
  \right)
  \left(
  \begin{array}{ccc}
    J_1 & J_2 & J_3 \\
    k_1 & k_2 & k_3
  \end{array}
  \right) & & \\
  \sum_{m m'}
  \left(
  \begin{array}{ccc}
    J' & 1 & J \\
    m' & 0 & m
  \end{array}
  \right) &=& \frac{1}{3}
\end{eqnarray}
one arrives from Eq.~(\ref{eq:alpha_dip}) to
\begin{widetext}
  \begin{small}
\begin{equation}
  \alpha_\mathrm{dip}(T) =
  \sum_{J' \nu', J \nu}
  \frac{(2J'+1)(2J+1)}{3}
  \left|
  \sum_{k'=k-1, k}^{k+1,\infty} 
  \left\langle \psi_{J' \nu' k'}| \mbm^\mathrm{M}_{k'-k}| \psi_{J \nu k} \right\rangle
  (-1)^{k'}
  \left(
  \begin{array}{ccc}
    J' & 1 & J \\
    -k' & k'-k & k
  \end{array}
  \right)  
  \right|^2
  g_{J \nu} \delta_{g_{J\nu}, g_{J'\nu'}}
  \frac{\ee^{-\beta E_{J' \nu'}} - \ee^{-\beta E_{J \nu}}}{E_{J\nu} - E_{J' \nu'}},
\label{eq:alpha_dip_DVR}
\end{equation}      
  \end{small}
\end{widetext}
where, in the DVR approach, one has
\begin{equation}
  \left\langle \psi_{J' \nu' k'}| \mbm^\mathrm{M}_{k'-k}| \psi_{J \nu k} \right\rangle
  = \sum_q \psi_{J' \nu' k'}(q) \mbm^\mathrm{M}_{k'-k}(q) \psi_{J \nu k}(q)
\label{eq:m_exp_val}
\end{equation}
since the spherical components of the dipole-moment operator are diagonal in the molecule-fixed
frame. In Eq.~(\ref{eq:alpha_dip_DVR}), the quantity
$
\left(
  \begin{array}{ccc}
    J' & 1 & J \\
    -k' & k'-k & k
  \end{array}
  \right)  
$
is a Wigner $3j$-symbol.
The diagonalization procedure outlined in Sec.~\ref{sec:DVR} provides, for any given angular
momentum $J$, the energies $E_{J \nu}$ and the wavefunctions $\psi_{J \nu k}(q)$ (as the eigenvalues
and eigenvectors of the three-body Hamiltonian in the molecule-fixed frame, respectively), enabling a
straightforward evaluation of the dipolar polarizability using Eqs.~(\ref{eq:m_exp_val}) and
(\ref{eq:alpha_dip_DVR}). In this paper, we obtained the degeneracy factors $g_{J \nu}$ from the
HITRAN2020 database.

\subsection{Rigid models}

Equation~(\ref{eq:alpha_dip_DVR}) is valid also in the case of rigid
molecular models of water. In this case, the eigenfunctions do not depend
on the coordinates $q$ describing the molecular vibrations, so that the
matrix element of Eq.~(\ref{eq:m_exp_val}) is replaced by
\begin{equation}
  \left\langle \psi_{J' \nu' k'}| \mbm^\mathrm{M}_{k'-k}| \psi_{J \nu k} \right\rangle
  = \psi_{J' \nu' k'} \mbm^\mathrm{M}_{k'-k} \psi_{J \nu k}
\label{eq:m_exp_val_rot}
\end{equation}
where now $\psi_{J \nu k}$ are the eigenfunctions of the rigid-rotor
Hamiltonian (see Sec.~\ref{sec:rigid_hamiltonian}).
In the case of a rigid molecule, the matrix elements of the spherical components of the
dipole-moment operator, $\mbm^\mathrm{M}_{k'-k}$, are constants.

\bibliography{ms}

\begin{thebibliography}{46}%
\makeatletter
\providecommand \@ifxundefined [1]{%
 \@ifx{#1\undefined}
}%
\providecommand \@ifnum [1]{%
 \ifnum #1\expandafter \@firstoftwo
 \else \expandafter \@secondoftwo
 \fi
}%
\providecommand \@ifx [1]{%
 \ifx #1\expandafter \@firstoftwo
 \else \expandafter \@secondoftwo
 \fi
}%
\providecommand \natexlab [1]{#1}%
\providecommand \enquote  [1]{``#1''}%
\providecommand \bibnamefont  [1]{#1}%
\providecommand \bibfnamefont [1]{#1}%
\providecommand \citenamefont [1]{#1}%
\providecommand \href@noop [0]{\@secondoftwo}%
\providecommand \href [0]{\begingroup \@sanitize@url \@href}%
\providecommand \@href[1]{\@@startlink{#1}\@@href}%
\providecommand \@@href[1]{\endgroup#1\@@endlink}%
\providecommand \@sanitize@url [0]{\catcode `\\12\catcode `\$12\catcode
  `\&12\catcode `\#12\catcode `\^12\catcode `\_12\catcode `\%12\relax}%
\providecommand \@@startlink[1]{}%
\providecommand \@@endlink[0]{}%
\providecommand \url  [0]{\begingroup\@sanitize@url \@url }%
\providecommand \@url [1]{\endgroup\@href {#1}{\urlprefix }}%
\providecommand \urlprefix  [0]{URL }%
\providecommand \Eprint [0]{\href }%
\providecommand \doibase [0]{https://doi.org/}%
\providecommand \selectlanguage [0]{\@gobble}%
\providecommand \bibinfo  [0]{\@secondoftwo}%
\providecommand \bibfield  [0]{\@secondoftwo}%
\providecommand \translation [1]{[#1]}%
\providecommand \BibitemOpen [0]{}%
\providecommand \bibitemStop [0]{}%
\providecommand \bibitemNoStop [0]{.\EOS\space}%
\providecommand \EOS [0]{\spacefactor3000\relax}%
\providecommand \BibitemShut  [1]{\csname bibitem#1\endcsname}%
\let\auto@bib@innerbib\@empty
\bibitem [{\citenamefont {Cuccaro}\ \emph {et~al.}(2012)\citenamefont
  {Cuccaro}, \citenamefont {Gavioso}, \citenamefont {Benedetto}, \citenamefont
  {{Madonna Ripa}}, \citenamefont {Fernicola},\ and\ \citenamefont
  {Guianvarc’h}}]{Cuccaro_2012}%
  \BibitemOpen
  \bibfield  {author} {\bibinfo {author} {\bibfnamefont {R.}~\bibnamefont
  {Cuccaro}}, \bibinfo {author} {\bibfnamefont {R.~M.}\ \bibnamefont
  {Gavioso}}, \bibinfo {author} {\bibfnamefont {G.}~\bibnamefont {Benedetto}},
  \bibinfo {author} {\bibfnamefont {D.}~\bibnamefont {{Madonna Ripa}}},
  \bibinfo {author} {\bibfnamefont {V.}~\bibnamefont {Fernicola}},\ and\
  \bibinfo {author} {\bibfnamefont {C.}~\bibnamefont {Guianvarc’h}},\
  }\bibfield  {title} {\enquote {\bibinfo {title} {Microwave determination of
  water mole fraction in humid gas mixtures},}\ }\href
  {https://doi.org/10.1007/s10765-011-1007-x} {\bibfield  {journal} {\bibinfo
  {journal} {Int. J. Thermophys.}\ }\textbf {\bibinfo {volume} {33}},\ \bibinfo
  {pages} {1352} (\bibinfo {year} {2012})}\BibitemShut {NoStop}%
\bibitem [{\citenamefont {Gavioso}\ \emph {et~al.}(2014)\citenamefont
  {Gavioso}, \citenamefont {{Madonna Ripa}}, \citenamefont {Benyon},
  \citenamefont {Gallegos}, \citenamefont {Perez-Sanz}, \citenamefont
  {Corbellini}, \citenamefont {Avila},\ and\ \citenamefont
  {Benito}}]{Gavioso_2014}%
  \BibitemOpen
  \bibfield  {author} {\bibinfo {author} {\bibfnamefont {R.~M.}\ \bibnamefont
  {Gavioso}}, \bibinfo {author} {\bibfnamefont {D.}~\bibnamefont {{Madonna
  Ripa}}}, \bibinfo {author} {\bibfnamefont {R.}~\bibnamefont {Benyon}},
  \bibinfo {author} {\bibfnamefont {J.~G.}\ \bibnamefont {Gallegos}}, \bibinfo
  {author} {\bibfnamefont {F.}~\bibnamefont {Perez-Sanz}}, \bibinfo {author}
  {\bibfnamefont {S.}~\bibnamefont {Corbellini}}, \bibinfo {author}
  {\bibfnamefont {S.}~\bibnamefont {Avila}},\ and\ \bibinfo {author}
  {\bibfnamefont {A.~M.}\ \bibnamefont {Benito}},\ }\bibfield  {title}
  {\enquote {\bibinfo {title} {Measuring humidity in methane and natural gas
  with a microwave technique},}\ }\href
  {https://doi.org/10.1007/s10765-014-1566-8} {\bibfield  {journal} {\bibinfo
  {journal} {Int. J. Thermophys.}\ }\textbf {\bibinfo {volume} {35}},\ \bibinfo
  {pages} {748} (\bibinfo {year} {2014})}\BibitemShut {NoStop}%
\bibitem [{\citenamefont {Lao}\ \emph {et~al.}(2018)\citenamefont {Lao},
  \citenamefont {Jia}, \citenamefont {Maitra},\ and\ \citenamefont
  {DiStasio}}]{DPS_2018}%
  \BibitemOpen
  \bibfield  {author} {\bibinfo {author} {\bibfnamefont {K.~U.}\ \bibnamefont
  {Lao}}, \bibinfo {author} {\bibfnamefont {J.}~\bibnamefont {Jia}}, \bibinfo
  {author} {\bibfnamefont {R.}~\bibnamefont {Maitra}},\ and\ \bibinfo {author}
  {\bibfnamefont {R.~A.}\ \bibnamefont {DiStasio}},\ }\bibfield  {title}
  {\enquote {\bibinfo {title} {On the geometric dependence of the molecular
  dipole polarizability in water: A benchmark study of higher-order electron
  correlation, basis set incompleteness error, core electron effects, and
  zero-point vibrational contributions},}\ }\href@noop {} {\bibfield  {journal}
  {\bibinfo  {journal} {J. Chem. Phys.}\ }\textbf {\bibinfo {volume} {149}},\
  \bibinfo {pages} {204303} (\bibinfo {year} {2018})}\BibitemShut {NoStop}%
\bibitem [{\citenamefont {Sch\"{o}del}, \citenamefont {Walkov},\ and\
  \citenamefont {Abou-Zeid}(2006)}]{Schoedel_2006}%
  \BibitemOpen
  \bibfield  {author} {\bibinfo {author} {\bibfnamefont {R.}~\bibnamefont
  {Sch\"{o}del}}, \bibinfo {author} {\bibfnamefont {A.}~\bibnamefont
  {Walkov}},\ and\ \bibinfo {author} {\bibfnamefont {A.}~\bibnamefont
  {Abou-Zeid}},\ }\bibfield  {title} {\enquote {\bibinfo {title} {High-accuracy
  determination of water vapor refractivity by length interferometry},}\ }\href
  {https://doi.org/10.1364/OL.31.001979} {\bibfield  {journal} {\bibinfo
  {journal} {Opt. Lett.}\ }\textbf {\bibinfo {volume} {31}},\ \bibinfo {pages}
  {1979} (\bibinfo {year} {2006})}\BibitemShut {NoStop}%
\bibitem [{\citenamefont {Egan}(2022)}]{Egan_2022}%
  \BibitemOpen
  \bibfield  {author} {\bibinfo {author} {\bibfnamefont {P.~F.}\ \bibnamefont
  {Egan}},\ }\bibfield  {title} {\enquote {\bibinfo {title} {Capability of
  commercial trackers as compensators for the absolute refractive index of
  air},}\ }\href
  {https://doi.org/https://doi.org/10.1016/j.precisioneng.2022.04.011}
  {\bibfield  {journal} {\bibinfo  {journal} {Precision Eng.}\ }\textbf
  {\bibinfo {volume} {77}},\ \bibinfo {pages} {46} (\bibinfo {year}
  {2022})}\BibitemShut {NoStop}%
\bibitem [{\citenamefont {Shostak}, \citenamefont {Ebenstein},\ and\
  \citenamefont {Muenter}(1991)}]{Shostak_1991}%
  \BibitemOpen
  \bibfield  {author} {\bibinfo {author} {\bibfnamefont {S.~L.}\ \bibnamefont
  {Shostak}}, \bibinfo {author} {\bibfnamefont {W.~L.}\ \bibnamefont
  {Ebenstein}},\ and\ \bibinfo {author} {\bibfnamefont {J.~S.}\ \bibnamefont
  {Muenter}},\ }\bibfield  {title} {\enquote {\bibinfo {title} {The dipole
  moment of water. {I}. {D}ipole moments and hyperfine properties of
  {H}${}_2${O} and {HDO} in the ground and excited vibrational states},}\
  }\href {https://doi.org/10.1063/1.460471} {\bibfield  {journal} {\bibinfo
  {journal} {J. Chem. Phys.}\ }\textbf {\bibinfo {volume} {94}},\ \bibinfo
  {pages} {5875} (\bibinfo {year} {1991})}\BibitemShut {NoStop}%
\bibitem [{\citenamefont {MacRury}\ and\ \citenamefont
  {Steele}(1974)}]{MacRury74}%
  \BibitemOpen
  \bibfield  {author} {\bibinfo {author} {\bibfnamefont {T.~B.}\ \bibnamefont
  {MacRury}}\ and\ \bibinfo {author} {\bibfnamefont {W.~A.}\ \bibnamefont
  {Steele}},\ }\bibfield  {title} {\enquote {\bibinfo {title} {Quantum effects
  on the dielectric virial coefficients of polar gases},}\ }\href@noop {}
  {\bibfield  {journal} {\bibinfo  {journal} {J. Chem. Phys.}\ }\textbf
  {\bibinfo {volume} {61}},\ \bibinfo {pages} {3352} (\bibinfo {year}
  {1974})}\BibitemShut {NoStop}%
\bibitem [{\citenamefont {Gray}, \citenamefont {Gubbins},\ and\ \citenamefont
  {Joslin}(2011)}]{GGJ2}%
  \BibitemOpen
  \bibfield  {author} {\bibinfo {author} {\bibfnamefont {C.~G.}\ \bibnamefont
  {Gray}}, \bibinfo {author} {\bibfnamefont {K.~E.}\ \bibnamefont {Gubbins}},\
  and\ \bibinfo {author} {\bibfnamefont {C.~G.}\ \bibnamefont {Joslin}},\
  }\href@noop {} {\emph {\bibinfo {title} {Theory of Molecular Fluids}}},\
  Vol.\ \bibinfo {volume} {2: {A}pplications}\ (\bibinfo  {publisher} {Oxford
  Science Publications},\ \bibinfo {year} {2011})\BibitemShut {NoStop}%
\bibitem [{\citenamefont {Hill}(1958)}]{Hill58}%
  \BibitemOpen
  \bibfield  {author} {\bibinfo {author} {\bibfnamefont {T.~L.}\ \bibnamefont
  {Hill}},\ }\bibfield  {title} {\enquote {\bibinfo {title} {Theory of the
  dielectric constant of imperfect gases and dilute solutions},}\ }\href@noop
  {} {\bibfield  {journal} {\bibinfo  {journal} {J. Chem. Phys.}\ }\textbf
  {\bibinfo {volume} {28}},\ \bibinfo {pages} {61} (\bibinfo {year}
  {1958})}\BibitemShut {NoStop}%
\bibitem [{\citenamefont {Garberoglio}, \citenamefont {Harvey},\ and\
  \citenamefont {Jeziorski}(2021)}]{Garberoglio_2021}%
  \BibitemOpen
  \bibfield  {author} {\bibinfo {author} {\bibfnamefont {G.}~\bibnamefont
  {Garberoglio}}, \bibinfo {author} {\bibfnamefont {A.~H.}\ \bibnamefont
  {Harvey}},\ and\ \bibinfo {author} {\bibfnamefont {B.}~\bibnamefont
  {Jeziorski}},\ }\bibfield  {title} {\enquote {\bibinfo {title} {Path-integral
  calculation of the third dielectric virial coefficient of noble gases},}\
  }\href {https://doi.org/10.1063/5.0077684} {\bibfield  {journal} {\bibinfo
  {journal} {J. Chem. Phys.}\ }\textbf {\bibinfo {volume} {155}},\ \bibinfo
  {pages} {234103} (\bibinfo {year} {2021})}\BibitemShut {NoStop}%
\bibitem [{\citenamefont {Zare}(1988)}]{Zare}%
  \BibitemOpen
  \bibfield  {author} {\bibinfo {author} {\bibfnamefont {R.~N.}\ \bibnamefont
  {Zare}},\ }\href@noop {} {\emph {\bibinfo {title} {Angular {M}omentum:
  {U}nderstanding {S}patial {A}spects in {C}hemistry and {P}hysics}}}\
  (\bibinfo  {publisher} {Wiley},\ \bibinfo {address} {New York},\ \bibinfo
  {year} {1988})\BibitemShut {NoStop}%
\bibitem [{\citenamefont {Czak{\'o}}, \citenamefont {M{\'a}tyus},\ and\
  \citenamefont {Cs{\'a}sz{\'a}r}(2009)}]{Czako:09}%
  \BibitemOpen
  \bibfield  {author} {\bibinfo {author} {\bibfnamefont {G.}~\bibnamefont
  {Czak{\'o}}}, \bibinfo {author} {\bibfnamefont {E.}~\bibnamefont
  {M{\'a}tyus}},\ and\ \bibinfo {author} {\bibfnamefont {A.~G.}\ \bibnamefont
  {Cs{\'a}sz{\'a}r}},\ }\bibfield  {title} {\enquote {\bibinfo {title}
  {Bridging theory with experiment: A benchmark study of thermally averaged
  structural and effective spectroscopic parameters of the water molecule},}\
  }\href@noop {} {\bibfield  {journal} {\bibinfo  {journal} {J. Phys. Chem. A}\
  }\textbf {\bibinfo {volume} {113}},\ \bibinfo {pages} {11665} (\bibinfo
  {year} {2009})}\BibitemShut {NoStop}%
\bibitem [{\citenamefont {Illinger}\ and\ \citenamefont
  {Smyth}(1960)}]{Illinger60.1}%
  \BibitemOpen
  \bibfield  {author} {\bibinfo {author} {\bibfnamefont {K.~H.}\ \bibnamefont
  {Illinger}}\ and\ \bibinfo {author} {\bibfnamefont {C.~P.}\ \bibnamefont
  {Smyth}},\ }\bibfield  {title} {\enquote {\bibinfo {title} {Atomic
  polarization. {I}. {V}ibrational polarization of gases},}\ }\href@noop {}
  {\bibfield  {journal} {\bibinfo  {journal} {J. Chem. Phys.}\ }\textbf
  {\bibinfo {volume} {32}},\ \bibinfo {pages} {787} (\bibinfo {year}
  {1960})}\BibitemShut {NoStop}%
\bibitem [{\citenamefont {Bishop}(1990)}]{Bishop90}%
  \BibitemOpen
  \bibfield  {author} {\bibinfo {author} {\bibfnamefont {D.~M.}\ \bibnamefont
  {Bishop}},\ }\bibfield  {title} {\enquote {\bibinfo {title} {Molecular
  vibrational and rotational motion in static and dynamic electric fields},}\
  }\href@noop {} {\bibfield  {journal} {\bibinfo  {journal} {Rev. Mod. Phys.}\
  }\textbf {\bibinfo {volume} {62}},\ \bibinfo {pages} {343} (\bibinfo {year}
  {1990})}\BibitemShut {NoStop}%
\bibitem [{\citenamefont {Hilborn}(1982)}]{Hilborn82}%
  \BibitemOpen
  \bibfield  {author} {\bibinfo {author} {\bibfnamefont {R.~C.}\ \bibnamefont
  {Hilborn}},\ }\bibfield  {title} {\enquote {\bibinfo {title} {Einstein
  coefficients, cross sections, $f$-values, dipole moments, and all that},}\
  }\href@noop {} {\bibfield  {journal} {\bibinfo  {journal} {Am. J. Phys.}\
  }\textbf {\bibinfo {volume} {50}},\ \bibinfo {pages} {982} (\bibinfo {year}
  {1982})}\BibitemShut {NoStop}%
\bibitem [{\citenamefont {Gordon}\ \emph {et~al.}(2021)\citenamefont {Gordon}
  \emph {et~al.}}]{HITRAN2020}%
  \BibitemOpen
  \bibfield  {author} {\bibinfo {author} {\bibfnamefont {I.~E.}\ \bibnamefont
  {Gordon}} \emph {et~al.},\ }\bibfield  {title} {\enquote {\bibinfo {title}
  {The {HITRAN2020} molecular spectroscopic database},}\ }\href@noop {}
  {\bibfield  {journal} {\bibinfo  {journal} {J. Quant. Spectrosc. Radiat.
  Trans.}\ }\textbf {\bibinfo {volume} {277}},\ \bibinfo {pages} {107949}
  (\bibinfo {year} {2021})}\BibitemShut {NoStop}%
\bibitem [{\citenamefont {Gamache}\ and\ \citenamefont
  {Rothman}(1992)}]{hitran_lte}%
  \BibitemOpen
  \bibfield  {author} {\bibinfo {author} {\bibfnamefont {R.~R.}\ \bibnamefont
  {Gamache}}\ and\ \bibinfo {author} {\bibfnamefont {L.~S.}\ \bibnamefont
  {Rothman}},\ }\bibfield  {title} {\enquote {\bibinfo {title} {Extension of
  the {HITRAN} database to non-{LTE} applications},}\ }\href@noop {} {\bibfield
   {journal} {\bibinfo  {journal} {J. Quant. Spectrosc. Radiat. Trans.}\
  }\textbf {\bibinfo {volume} {48}},\ \bibinfo {pages} {519} (\bibinfo {year}
  {1992})}\BibitemShut {NoStop}%
\bibitem [{\citenamefont {{\v{S}}ime{\v{c}}kov{\'a}}\ \emph
  {et~al.}(2006)\citenamefont {{\v{S}}ime{\v{c}}kov{\'a}}, \citenamefont
  {Jacquemart}, \citenamefont {Rothman}, \citenamefont {Gamache},\ and\
  \citenamefont {Goldman}}]{hitran_a}%
  \BibitemOpen
  \bibfield  {author} {\bibinfo {author} {\bibfnamefont {M.}~\bibnamefont
  {{\v{S}}ime{\v{c}}kov{\'a}}}, \bibinfo {author} {\bibfnamefont
  {D.}~\bibnamefont {Jacquemart}}, \bibinfo {author} {\bibfnamefont {L.~S.}\
  \bibnamefont {Rothman}}, \bibinfo {author} {\bibfnamefont {R.~R.}\
  \bibnamefont {Gamache}},\ and\ \bibinfo {author} {\bibfnamefont
  {A.}~\bibnamefont {Goldman}},\ }\bibfield  {title} {\enquote {\bibinfo
  {title} {Einstein {A}-coefficients and statistical weights for molecular
  absorption transitions in the {HITRAN} database},}\ }\href@noop {} {\bibfield
   {journal} {\bibinfo  {journal} {J. Quant. Spectrosc. Radiat. Trans.}\
  }\textbf {\bibinfo {volume} {98}},\ \bibinfo {pages} {130} (\bibinfo {year}
  {2006})}\BibitemShut {NoStop}%
\bibitem [{\citenamefont {Mizus}\ \emph {et~al.}(2018)\citenamefont {Mizus},
  \citenamefont {Kyuberis}, \citenamefont {Zobov}, \citenamefont {Makhnev},
  \citenamefont {Polyansky},\ and\ \citenamefont {Tennyson}}]{PES15K}%
  \BibitemOpen
  \bibfield  {author} {\bibinfo {author} {\bibfnamefont {I.~I.}\ \bibnamefont
  {Mizus}}, \bibinfo {author} {\bibfnamefont {A.~A.}\ \bibnamefont {Kyuberis}},
  \bibinfo {author} {\bibfnamefont {N.~F.}\ \bibnamefont {Zobov}}, \bibinfo
  {author} {\bibfnamefont {V.~Y.}\ \bibnamefont {Makhnev}}, \bibinfo {author}
  {\bibfnamefont {O.~L.}\ \bibnamefont {Polyansky}},\ and\ \bibinfo {author}
  {\bibfnamefont {J.}~\bibnamefont {Tennyson}},\ }\bibfield  {title} {\enquote
  {\bibinfo {title} {High-accuracy water potential energy surface for the
  calculation of infrared spectra},}\ }\href@noop {} {\bibfield  {journal}
  {\bibinfo  {journal} {Phil. Trans. Royal Soc. A}\ }\textbf {\bibinfo {volume}
  {376}},\ \bibinfo {pages} {20170149} (\bibinfo {year} {2018})}\BibitemShut
  {NoStop}%
\bibitem [{\citenamefont {Conway}\ \emph {et~al.}(2018)\citenamefont {Conway},
  \citenamefont {Kyuberis}, \citenamefont {Polyansky}, \citenamefont
  {Tennyson},\ and\ \citenamefont {Zobov}}]{DMS_2018}%
  \BibitemOpen
  \bibfield  {author} {\bibinfo {author} {\bibfnamefont {E.~K.}\ \bibnamefont
  {Conway}}, \bibinfo {author} {\bibfnamefont {A.~A.}\ \bibnamefont
  {Kyuberis}}, \bibinfo {author} {\bibfnamefont {O.~L.}\ \bibnamefont
  {Polyansky}}, \bibinfo {author} {\bibfnamefont {J.}~\bibnamefont
  {Tennyson}},\ and\ \bibinfo {author} {\bibfnamefont {N.~F.}\ \bibnamefont
  {Zobov}},\ }\bibfield  {title} {\enquote {\bibinfo {title} {A highly accurate
  ab initio dipole moment surface for the ground electronic state of water
  vapour for spectra extending into the ultraviolet},}\ }\href@noop {}
  {\bibfield  {journal} {\bibinfo  {journal} {J. Chem. Phys.}\ }\textbf
  {\bibinfo {volume} {149}},\ \bibinfo {pages} {084307} (\bibinfo {year}
  {2018})}\BibitemShut {NoStop}%
\bibitem [{\citenamefont {Sutcliffe}\ and\ \citenamefont
  {Tennyson}(1987)}]{Sutcliffe87}%
  \BibitemOpen
  \bibfield  {author} {\bibinfo {author} {\bibfnamefont {B.~T.}\ \bibnamefont
  {Sutcliffe}}\ and\ \bibinfo {author} {\bibfnamefont {J.}~\bibnamefont
  {Tennyson}},\ }\bibfield  {title} {\enquote {\bibinfo {title} {Variational
  methods for the calculation of rovibrational energy levels of small
  molecules},}\ }\href@noop {} {\bibfield  {journal} {\bibinfo  {journal} {J.
  Chem. Soc., Faraday Trans. 2}\ }\textbf {\bibinfo {volume} {83}},\ \bibinfo
  {pages} {1663} (\bibinfo {year} {1987})}\BibitemShut {NoStop}%
\bibitem [{\citenamefont {Light}\ and\ \citenamefont
  {Carrington~Jr}(2000)}]{Light2k}%
  \BibitemOpen
  \bibfield  {author} {\bibinfo {author} {\bibfnamefont {J.~C.}\ \bibnamefont
  {Light}}\ and\ \bibinfo {author} {\bibfnamefont {T.}~\bibnamefont
  {Carrington~Jr}},\ }\bibfield  {title} {\enquote {\bibinfo {title}
  {Discrete-variable representations and their utilization},}\ }\href@noop {}
  {\bibfield  {journal} {\bibinfo  {journal} {Adv. Chem. Phys.}\ }\textbf
  {\bibinfo {volume} {114}},\ \bibinfo {pages} {263} (\bibinfo {year}
  {2000})}\BibitemShut {NoStop}%
\bibitem [{\citenamefont {Szalay}(1993)}]{Szalay93}%
  \BibitemOpen
  \bibfield  {author} {\bibinfo {author} {\bibfnamefont {V.}~\bibnamefont
  {Szalay}},\ }\bibfield  {title} {\enquote {\bibinfo {title} {Discrete
  variable representations of differential operators},}\ }\href@noop {}
  {\bibfield  {journal} {\bibinfo  {journal} {J. Chem. Phys.}\ }\textbf
  {\bibinfo {volume} {99}},\ \bibinfo {pages} {1978} (\bibinfo {year}
  {1993})}\BibitemShut {NoStop}%
\bibitem [{\citenamefont {Baye}(2015)}]{Baye15}%
  \BibitemOpen
  \bibfield  {author} {\bibinfo {author} {\bibfnamefont {D.}~\bibnamefont
  {Baye}},\ }\bibfield  {title} {\enquote {\bibinfo {title} {The
  {L}agrange-mesh method},}\ }\href
  {https://doi.org/https://doi.org/10.1016/j.physrep.2014.11.006} {\bibfield
  {journal} {\bibinfo  {journal} {Phys. Rep.}\ }\textbf {\bibinfo {volume}
  {565}},\ \bibinfo {pages} {1} (\bibinfo {year} {2015})}\BibitemShut {NoStop}%
\bibitem [{\citenamefont {Tennyson}\ \emph {et~al.}(2004)\citenamefont
  {Tennyson}, \citenamefont {Kostin}, \citenamefont {Barletta}, \citenamefont
  {Harris}, \citenamefont {Polyansky}, \citenamefont {Ramanlal},\ and\
  \citenamefont {Zobov}}]{DVR3D}%
  \BibitemOpen
  \bibfield  {author} {\bibinfo {author} {\bibfnamefont {J.}~\bibnamefont
  {Tennyson}}, \bibinfo {author} {\bibfnamefont {M.~A.}\ \bibnamefont
  {Kostin}}, \bibinfo {author} {\bibfnamefont {P.}~\bibnamefont {Barletta}},
  \bibinfo {author} {\bibfnamefont {G.~J.}\ \bibnamefont {Harris}}, \bibinfo
  {author} {\bibfnamefont {O.~L.}\ \bibnamefont {Polyansky}}, \bibinfo {author}
  {\bibfnamefont {J.}~\bibnamefont {Ramanlal}},\ and\ \bibinfo {author}
  {\bibfnamefont {N.~F.}\ \bibnamefont {Zobov}},\ }\bibfield  {title} {\enquote
  {\bibinfo {title} {{DVR3D}: a program suite for the calculation of
  rotation–vibration spectra of triatomic molecules},}\ }\href@noop {}
  {\bibfield  {journal} {\bibinfo  {journal} {Comput. Phys. Commun.}\ }\textbf
  {\bibinfo {volume} {163}},\ \bibinfo {pages} {85} (\bibinfo {year}
  {2004})}\BibitemShut {NoStop}%
\bibitem [{\citenamefont {Czak{\'o}}\ \emph {et~al.}(2004)\citenamefont
  {Czak{\'o}}, \citenamefont {Furtenbacher}, \citenamefont {Cs{\'a}sz{\'a}r},\
  and\ \citenamefont {Szalay}}]{Czako_DVR}%
  \BibitemOpen
  \bibfield  {author} {\bibinfo {author} {\bibfnamefont {G.}~\bibnamefont
  {Czak{\'o}}}, \bibinfo {author} {\bibfnamefont {T.}~\bibnamefont
  {Furtenbacher}}, \bibinfo {author} {\bibfnamefont {A.~G.}\ \bibnamefont
  {Cs{\'a}sz{\'a}r}},\ and\ \bibinfo {author} {\bibfnamefont {V.}~\bibnamefont
  {Szalay}},\ }\bibfield  {title} {\enquote {\bibinfo {title} {Variational
  vibrational calculations using high-order anharmonic force fields},}\
  }\href@noop {} {\bibfield  {journal} {\bibinfo  {journal} {Mol. Phys.}\
  }\textbf {\bibinfo {volume} {102}},\ \bibinfo {pages} {2411} (\bibinfo {year}
  {2004})}\BibitemShut {NoStop}%
\bibitem [{\citenamefont {M{\'a}tyus}\ \emph {et~al.}(2010)\citenamefont
  {M{\'a}tyus}, \citenamefont {F{\'a}bri}, \citenamefont {Szidarovszky},
  \citenamefont {Czak{\'o}}, \citenamefont {Allen},\ and\ \citenamefont
  {Cs{\'a}sz{\'a}r}}]{Edit10}%
  \BibitemOpen
  \bibfield  {author} {\bibinfo {author} {\bibfnamefont {E.}~\bibnamefont
  {M{\'a}tyus}}, \bibinfo {author} {\bibfnamefont {C.}~\bibnamefont
  {F{\'a}bri}}, \bibinfo {author} {\bibfnamefont {T.}~\bibnamefont
  {Szidarovszky}}, \bibinfo {author} {\bibfnamefont {G.}~\bibnamefont
  {Czak{\'o}}}, \bibinfo {author} {\bibfnamefont {W.~D.}\ \bibnamefont
  {Allen}},\ and\ \bibinfo {author} {\bibfnamefont {A.~G.}\ \bibnamefont
  {Cs{\'a}sz{\'a}r}},\ }\bibfield  {title} {\enquote {\bibinfo {title}
  {Assigning quantum labels to variationally computed rotational-vibrational
  eigenstates of polyatomic molecules},}\ }\href@noop {} {\bibfield  {journal}
  {\bibinfo  {journal} {J. Chem. Phys.}\ }\textbf {\bibinfo {volume} {133}},\
  \bibinfo {pages} {034113} (\bibinfo {year} {2010})}\BibitemShut {NoStop}%
\bibitem [{\citenamefont {Le~Sueur}\ \emph {et~al.}(1992)\citenamefont
  {Le~Sueur}, \citenamefont {Miller}, \citenamefont {Tennyson},\ and\
  \citenamefont {Sutcliffe}}]{LeSueur92}%
  \BibitemOpen
  \bibfield  {author} {\bibinfo {author} {\bibfnamefont {C.~R.}\ \bibnamefont
  {Le~Sueur}}, \bibinfo {author} {\bibfnamefont {S.}~\bibnamefont {Miller}},
  \bibinfo {author} {\bibfnamefont {J.}~\bibnamefont {Tennyson}},\ and\
  \bibinfo {author} {\bibfnamefont {B.~T.}\ \bibnamefont {Sutcliffe}},\
  }\bibfield  {title} {\enquote {\bibinfo {title} {On the use of variational
  wavefunctions in calculating vibrational band intensities},}\ }\href@noop {}
  {\bibfield  {journal} {\bibinfo  {journal} {Mol. Phys.}\ }\textbf {\bibinfo
  {volume} {76}},\ \bibinfo {pages} {1147} (\bibinfo {year}
  {1992})}\BibitemShut {NoStop}%
\bibitem [{\citenamefont {Feynman}\ and\ \citenamefont {Hibbs}(1965)}]{FH}%
  \BibitemOpen
  \bibfield  {author} {\bibinfo {author} {\bibfnamefont {R.~P.}\ \bibnamefont
  {Feynman}}\ and\ \bibinfo {author} {\bibfnamefont {A.}~\bibnamefont
  {Hibbs}},\ }\href@noop {} {\emph {\bibinfo {title} {Quantum Mechanics and
  Path Integrals}}}\ (\bibinfo  {publisher} {McGraw-Hill},\ \bibinfo {address}
  {New York},\ \bibinfo {year} {1965})\BibitemShut {NoStop}%
\bibitem [{\citenamefont {Noya}\ \emph {et~al.}(2011)\citenamefont {Noya},
  \citenamefont {Ses{\'e}}, \citenamefont {Ram{\'\i}rez}, \citenamefont
  {McBride}, \citenamefont {Conde},\ and\ \citenamefont {Vega}}]{Noya11_1}%
  \BibitemOpen
  \bibfield  {author} {\bibinfo {author} {\bibfnamefont {E.~G.}\ \bibnamefont
  {Noya}}, \bibinfo {author} {\bibfnamefont {L.~M.}\ \bibnamefont {Ses{\'e}}},
  \bibinfo {author} {\bibfnamefont {R.}~\bibnamefont {Ram{\'\i}rez}}, \bibinfo
  {author} {\bibfnamefont {C.}~\bibnamefont {McBride}}, \bibinfo {author}
  {\bibfnamefont {M.~M.}\ \bibnamefont {Conde}},\ and\ \bibinfo {author}
  {\bibfnamefont {C.}~\bibnamefont {Vega}},\ }\bibfield  {title} {\enquote
  {\bibinfo {title} {Path integral {M}onte {C}arlo simulations for rigid rotors
  and their application to water},}\ }\href
  {https://doi.org/10.1080/00268976.2010.528202} {\bibfield  {journal}
  {\bibinfo  {journal} {Mol. Phys.}\ }\textbf {\bibinfo {volume} {109}},\
  \bibinfo {pages} {149} (\bibinfo {year} {2011})}\BibitemShut {NoStop}%
\bibitem [{\citenamefont {Noya}, \citenamefont {Vega},\ and\ \citenamefont
  {McBride}(2011)}]{Noya11_2}%
  \BibitemOpen
  \bibfield  {author} {\bibinfo {author} {\bibfnamefont {E.~G.}\ \bibnamefont
  {Noya}}, \bibinfo {author} {\bibfnamefont {C.}~\bibnamefont {Vega}},\ and\
  \bibinfo {author} {\bibfnamefont {C.}~\bibnamefont {McBride}},\ }\bibfield
  {title} {\enquote {\bibinfo {title} {A quantum propagator for path-integral
  simulations of rigid molecules},}\ }\href@noop {} {\bibfield  {journal}
  {\bibinfo  {journal} {J. Chem. Phys.}\ }\textbf {\bibinfo {volume} {134}},\
  \bibinfo {pages} {054117} (\bibinfo {year} {2011})}\BibitemShut {NoStop}%
\bibitem [{\citenamefont {Garberoglio}\ \emph {et~al.}(2018)\citenamefont
  {Garberoglio}, \citenamefont {Jankowski}, \citenamefont {Szalewicz},\ and\
  \citenamefont {Harvey}}]{Garberoglio18}%
  \BibitemOpen
  \bibfield  {author} {\bibinfo {author} {\bibfnamefont {G.}~\bibnamefont
  {Garberoglio}}, \bibinfo {author} {\bibfnamefont {P.}~\bibnamefont
  {Jankowski}}, \bibinfo {author} {\bibfnamefont {K.}~\bibnamefont
  {Szalewicz}},\ and\ \bibinfo {author} {\bibfnamefont {A.~H.}\ \bibnamefont
  {Harvey}},\ }\bibfield  {title} {\enquote {\bibinfo {title} {Fully quantum
  calculation of the second and third virial coefficients of water and its
  isotopologues from ab initio potentials},}\ }\href@noop {} {\bibfield
  {journal} {\bibinfo  {journal} {Faraday Discuss.}\ }\textbf {\bibinfo
  {volume} {148}},\ \bibinfo {pages} {174501} (\bibinfo {year}
  {2018})}\BibitemShut {NoStop}%
\bibitem [{\citenamefont {Bishop}, \citenamefont {Pipin},\ and\ \citenamefont
  {Silverman}(1986)}]{Bishop86}%
  \BibitemOpen
  \bibfield  {author} {\bibinfo {author} {\bibfnamefont {D.~M.}\ \bibnamefont
  {Bishop}}, \bibinfo {author} {\bibfnamefont {J.}~\bibnamefont {Pipin}},\ and\
  \bibinfo {author} {\bibfnamefont {J.~N.}\ \bibnamefont {Silverman}},\
  }\bibfield  {title} {\enquote {\bibinfo {title} {Methods for introducing
  vibrational effects in the calculation of electric dipole polarizabilities
  and hyperpolarizabilities (with reference to {H}${}^+_2$)},}\ }\href@noop {}
  {\bibfield  {journal} {\bibinfo  {journal} {Mol. Phys.}\ }\textbf {\bibinfo
  {volume} {59}},\ \bibinfo {pages} {165} (\bibinfo {year} {1986})}\BibitemShut
  {NoStop}%
\bibitem [{\citenamefont {Hellmann}\ and\ \citenamefont
  {Harvey}(2020)}]{HDO-geo}%
  \BibitemOpen
  \bibfield  {author} {\bibinfo {author} {\bibfnamefont {R.}~\bibnamefont
  {Hellmann}}\ and\ \bibinfo {author} {\bibfnamefont {A.~H.}\ \bibnamefont
  {Harvey}},\ }\bibfield  {title} {\enquote {\bibinfo {title} {First-principles
  diffusivity ratios for kinetic isotope fractionation of water in air},}\
  }\href@noop {} {\bibfield  {journal} {\bibinfo  {journal} {Geophys. Res.
  Lett.}\ }\textbf {\bibinfo {volume} {47}},\ \bibinfo {pages} {e2020GL089999}
  (\bibinfo {year} {2020})}\BibitemShut {NoStop}%
\bibitem [{\citenamefont {Tuckerman}\ \emph {et~al.}(1993)\citenamefont
  {Tuckerman}, \citenamefont {Berne}, \citenamefont {Martyna},\ and\
  \citenamefont {Klein}}]{hmc2}%
  \BibitemOpen
  \bibfield  {author} {\bibinfo {author} {\bibfnamefont {M.~E.}\ \bibnamefont
  {Tuckerman}}, \bibinfo {author} {\bibfnamefont {B.~J.}\ \bibnamefont
  {Berne}}, \bibinfo {author} {\bibfnamefont {G.~J.}\ \bibnamefont {Martyna}},\
  and\ \bibinfo {author} {\bibfnamefont {M.~L.}\ \bibnamefont {Klein}},\
  }\bibfield  {title} {\enquote {\bibinfo {title} {Efficient molecular dynamics
  and hybrid {M}onte {C}arlo algorithms for path integrals},}\ }\href@noop {}
  {\bibfield  {journal} {\bibinfo  {journal} {J. Chem. Phys.}\ }\textbf
  {\bibinfo {volume} {99}},\ \bibinfo {pages} {2796} (\bibinfo {year}
  {1993})}\BibitemShut {NoStop}%
\bibitem [{\citenamefont {Zeiss}\ and\ \citenamefont {Meath}(1977)}]{Zeiss77}%
  \BibitemOpen
  \bibfield  {author} {\bibinfo {author} {\bibfnamefont {G.~D.}\ \bibnamefont
  {Zeiss}}\ and\ \bibinfo {author} {\bibfnamefont {W.~J.}\ \bibnamefont
  {Meath}},\ }\bibfield  {title} {\enquote {\bibinfo {title} {Dispersion energy
  constants ${C}_6({A}, {B})$, dipole oscillator strength sums and
  refractivities for {L}i, {N}, {O}, {H}${}_2$, {N}${}_2$, {O}${}_2$,
  {NH}${}_3$, {H}${}_2${O}, {NO} and {N}${}_2${O}},}\ }\href
  {https://doi.org/10.1080/00268977700100991} {\bibfield  {journal} {\bibinfo
  {journal} {Mol. Phys.}\ }\textbf {\bibinfo {volume} {33}},\ \bibinfo {pages}
  {1155} (\bibinfo {year} {1977})}\BibitemShut {NoStop}%
\bibitem [{\citenamefont {Avila}(2005)}]{Avila05}%
  \BibitemOpen
  \bibfield  {author} {\bibinfo {author} {\bibfnamefont {G.}~\bibnamefont
  {Avila}},\ }\bibfield  {title} {\enquote {\bibinfo {title} {Ab initio dipole
  polarizability surfaces of water molecule: Static and dynamic at 514.5~nm},}\
  }\href@noop {} {\bibfield  {journal} {\bibinfo  {journal} {J. Chem. Phys.}\
  }\textbf {\bibinfo {volume} {122}},\ \bibinfo {pages} {144310} (\bibinfo
  {year} {2005})}\BibitemShut {NoStop}%
\bibitem [{\citenamefont {Bishop}\ and\ \citenamefont
  {Cheung}(1982)}]{Bishop_1982}%
  \BibitemOpen
  \bibfield  {author} {\bibinfo {author} {\bibfnamefont {D.~M.}\ \bibnamefont
  {Bishop}}\ and\ \bibinfo {author} {\bibfnamefont {L.~M.}\ \bibnamefont
  {Cheung}},\ }\bibfield  {title} {\enquote {\bibinfo {title} {Vibrational
  contributions to molecular dipole polarizabilities},}\ }\href
  {https://doi.org/10.1063/1.555658} {\bibfield  {journal} {\bibinfo  {journal}
  {J. Phys. Chem. Ref. Data}\ }\textbf {\bibinfo {volume} {11}},\ \bibinfo
  {pages} {119--133} (\bibinfo {year} {1982})}\BibitemShut {NoStop}%
\bibitem [{\citenamefont {Ruud}, \citenamefont {Jonsson},\ and\ \citenamefont
  {Taylor}(2000)}]{Ruud_2000}%
  \BibitemOpen
  \bibfield  {author} {\bibinfo {author} {\bibfnamefont {K.}~\bibnamefont
  {Ruud}}, \bibinfo {author} {\bibfnamefont {D.}~\bibnamefont {Jonsson}},\ and\
  \bibinfo {author} {\bibfnamefont {P.~R.}\ \bibnamefont {Taylor}},\ }\bibfield
   {title} {\enquote {\bibinfo {title} {Vibrational effects on electric and
  magnetic susceptibilities: application to the properties of the water
  molecule},}\ }\href {https://doi.org/10.1039/B000917M} {\bibfield  {journal}
  {\bibinfo  {journal} {Phys. Chem. Chem. Phys.}\ }\textbf {\bibinfo {volume}
  {2}},\ \bibinfo {pages} {2161--2171} (\bibinfo {year} {2000})}\BibitemShut
  {NoStop}%
\bibitem [{\citenamefont {Hohm}(2013)}]{Hohm_2013}%
  \BibitemOpen
  \bibfield  {author} {\bibinfo {author} {\bibfnamefont {U.}~\bibnamefont
  {Hohm}},\ }\bibfield  {title} {\enquote {\bibinfo {title} {Experimental
  static dipole–dipole polarizabilities of molecules},}\ }\href
  {https://doi.org/https://doi.org/10.1016/j.molstruc.2013.10.003} {\bibfield
  {journal} {\bibinfo  {journal} {J. Mol. Struct.}\ }\textbf {\bibinfo {volume}
  {1054-1055}},\ \bibinfo {pages} {282--292} (\bibinfo {year}
  {2013})}\BibitemShut {NoStop}%
\bibitem [{\citenamefont {Dyke}\ and\ \citenamefont
  {Muenter}(1973)}]{Dyke_1973}%
  \BibitemOpen
  \bibfield  {author} {\bibinfo {author} {\bibfnamefont {T.~R.}\ \bibnamefont
  {Dyke}}\ and\ \bibinfo {author} {\bibfnamefont {J.~S.}\ \bibnamefont
  {Muenter}},\ }\bibfield  {title} {\enquote {\bibinfo {title} {Electric dipole
  moments of low {$J$} states of {H$_2$O} and {D$_2$O}},}\ }\href
  {https://doi.org/10.1063/1.1680453} {\bibfield  {journal} {\bibinfo
  {journal} {J. Chem. Phys.}\ }\textbf {\bibinfo {volume} {59}},\ \bibinfo
  {pages} {3125} (\bibinfo {year} {1973})}\BibitemShut {NoStop}%
\bibitem [{\citenamefont {Fern{\'a}ndez}\ \emph {et~al.}(1997)\citenamefont
  {Fern{\'a}ndez}, \citenamefont {Goodwin}, \citenamefont {Lemmon},
  \citenamefont {{Levelt Sengers}},\ and\ \citenamefont
  {Williams}}]{Dielec_1997}%
  \BibitemOpen
  \bibfield  {author} {\bibinfo {author} {\bibfnamefont {D.~P.}\ \bibnamefont
  {Fern{\'a}ndez}}, \bibinfo {author} {\bibfnamefont {A.~R.~H.}\ \bibnamefont
  {Goodwin}}, \bibinfo {author} {\bibfnamefont {E.~W.}\ \bibnamefont {Lemmon}},
  \bibinfo {author} {\bibfnamefont {J.~M.~H.}\ \bibnamefont {{Levelt
  Sengers}}},\ and\ \bibinfo {author} {\bibfnamefont {R.~C.}\ \bibnamefont
  {Williams}},\ }\bibfield  {title} {\enquote {\bibinfo {title} {A formulation
  for the static permittivity of water and steam at temperatures from 238 {K}
  to 873 {K} at pressures up to 1200 {MPa}, including derivatives and
  {D}ebye--{H}ückel coefficients},}\ }\href {https://doi.org/10.1063/1.555997}
  {\bibfield  {journal} {\bibinfo  {journal} {J. Phys. Chem. Ref. Data}\
  }\textbf {\bibinfo {volume} {26}},\ \bibinfo {pages} {1125} (\bibinfo {year}
  {1997})}\BibitemShut {NoStop}%
\bibitem [{\citenamefont {Harvey}, \citenamefont {Hrub{\'y}},\ and\
  \citenamefont {Meier}(2023)}]{Harvey2023}%
  \BibitemOpen
  \bibfield  {author} {\bibinfo {author} {\bibfnamefont {A.~H.}\ \bibnamefont
  {Harvey}}, \bibinfo {author} {\bibfnamefont {J.}~\bibnamefont {Hrub{\'y}}},\
  and\ \bibinfo {author} {\bibfnamefont {K.}~\bibnamefont {Meier}},\ }\bibfield
   {title} {\enquote {\bibinfo {title} {{Improved and Always Improving:
  Reference Formulations for Thermophysical Properties of Water}},}\
  }\href@noop {} {\bibfield  {journal} {\bibinfo  {journal} {J. Phys. Chem.
  Ref. Data}\ }\textbf {\bibinfo {volume} {52}},\ \bibinfo {pages} {011501}
  (\bibinfo {year} {2023})}\BibitemShut {NoStop}%
\bibitem [{\citenamefont {Harvey}, \citenamefont {Gallagher},\ and\
  \citenamefont {{Levelt Sengers}}(1998)}]{Harvey23_refractive}%
  \BibitemOpen
  \bibfield  {author} {\bibinfo {author} {\bibfnamefont {A.~H.}\ \bibnamefont
  {Harvey}}, \bibinfo {author} {\bibfnamefont {J.~S.}\ \bibnamefont
  {Gallagher}},\ and\ \bibinfo {author} {\bibfnamefont {J.~M.~H.}\ \bibnamefont
  {{Levelt Sengers}}},\ }\bibfield  {title} {\enquote {\bibinfo {title}
  {{Revised Formulation for the Refractive Index of Water and Steam as a
  Function of Wavelength, Temperature and Density}},}\ }\href@noop {}
  {\bibfield  {journal} {\bibinfo  {journal} {J. Phys. Chem. Ref. Data}\
  }\textbf {\bibinfo {volume} {27}},\ \bibinfo {pages} {761} (\bibinfo {year}
  {1998})}\BibitemShut {NoStop}%
\bibitem [{\citenamefont {Yang}, \citenamefont {Schultz},\ and\ \citenamefont
  {Kofke}(2017)}]{Yang_2017}%
  \BibitemOpen
  \bibfield  {author} {\bibinfo {author} {\bibfnamefont {S.}~\bibnamefont
  {Yang}}, \bibinfo {author} {\bibfnamefont {A.~J.}\ \bibnamefont {Schultz}},\
  and\ \bibinfo {author} {\bibfnamefont {D.~A.}\ \bibnamefont {Kofke}},\
  }\bibfield  {title} {\enquote {\bibinfo {title} {Evaluation of second and
  third dielectric virial coefficients for non-polarisable molecular models},}\
  }\href {https://doi.org/10.1080/00268976.2017.1301585} {\bibfield  {journal}
  {\bibinfo  {journal} {Mol. Phys.}\ }\textbf {\bibinfo {volume} {115}},\
  \bibinfo {pages} {991} (\bibinfo {year} {2017})}\BibitemShut {NoStop}%
\bibitem [{\citenamefont {Stone}, \citenamefont {Tantirungrotechai},\ and\
  \citenamefont {Buckingham}(2000)}]{Stone_2000}%
  \BibitemOpen
  \bibfield  {author} {\bibinfo {author} {\bibfnamefont {A.~J.}\ \bibnamefont
  {Stone}}, \bibinfo {author} {\bibfnamefont {Y.}~\bibnamefont
  {Tantirungrotechai}},\ and\ \bibinfo {author} {\bibfnamefont {A.~D.}\
  \bibnamefont {Buckingham}},\ }\bibfield  {title} {\enquote {\bibinfo {title}
  {The dielectric virial coefficient and model intermolecular potentials},}\
  }\href {https://doi.org/10.1039/A905990C} {\bibfield  {journal} {\bibinfo
  {journal} {Phys. Chem. Chem. Phys.}\ }\textbf {\bibinfo {volume} {2}},\
  \bibinfo {pages} {429} (\bibinfo {year} {2000})}\BibitemShut {NoStop}%
\end{thebibliography}%


\begin{thebibliography}{4}%
\makeatletter
\providecommand \@ifxundefined [1]{%
 \@ifx{#1\undefined}
}%
\providecommand \@ifnum [1]{%
 \ifnum #1\expandafter \@firstoftwo
 \else \expandafter \@secondoftwo
 \fi
}%
\providecommand \@ifx [1]{%
 \ifx #1\expandafter \@firstoftwo
 \else \expandafter \@secondoftwo
 \fi
}%
\providecommand \natexlab [1]{#1}%
\providecommand \enquote  [1]{``#1''}%
\providecommand \bibnamefont  [1]{#1}%
\providecommand \bibfnamefont [1]{#1}%
\providecommand \citenamefont [1]{#1}%
\providecommand \href@noop [0]{\@secondoftwo}%
\providecommand \href [0]{\begingroup \@sanitize@url \@href}%
\providecommand \@href[1]{\@@startlink{#1}\@@href}%
\providecommand \@@href[1]{\endgroup#1\@@endlink}%
\providecommand \@sanitize@url [0]{\catcode `\\12\catcode `\$12\catcode
  `\&12\catcode `\#12\catcode `\^12\catcode `\_12\catcode `\%12\relax}%
\providecommand \@@startlink[1]{}%
\providecommand \@@endlink[0]{}%
\providecommand \url  [0]{\begingroup\@sanitize@url \@url }%
\providecommand \@url [1]{\endgroup\@href {#1}{\urlprefix }}%
\providecommand \urlprefix  [0]{URL }%
\providecommand \Eprint [0]{\href }%
\providecommand \doibase [0]{https://doi.org/}%
\providecommand \selectlanguage [0]{\@gobble}%
\providecommand \bibinfo  [0]{\@secondoftwo}%
\providecommand \bibfield  [0]{\@secondoftwo}%
\providecommand \translation [1]{[#1]}%
\providecommand \BibitemOpen [0]{}%
\providecommand \bibitemStop [0]{}%
\providecommand \bibitemNoStop [0]{.\EOS\space}%
\providecommand \EOS [0]{\spacefactor3000\relax}%
\providecommand \BibitemShut  [1]{\csname bibitem#1\endcsname}%
\let\auto@bib@innerbib\@empty
\bibitem [{\citenamefont {Wilcox}(1967)}]{Wilcox67}%
  \BibitemOpen
  \bibfield  {author} {\bibinfo {author} {\bibfnamefont {R.~M.}\ \bibnamefont
  {Wilcox}},\ }\bibfield  {title} {\enquote {\bibinfo {title} {Exponential
  operators and parameter differentiation in quantum physics},}\ }\href@noop {}
  {\bibfield  {journal} {\bibinfo  {journal} {J. Math. Phys.}\ }\textbf
  {\bibinfo {volume} {8}},\ \bibinfo {pages} {962} (\bibinfo {year}
  {1967})}\BibitemShut {NoStop}%
\bibitem [{\citenamefont {Hellmann}\ and\ \citenamefont
  {Harvey}(2020)}]{HDO-geo}%
  \BibitemOpen
  \bibfield  {author} {\bibinfo {author} {\bibfnamefont {R.}~\bibnamefont
  {Hellmann}}\ and\ \bibinfo {author} {\bibfnamefont {A.~H.}\ \bibnamefont
  {Harvey}},\ }\bibfield  {title} {\enquote {\bibinfo {title} {First-principles
  diffusivity ratios for kinetic isotope fractionation of water in air},}\
  }\href@noop {} {\bibfield  {journal} {\bibinfo  {journal} {Geophys. Res.
  Lett.}\ }\textbf {\bibinfo {volume} {47}},\ \bibinfo {pages} {e2020GL089999}
  (\bibinfo {year} {2020})}\BibitemShut {NoStop}%
\bibitem [{\citenamefont {Gamache}(2022)}]{RG_private}%
  \BibitemOpen
  \bibfield  {author} {\bibinfo {author} {\bibfnamefont {R.}~\bibnamefont
  {Gamache}},\ }\href@noop {} {\enquote {\bibinfo {title} {Private
  communication},}\ } (\bibinfo {year} {August 2022})\BibitemShut {NoStop}%
\bibitem [{\citenamefont {Czak{\'o}}, \citenamefont {M{\'a}tyus},\ and\
  \citenamefont {Cs{\'a}sz{\'a}r}(2009)}]{Czako:09}%
  \BibitemOpen
  \bibfield  {author} {\bibinfo {author} {\bibfnamefont {G.}~\bibnamefont
  {Czak{\'o}}}, \bibinfo {author} {\bibfnamefont {E.}~\bibnamefont
  {M{\'a}tyus}},\ and\ \bibinfo {author} {\bibfnamefont {A.~G.}\ \bibnamefont
  {Cs{\'a}sz{\'a}r}},\ }\bibfield  {title} {\enquote {\bibinfo {title}
  {Bridging theory with experiment: A benchmark study of thermally averaged
  structural and effective spectroscopic parameters of the water molecule},}\
  }\href@noop {} {\bibfield  {journal} {\bibinfo  {journal} {J. Phys. Chem. A}\
  }\textbf {\bibinfo {volume} {113}},\ \bibinfo {pages} {11665} (\bibinfo
  {year} {2009})}\BibitemShut {NoStop}%
\end{thebibliography}%

\end{document}


\title{Supplementary Material for: Comprehensive Quantum Calculation of the First Dielectric
  Virial Coefficient of Water} 

\author{Giovanni Garberoglio}
\email{garberoglio@ectstar.eu}
\affiliation{European Centre for Theoretical Studies in Nuclear Physics
and Related Areas (FBK-ECT*), Trento I-38123, Italy.}

\author{Chiara Lissoni}
\thanks{These authors contributed equally.}
\author{Luca Spagnoli}
\thanks{These authors contributed equally.}
\affiliation{Physics Department, University of Trento, Trento I-38123, Italy.}

\author{Allan H. Harvey}
\email{allan.harvey@nist.gov}
\affiliation{Applied Chemicals and Materials Division, National
  Institute of Standards and Technology, Boulder, CO 80305, USA.}

\date{\today}

\maketitle

\tableofcontents

\newpage

\section{Differentiation of the molecular polarizability}

As discussed in the main text, the first dielectric virial coefficient is given by
\begin{equation}
  A_\varepsilon = \frac{4 \pi}{3} \NA \left. \frac{\dd
    p(F)}{\dd F}\right|_{F=0}
  \label{eq:Aeps}
\end{equation}
with
\begin{eqnarray}
  p(F) &=& \frac{ \mathrm{Tr}\left[
    (\mbm \cdot \mbe + \mbe \cdot \mbal \cdot \mbe F) ~  \exp\left(-\beta H(F) \right)
      \right]}{\mathrm{Tr}\left[\ee^{-\beta H(F)}\right]} \label{eq:p} \\
    H(F) &=& H_0 - F ~ \mbm \cdot \mbe - \frac{F^2}{2} ~ \mbe \cdot
  \mbal \cdot \mbe, \label{eq:H}
\end{eqnarray}

In order to perform the differentiation with respect to the external electric field $F$ in
Eq.~(\ref{eq:Aeps}), it is convenient to write
\begin{eqnarray}
  p(F) &=& \frac{N(F)}{D(F)} \\
  N(F) &=& \mathrm{Tr}\left[
    (\mbm \cdot \mbe + \mbe \cdot \mbal \cdot \mbe F) ~  \exp\left(-\beta H(F) \right)
    \right] \label{eq:N}\\
  D(F) &=& \mathrm{Tr}\left[\ee^{-\beta H(F)}\right], \label{eq:D}
\end{eqnarray}
so that
\begin{equation}
  \frac{\dd p(F)}{\dd F} = \frac{N'(F) D(F) - D'(F) N(F)}{D^2(F)},
\end{equation}
where the prime indicates differentiation with respect to $F$. This expression is valid even when
considering that in general all the quantities appearing in Eqs.~(\ref{eq:N}) and (\ref{eq:D}) are
quantum mechanical operators which do not commute among themselves.
The derivatives of the numerator $N(F)$ and the denominator $D(F)$ can be written as
\begin{eqnarray}
  N'(F) &=& \mathrm{Tr}\left[ \mbe \cdot \mbal \cdot \mbe ~  \exp\left(-\beta H(F) \right) +
    (\mbm \cdot \mbe + \mbe \cdot \mbal \cdot \mbe F) ~  \frac{\dd}{\dd F}\exp\left(-\beta H(F) \right)
      \right] \label{eq:Nprime}\\
  D'(F) &=& \mathrm{Tr}\left[ \frac{\dd}{\dd F} \ee^{-\beta H(F)}\right].
\end{eqnarray}

When taking the derivative of $\ee^{-\beta H(F)}$ in the quantum regime, it is convenient to use the
identity~\cite{Wilcox67} 
\begin{equation}
  \frac{\dd}{\dd F} \ee^{A(F)} = \int_0^1  \ee^{\lambda A(F)} \frac{\dd
    A(F)}{\dd F} \ee^{(1-\lambda) A(F)} ~ \dd\lambda,
  \label{eq:derivative}
\end{equation}
which takes into account the presence of non-commuting operators in $A(F)$.  When
Eq.~(\ref{eq:derivative}) is used to evaluate the term $D'(F)$, we get, due to the cyclic property of
the trace
\begin{equation}
  D'(F \to 0) = \mathrm{Tr}\left[ \beta \mbm \cdot \mbe ~ \ee^{-\beta H(0)}\right] = 0,
\end{equation}
which is zero because of rotational invariance. We are then left with
\begin{equation}
  \left. \frac{\dd p(F)}{\dd F}\right|_{F\to0} = \frac{N'(0)}{D(0)}.
  \label{eq:dpdF0}
\end{equation}

The first term on the right-hand side of Eq.~(\ref{eq:Nprime}) becomes
\begin{equation}
  \lim_{F \to 0} \mathrm{Tr}\left[ \mbe \cdot \mbal \cdot \mbe ~  \exp\left(-\beta H(F) \right)
    \right] = \mathrm{Tr}\left[ \mbe \cdot \mbal \cdot \mbe ~  \exp\left(-\beta H(0) \right)
    \right],
  \label{eq:pol}
\end{equation}
which, divided by $D(0)$, results in the average value of the electronic polarizability at
zero-field; a result valid both in classical and quantum statistical mechanics.
Using Eq.~(\ref{eq:derivative}), the second term on the right-hand side of Eq.~(\ref{eq:Nprime})
becomes, in the $F \to 0$ limit,
\begin{equation}
  \tr\left[ \mbm \cdot \mbe \frac{\dd}{\dd F} \ee^{-\beta H(F)}\right] 
  \underset{F\to0}{=}  
  \mathrm{Tr}\left[ \mbm \cdot \mbe \int_0^1 \ee^{-\lambda \beta H_0} ~ \beta \mbm
    \cdot \mbe ~ \ee^{-(1-\lambda)\beta H_0} ~ \dd \lambda
    \right].
  \label{eq:me2}
\end{equation}

In the classical limit, where we can assume that the operators $\mbm \cdot \mbe$ and $H_0$ commute,
Eq.~(\ref{eq:me2}) divided by $D(0)$ results in the average value $\beta (\mbm \cdot \mbe)^2 =
\frac{\beta}{3} |\mbm|^2$, so that one can write
\begin{equation}
A_\varepsilon^\mathrm{(cl)} = \frac{4 \pi}{3} \NA \left\langle
  \aiso + \frac{\beta |\mbm|^2}{3} 
  \right\rangle,
  \label{eq:Acl}  
\end{equation}

In the general case, we can write
\begin{eqnarray}
  \tr\left[ \mbm \cdot \mbe \frac{\dd}{\dd F} \ee^{-\beta H(F)}\right] &
  \underset{F\to0}{=}&
  \sum_{i,j} \langle i | \mbm \cdot \mbe | j \rangle
  \int_0^1 \langle j | \ee^{-\lambda \beta E_j}
  \beta \mbm \cdot \mbe \ee^{-(1-\lambda) \beta E_i} | i \rangle \dd\lambda \\
  &=& \sum_{i,j} \langle i | \mbm \cdot \mbe | j \rangle \ee^{-\beta E_i}
  \int_0^1 \langle j | \ee^{-\lambda \beta (E_j-E_i)}
  \beta \mbm \cdot \mbe | i \rangle \dd\lambda \\
  &=& \sum_{i \neq j}
  |\langle i | \mbm \cdot \mbe | j \rangle|^2
  \frac{\ee^{-\beta E_i} - \ee^{-\beta E_j}}{E_j-E_i},
  \label{eq:dip}
\end{eqnarray}  
where we have inserted one completeness relation,
\begin{equation}
  1 = \sum_j |j\rangle\langle j|,
\end{equation}
and we have used the fact that $\langle i | \mbm \cdot \mbe | i \rangle = 0$ due to rotational
invariance. Using Eqs.~(\ref{eq:dpdF0}), (\ref{eq:pol}), and (\ref{eq:dip}), and defining
\begin{eqnarray}
  Q_1(\beta) &=& \sum_i \ee^{-\beta E_i} \\
  \alpha_\mathrm{el} &=& \frac{\sum_i \left\langle i | \tr(\mbal) | i \right\rangle \ee^{-\beta
      E_i}}{3 Q_1(\beta)} \\
  \alpha_\mathrm{dip} &=&  \frac{1}{Q_1(\beta)}
  \sum_{i \neq j}
  |\langle i | \mbm \cdot \mbe | j \rangle|^2
  \frac{\ee^{-\beta E_i} - \ee^{-\beta E_j}}{E_j-E_i},
\end{eqnarray}
we can finally write
\begin{equation}
  A_\varepsilon = \frac{4 \pi}{3} \NA \left( \alpha_\mathrm{el} + \alpha_\mathrm{dip} \right).
\end{equation}

\section{Contribution to $A_{\varepsilon}^\mathrm{(dip)}$ for H${}_2$O}

{\squeezetable
\begin{table*}[h]
  \caption{\label{tab:Aeps_dip_H2O_comp}
    Vibrational and rotational contribution to $\Aeps^\mathrm{(dip)}$ of
    H${}_2$O from HITRAN2020. All of the uncertainties are reported at $k=2$ coverage.}
  \begin{ruledtabular}
  \begin{tabular}{dcc}
    \multicolumn{1}{c}{Temperature} & $A_{\varepsilon}^\mathrm{(dip,vib)}$ &
    $A_{\varepsilon}^\mathrm{(dip,rot)}$ \\
      \multicolumn{1}{c}{(K)} & (cm${}^3$/mol) & (cm${}^3$/mol) \\
      \hline
50	& $	0.1002	\pm	0.0015	$&$	349.9	\pm	12.2	$ \\
75	& $	0.1007	\pm	0.0015	$&$	247.4	\pm	8.7	$ \\
100	& $	0.1011	\pm	0.0015	$&$	191.1	\pm	6.7	$ \\
125	& $	0.1016	\pm	0.0015	$&$	155.7	\pm	5.4	$ \\
150	& $	0.1021	\pm	0.0015	$&$	131.3	\pm	4.6	$ \\
175	& $	0.1026	\pm	0.0015	$&$	113.5	\pm	4.0	$ \\
200	& $	0.1030	\pm	0.0015	$&$	100.0	\pm	3.5	$ \\
225	& $	0.1035	\pm	0.0015	$&$	89.3	\pm	3.1	$ \\
250	& $	0.1039	\pm	0.0015	$&$	80.7	\pm	2.8	$ \\
273.16	& $	0.1043	\pm	0.0015	$&$	74.1	\pm	2.6	$ \\
293.15	& $	0.1047	\pm	0.0015	$&$	69.2	\pm	2.4	$ \\
300	& $	0.1048	\pm	0.0015	$&$	67.7	\pm	2.4	$ \\
325	& $	0.1052	\pm	0.0016	$&$	62.6	\pm	2.2	$ \\
350	& $	0.1056	\pm	0.0016	$&$	58.3	\pm	2.0	$ \\
375	& $	0.1061	\pm	0.0016	$&$	54.5	\pm	1.9	$ \\
400	& $	0.1065	\pm	0.0016	$&$	51.2	\pm	1.8	$ \\
450	& $	0.1073	\pm	0.0016	$&$	45.6	\pm	1.6	$ \\
500	& $	0.1082	\pm	0.0017	$&$	41.1	\pm	1.4	$ \\
550	& $	0.1092	\pm	0.0017	$&$	37.5	\pm	1.3	$ \\
600	& $	0.1103	\pm	0.0019	$&$	34.4	\pm	1.2	$ \\
650	& $	0.112	\pm	0.002	$&$	31.8	\pm	1.1	$ \\
700	& $	0.113	\pm	0.002	$&$	29.6	\pm	1.0	$ \\
750	& $	0.115	\pm	0.003	$&$	27.6	\pm	1.0	$ \\
800	& $	0.118	\pm	0.003	$&$	25.9	\pm	0.9	$ \\
900	& $	0.123	\pm	0.004	$&$	23.0	\pm	0.8	$ \\
1000	& $	0.130	\pm	0.006	$&$	20.7	\pm	0.8	$ \\
1250	& $	0.148	\pm	0.011	$&$	16.4	\pm	0.7	$ \\
1500	& $	0.161	\pm	0.015	$&$	13.3	\pm	0.7	$ \\
1750	& $	0.165	\pm	0.018	$&$	10.9	\pm	0.6	$ \\
2000	& $	0.161	\pm	0.020	$&$	9.0	\pm	0.6	$
  \end{tabular}
  \end{ruledtabular}
\end{table*}  
}
  
\clearpage

\section{Results for \HDO}

In this case, the partition function computed from our data is closer to the reference HITRAN2020 values
than for H${}_2{}^{16}$O. Since we assume that the relative uncertainty of our calculations is at
least as large the relative difference between the partition functions, the uncertainty assigned to
\HDO\ calculations is smaller than for other isotopologues.

\subsection{Electronic polarizability}

\begin{figure}[h]
  \center\includegraphics[width=0.9\linewidth]{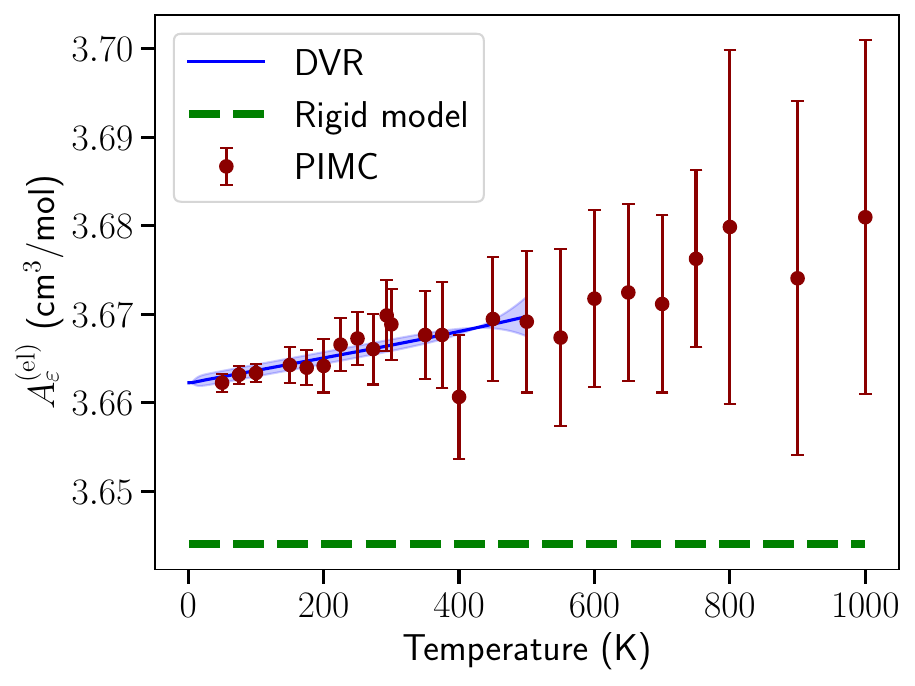}    
  \caption{The electronic polarizability contribution to $A_\varepsilon$,
    $A_{\varepsilon}^\mathrm{(el)}$, of HD${}^{16}$O as a 
    function of temperature. The dashed green line is the constant value
    corresponding to the rigid model of HD${}^{16}$O reported in
    Ref.~\onlinecite{HDO-geo} and which results in
    $A_{\varepsilon}^\mathrm{(el)} = 3.64403$~cm${}^3$/mol; the blue line
    is the result of our DVR calculations, together with an estimated
    uncertainty coming from the limited number of angular momenta that have
    been considered, reported as a blue area. The circles are the results
    of the path-integral simulations. All uncertainties are reported at the
    $k=2$ coverage value and do not include the propagation of the unknown
    uncertainty of the water electronic polarizability surface.}
  \label{fig:Aeps_pol_HDO}
\end{figure}

The results for the electronic polarizability contribution to $A_\varepsilon$ are reported in
Table~\ref{tab:Aeps_pol_HDO} and Fig.~\ref{fig:Aeps_pol_HDO}.

{\squeezetable 
\begin{table*}[h]
  \caption{\label{tab:Aeps_pol_HDO} The values of $\Aeps^\mathrm{(el)}$ for HD${}^{16}$O. All of the
    uncertainties are reported at $k=2$ coverage and do not include the propagation of the unknown
    uncertainty of the water electronic polarizability surface.}
  \begin{ruledtabular}
  \begin{tabular}{dcc}
    \multicolumn{1}{c}{Temperature} & $A_{\varepsilon}^\mathrm{(el)}$ (path-integral) &
    $A_{\varepsilon}^\mathrm{(el)}$ (DVR) \\
      \multicolumn{1}{c}{(K)} & (cm${}^3$/mol) & (cm${}^3$/mol) \\
      \hline
1	&$		–		$&$	3.66227	\pm	0.00001	$ \\
10	&$		–		$&$	3.66235	\pm	0.00016	$ \\
25	&$		–		$&$	3.6626	\pm	0.0003	$ \\
50	&$	3.662	\pm	0.001	$&$	3.6629	\pm	0.0003	$ \\
75	&$	3.663	\pm	0.001	$&$	3.6633	\pm	0.0003	$ \\
100	&$	3.663	\pm	0.001	$&$	3.6636	\pm	0.0003	$ \\
125	&$	3.664	\pm	0.001	$&$	3.6640	\pm	0.0003	$ \\
150	&$	3.664	\pm	0.002	$&$	3.6644	\pm	0.0003	$ \\
175	&$	3.664	\pm	0.002	$&$	3.6647	\pm	0.0003	$ \\
200	&$	3.664	\pm	0.003	$&$	3.6651	\pm	0.0003	$ \\
225	&$	3.667	\pm	0.003	$&$	3.6654	\pm	0.0003	$ \\
250	&$	3.667	\pm	0.003	$&$	3.6658	\pm	0.0003	$ \\
273.16	&$	3.666	\pm	0.004	$&$	3.6661	\pm	0.0003	$ \\
293.15	&$	3.670	\pm	0.004	$&$	3.6664	\pm	0.0003	$ \\
300	&$	3.669	\pm	0.004	$&$	3.6665	\pm	0.0003	$ \\
325	&$	3.669	\pm	0.004	$&$	3.6669	\pm	0.0003	$ \\
350	&$	3.668	\pm	0.005	$&$	3.6673	\pm	0.0003	$ \\
375	&$	3.668	\pm	0.006	$&$	3.6677	\pm	0.0002	$ \\
400	&$	3.661	\pm	0.007	$&$	3.6681	\pm	0.0002	$ \\
450	&$	3.669	\pm	0.007	$&$	3.6689	\pm	0.0002	$ \\
500	&$	3.669	\pm	0.008	$&$	3.670	\pm	0.001	$ \\
550	&$	3.667	\pm	0.010	$&$		–		$ \\
600	&$	3.67	\pm	0.01	$&$		–		$ \\
650	&$	3.67	\pm	0.01	$&$		–		$ \\
700	&$	3.67	\pm	0.01	$&$		–		$ \\
750	&$	3.68	\pm	0.01	$&$		–		$ \\
800	&$	3.68	\pm	0.02	$&$		–		$ \\
900	&$	3.67	\pm	0.02	$&$		–		$ \\
1000	&$	3.68	\pm	0.02	$&$		–		$ \\
1250	&$	3.68	\pm	0.03	$&$		–		$ \\
1500	&$	3.68	\pm	0.04	$&$		–		$ \\
1750	&$	3.69	\pm	0.04	$&$		–		$ \\
2000	&$	3.71	\pm	0.06	$&$		–		$ 
  \end{tabular}
  \end{ruledtabular}
\end{table*}  
}

\clearpage

\subsection{Dipolar polarizability}

\begin{figure}[h]
  \center\includegraphics[width=0.9\linewidth]{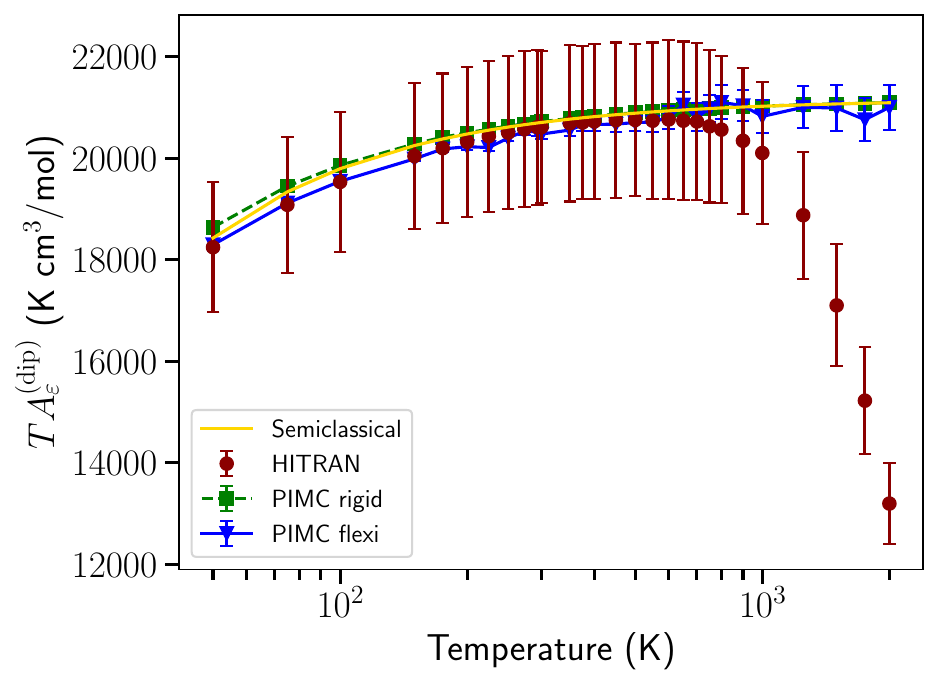}  
  \caption{The dipolar contribution to $A_\varepsilon$,
    $A_{\varepsilon}^\mathrm{(dip)}$, of HD${}^{16}$O as a function of
    temperature. All uncertainties are reported at the
    $k=2$ coverage value and do not include the propagation of the unknown
    uncertainty of the water dipole-moment surface. Lines are a guide for the eye.}
  \label{fig:Aeps_dip_HDO}
\end{figure}

The results for the dipolar contribution to $A_\varepsilon$ for \HDO\ are reported in
Table~\ref{tab:Aeps_dip_HDO} and Fig.~\ref{fig:Aeps_dip_HDO}.
Table~\ref{tab:Aeps_dip_HDO_comp} reportes the vibrational and rotational contribution to
$A_{\varepsilon}^\mathrm{(dip)}$.

{\squeezetable 
\begin{table*}[h]
  \caption{\label{tab:Aeps_dip_HDO}
    The values of $A_\varepsilon^\mathrm{(dip)}$
    for HD${}^{16}$O using various models, and its total value (last
    column) from 
    path-integral simulations.
    All of the uncertainties are reported at $k=2$ coverage and do
    not include the propagation of the unknown
    uncertainty of the water dipole-moment surface.}
  \begin{ruledtabular}
  \begin{tabular}{ddccc|c}
    \multicolumn{1}{c}{Temperature} &
    \multicolumn{1}{c}{$A_{\varepsilon}^\mathrm{(dip)}  $ (semiclassical)} &
    \multicolumn{1}{c}{$A_{\varepsilon}^\mathrm{(dip)}$ (HITRAN2020)} &
    $A_{\varepsilon}^\mathrm{(dip)}$ (rigid) &
    $A_{\varepsilon}^\mathrm{(dip)}$ (flexible) &
    $A_{\varepsilon}$ (flexible) \\    
    \multicolumn{1}{c}{(K)} &
    \multicolumn{1}{c}{(cm${}^3$/mol)} &
    \multicolumn{1}{c}{(cm${}^3$/mol)} &
    (cm${}^3$/mol) &
    (cm${}^3$/mol) &    
    (cm${}^3$/mol) \\
    \hline
50	&	368.517	&	365	$\pm$	13	&	372.7	$\pm$	0.3	&	366	$\pm$	2	&	370	$\pm$	2	\\
75	&	257.815	&	254	$\pm$	9	&	259.28	$\pm$	0.11	&	254.9	$\pm$	0.9	&	258.5	$\pm$	0.9	\\
100	&	197.913	&	195	$\pm$	7	&	198.56	$\pm$	0.05	&	195.4	$\pm$	0.5	&	199.1	$\pm$	0.5	\\
125	&	160.515	&	159	$\pm$	6	&	160.87	$\pm$	0.03	&	158.5	$\pm$	0.2	&	162.2	$\pm$	0.2	\\
150	&	134.976	&	134	$\pm$	5	&	135.19	$\pm$	0.02	&	133.2	$\pm$	0.3	&	136.9	$\pm$	0.3	\\
175	&	116.437	&	115	$\pm$	4	&	116.63	$\pm$	0.02	&	115.3	$\pm$	0.3	&	119.0	$\pm$	0.3	\\
200	&	102.369	&	102	$\pm$	4	&	102.478	$\pm$	0.012	&	101.1	$\pm$	0.3	&	104.8	$\pm$	0.3	\\
225	&	91.333	&	91	$\pm$	3	&	91.417	$\pm$	0.010	&	89.8	$\pm$	0.3	&	93.5	$\pm$	0.3	\\
250	&	82.442	&	82	$\pm$	3	&	82.524	$\pm$	0.008	&	81.6	$\pm$	0.3	&	85.3	$\pm$	0.3	\\
273.16	&	75.622	&	75	$\pm$	3	&	75.671	$\pm$	0.007	&	75.2	$\pm$	0.3	&	78.8	$\pm$	0.3	\\
293.15	&	70.581	&	70	$\pm$	3	&	70.632	$\pm$	0.006	&	70.0	$\pm$	0.3	&	73.7	$\pm$	0.3	\\
300	&	69.005	&	69	$\pm$	3	&	69.050	$\pm$	0.005	&	68.2	$\pm$	0.3	&	71.9 $\pm$	0.3	\\
325	&	63.805	&	64	$\pm$	2	&	63.844	$\pm$	0.004	&	63.3	$\pm$	0.3	&	66.9	$\pm$	0.3	\\
350	&	59.333	&	59	$\pm$	2	&	59.371	$\pm$	0.004	&	58.7	$\pm$	0.3	&	62.4	$\pm$	0.3	\\
375	&	55.447	&	55	$\pm$	2	&	55.472	$\pm$	0.004	&	55.1	$\pm$	0.3	&	58.8	$\pm$	0.3	\\
400	&	52.038	&	51.8	$\pm$	1.9	&	52.070	$\pm$	0.003	&	51.7	$\pm$	0.3	&	55.3	$\pm$	0.3	\\
450	&	46.341	&	46.1	$\pm$	1.7	&	46.362	$\pm$	0.002	&	45.9	$\pm$	0.4	&	49.6	$\pm$	0.4	\\
500	&	41.767	&	41.5	$\pm$	1.5	&	41.785	$\pm$	0.002	&	41.4	$\pm$	0.3	&	45.1	$\pm$	0.3	\\
550	&	38.015	&	37.7	$\pm$	1.4	&	38.030	$\pm$	0.002	&	37.7	$\pm$	0.4	&	41.3	$\pm$	0.4	\\
600	&	34.882	&	34.6	$\pm$	1.3	&	34.8933	$\pm$	0.0014	&	34.7	$\pm$	0.4	&	38.4	$\pm$	0.4	\\
650	&	32.226	&	31.9	$\pm$	1.2	&	32.2374	$\pm$	0.0012	&	32.4	$\pm$	0.4	&	36.1	$\pm$	0.4	\\
700	&	29.945	&	29.6	$\pm$	1.1	&	29.9543	$\pm$	0.0010	&	29.7	$\pm$	0.4	&	33.4	$\pm$	0.4	\\
750	&	27.966	&	27.5	$\pm$	1.0	&	27.9740	$\pm$	0.0009	&	28.0	$\pm$	0.4	&	31.7	$\pm$	0.4	\\
800	&	26.233	&	25.7	$\pm$	0.9	&	26.2379	$\pm$	0.0008	&	26.4	$\pm$	0.4	&	30.1	$\pm$	0.4	\\
900	&	23.339	&	22.6	$\pm$	0.8	&	23.3429	$\pm$	0.0006	&	23.4	$\pm$	0.3	&	27.1	$\pm$	0.3	\\
1000	&	21.020	&	20.1	$\pm$	0.7	&	21.0218	$\pm$	0.0006	&	20.8	$\pm$	0.3	&	24.5	$\pm$	0.3	\\
1250	&	16.838	&	15.1	$\pm$	0.5	&	16.8401	$\pm$	0.0003	&	16.8	$\pm$	0.3	&	20.5	$\pm$	0.3	\\
1500	&	14.044	&	11.4	$\pm$	0.4	&	14.0452	$\pm$	0.0002	&	14.0	$\pm$	0.3	&	17.7	$\pm$	0.3	\\
1750	&	12.045	&	8.7	$\pm$	0.3	&	12.0456	$\pm$	0.0002	&	11.9	$\pm$	0.2	&	15.6	$\pm$	0.2	\\
2000	&	10.544	&	6.6	$\pm$	0.2	&	10.54550	$\pm$	0.00014	&	10.5	$\pm$	0.2	&	14.2	$\pm$	0.2
  \end{tabular}    
  \end{ruledtabular}  
\end{table*}      
}

{\squeezetable
\begin{table*}[h]
  \caption{\label{tab:Aeps_dip_HDO_comp}
    Vibrational and rotational contribution to $\Aeps^\mathrm{(dip)}$ of HDO from
    HITRAN2020. All of the uncertainties are reported at $k=2$ coverage.}
  \begin{ruledtabular}
  \begin{tabular}{dcc}
    \multicolumn{1}{c}{Temperature} & $A_{\varepsilon}^\mathrm{(dip,vib)}$ &
    $A_{\varepsilon}^\mathrm{(dip,rot)}$ \\
      \multicolumn{1}{c}{(K)} & (cm${}^3$/mol) & (cm${}^3$/mol) \\
      \hline
50	& $	0.112	\pm	0.004	$&$	364.8	\pm	12.8	$ \\
75	& $	0.112	\pm	0.004	$&$	254.3	\pm	8.9	$ \\
100	& $	0.113	\pm	0.004	$&$	195.2	\pm	6.9	$ \\
125	& $	0.113	\pm	0.004	$&$	158.5	\pm	5.6	$ \\
150	& $	0.114	\pm	0.004	$&$	133.5	\pm	4.8	$ \\
175	& $	0.114	\pm	0.004	$&$	115.3	\pm	4.2	$ \\
200	& $	0.114	\pm	0.004	$&$	101.5	\pm	3.7	$ \\
225	& $	0.115	\pm	0.004	$&$	90.7	\pm	3.3	$ \\
250	& $	0.115	\pm	0.004	$&$	81.9	\pm	3.0	$ \\
273.16	& $	0.115	\pm	0.004	$&$	75.1	\pm	2.8	$ \\
293.15	& $	0.116	\pm	0.004	$&$	70.1	\pm	2.6	$ \\
300	& $	0.116	\pm	0.004	$&$	68.6	\pm	2.5	$ \\
325	& $	0.116	\pm	0.004	$&$	63.4	\pm	2.3	$ \\
350	& $	0.117	\pm	0.004	$&$	59.0	\pm	2.2	$ \\
375	& $	0.117	\pm	0.004	$&$	55.1	\pm	2.0	$ \\
400	& $	0.117	\pm	0.004	$&$	51.7	\pm	1.9	$ \\
450	& $	0.118	\pm	0.004	$&$	46.0	\pm	1.7	$ \\
500	& $	0.119	\pm	0.004	$&$	41.4	\pm	1.5	$ \\
550	& $	0.119	\pm	0.004	$&$	37.6	\pm	1.4	$ \\
600	& $	0.120	\pm	0.004	$&$	34.5	\pm	1.3	$ \\
650	& $	0.120	\pm	0.004	$&$	31.8	\pm	1.2	$ \\
700	& $	0.120	\pm	0.004	$&$	29.4	\pm	1.1	$ \\
750	& $	0.120	\pm	0.004	$&$	27.4	\pm	1.0	$ \\
800	& $	0.120	\pm	0.004	$&$	25.6	\pm	0.9	$ \\
900	& $	0.118	\pm	0.004	$&$	22.5	\pm	0.8	$ \\
1000	& $	0.114	\pm	0.004	$&$	19.9	\pm	0.7	$ \\
1250	& $	0.102	\pm	0.004	$&$	15.0	\pm	0.5	$ \\
1500	& $	0.086	\pm	0.004	$&$	11.4	\pm	0.4	$ \\
1750	& $	0.071	\pm	0.003	$&$	8.6	\pm	0.3	$ \\
2000	& $	0.057	\pm	0.002	$&$	6.6	\pm	0.2	$
  \end{tabular}
  \end{ruledtabular}
\end{table*}  
}

\clearpage

\subsection{Dipole moment}

\begin{figure}
  \center\includegraphics[width=0.9\linewidth]{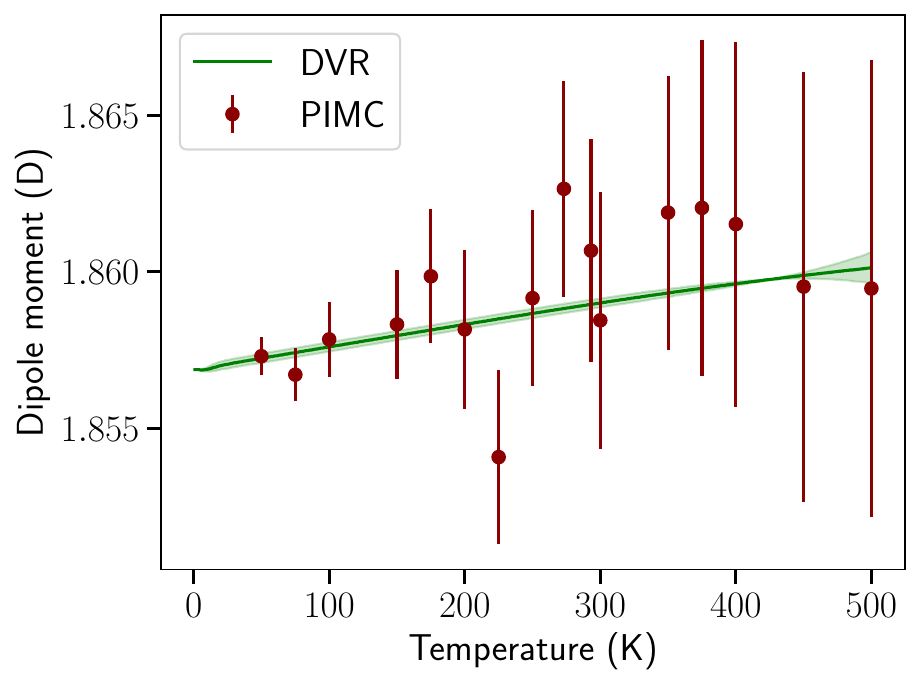}     
\caption{The average dipole moment $\mu(T)$ of HD${}^{16}$O from our DVR (solid line) and PIMC (points)
  calculations. The shaded area denotes an estimate of the uncertainty of the DVR calculation. The
  PIMC error bars represent the expanded $(k=2)$ statistical uncertainty of the Monte Carlo
  calculation.}
  \label{fig:mu_HDO}
\end{figure}

The results for the average dipole moment of \HDO\ reported in
Table~\ref{tab:dip_HDO} and Fig.~\ref{fig:mu_HDO}.

{\squeezetable  
\begin{table}[h]
  \caption{\label{tab:dip_HDO} The values of the average dipole moment $\mu(T)$ of HD${}^{16}$O (Debye) from
    our calculations. The PIMC uncertainties represent the expanded $(k=2)$ statistical uncertainty
    of the Monte Carlo calculation. Uncertainties do not include the propagation of the unknown
    uncertainty of the water dipole-moment surface.}
    \begin{ruledtabular}
  \begin{tabular}{dcc}
    \multicolumn{1}{c}{Temperature} &
    \multicolumn{1}{c}{$\mu(T)$ PIMC} &
    \multicolumn{1}{c}{$\mu(T)$ DVR} \\
    \multicolumn{1}{c}{(K)} &
    \multicolumn{1}{c}{(D)} &
    \multicolumn{1}{c}{(D)} \\
    \hline
    1	&$		–		$&$	1.85688	\pm	0.00010	$\\
10	&$		–		$&$	1.85689	\pm	0.00007	$\\
25	&$		–		$&$	1.85706	\pm	0.00014	$\\
50	&$	1.857	\pm	0.001	$&$	1.85725	\pm	0.00015	$\\
75	&$	1.857	\pm	0.001	$&$	1.85743	\pm	0.00015	$\\
100	&$	1.858	\pm	0.001	$&$	1.85761	\pm	0.00015	$\\
125	&$	1.858	\pm	0.001	$&$	1.85780	\pm	0.00014	$\\
150	&$	1.858	\pm	0.002	$&$	1.85797	\pm	0.00015	$\\
175	&$	1.860	\pm	0.002	$&$	1.85815	\pm	0.00015	$\\
200	&$	1.858	\pm	0.003	$&$	1.85833	\pm	0.00015	$\\
225	&$	1.854	\pm	0.003	$&$	1.85851	\pm	0.00015	$\\
250	&$	1.859	\pm	0.003	$&$	1.85868	\pm	0.00015	$\\
273.16	&$	1.863	\pm	0.003	$&$	1.85884	\pm	0.00015	$\\
293.15	&$	1.861	\pm	0.004	$&$	1.85897	\pm	0.00015	$\\
300	&$	1.858	\pm	0.004	$&$	1.85902	\pm	0.00015	$\\
325	&$	1.859	\pm	0.000	$&$	1.85920	\pm	0.00010	$\\
350	&$	1.862	\pm	0.004	$&$	1.85934	\pm	0.00015	$\\
375	&$	1.862	\pm	0.005	$&$	1.85949	\pm	0.00011	$\\
400	&$	1.862	\pm	0.006	$&$	1.85963	\pm	0.00007	$\\
450	&$	1.860	\pm	0.007	$&$	1.85990	\pm	0.00010	$\\
500	&$	1.859	\pm	0.007	$&$	1.86014	\pm	0.00050	$\\
550	&$	1.858	\pm	0.008	$&$		–		$\\
600	&$	1.862	\pm	0.010	$&$		–		$\\
650	&$	1.871	\pm	0.011	$&$		–		$\\
700	&$	1.860	\pm	0.012	$&$		–		$\\
750	&$	1.867	\pm	0.012	$&$		–		$\\
800	&$	1.872	\pm	0.015	$&$		–		$\\
900	&$	1.868	\pm	0.013	$&$		–		$\\
1000	&$	1.857	\pm	0.014	$&$		–		$\\
1250	&$	1.864	\pm	0.018	$&$		–		$\\
1500	&$	1.862	\pm	0.020	$&$		–		$\\
1750	&$	1.851	\pm	0.019	$&$		–		$\\
2000	&$	1.861	\pm	0.020	$&$		–		$
  \end{tabular}    
  \end{ruledtabular}  
\end{table}      
}

\clearpage

\section{Results for \DtO}

We do not have uncertainty estimates for the DVR calculations since we cannot reproduce the HITRAN2020
partition function due to an inconsistency of the HITRAN2020 database regarding the degeneracies of
D${}_2{}^{16}$O;~\cite{RG_private} it is possibly similar or worse than that of H${}_2{}^{16}$O.

\begin{figure}[h]
  \center\includegraphics[width=0.9\linewidth]{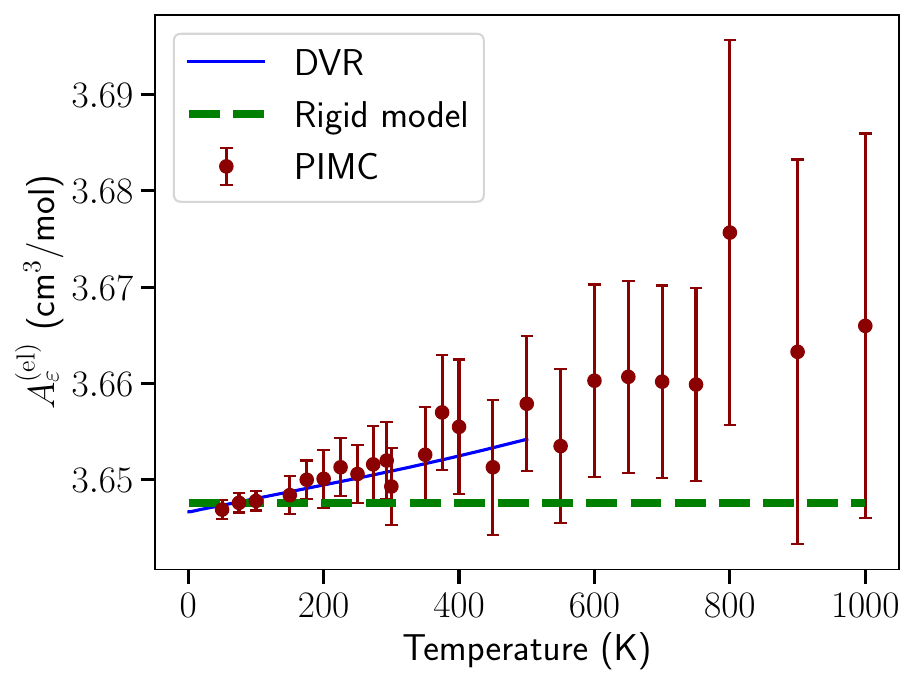}  
  \caption{The electronic polarizability contribution to
    $A_\varepsilon$, $A_{\varepsilon}^\mathrm{(el)}$, of D${}_2$O as a
    function of temperature. The dashed green line is the constant value
    corresponding to the rigid model of D${}_2$O reported in
    Ref.~\onlinecite{Czako:09} and which results in
    $A_{\varepsilon}^\mathrm{(el)} = 3.64757$~cm${}^3$/mol; the blue line
    is the result of our DVR calculations. The circles are the results
    of the path-integral simulations. All uncertainties are reported at the
    $k=2$ coverage value and do not include the propagation of the unknown
    uncertainty of the water electronic polarizability surface.}
  \label{fig:Aeps_pol_D2O}
\end{figure}

The results for the electronic polarizability contribution to $A_\varepsilon$ are reported in
Table~\ref{tab:Aeps_pol_D2O} and Fig.~\ref{fig:Aeps_pol_D2O}.

\subsection{Electronic polarizability}

{\squeezetable 
\begin{table*}[h]
  \caption{\label{tab:Aeps_pol_D2O}
    The values of $A_\varepsilon^\mathrm{(el)}$
    for D${}_2{}^{16}$O. All of the uncertainties are reported at $k=2$ coverage and do
    not include the propagation of the unknown
    uncertainty of the water electronic polarizability surface.}
  \begin{ruledtabular}
  \begin{tabular}{dcc}
    \multicolumn{1}{c}{Temperature} & $A_{\varepsilon}^\mathrm{(el)}$ (path-integral) &
    $A_{\varepsilon}^\mathrm{(el)}$ (DVR) \\
      \multicolumn{1}{c}{(K)} & (cm${}^3$/mol) & (cm${}^3$/mol) \\
      \hline
1	&$		–		$&$	3.6466	$ \\
10	&$		–		$&$	3.6468	$ \\
25	&$		–		$&$	3.6470	$ \\
50	&$	3.647	\pm	0.001	$&$	3.6473	$ \\
75	&$	3.648	\pm	0.001	$&$	3.6477	$ \\
100	&$	3.648	\pm	0.001	$&$	3.6480	$ \\
125	&$	3.648	\pm	0.001	$&$	3.6484	$ \\
150	&$	3.648	\pm	0.002	$&$	3.6487	$ \\
175	&$	3.650	\pm	0.002	$&$	3.6491	$ \\
200	&$	3.650	\pm	0.003	$&$	3.6494	$ \\
225	&$	3.651	\pm	0.003	$&$	3.6498	$ \\
250	&$	3.651	\pm	0.003	$&$	3.6501	$ \\
273.16	&$	3.652	\pm	0.004	$&$	3.6505	$ \\
293.15	&$	3.652	\pm	0.004	$&$	3.6508	$ \\
300	&$	3.649	\pm	0.004	$&$	3.6509	$ \\
325	&$	3.649	\pm	0.004	$&$	3.6512	$ \\
350	&$	3.653	\pm	0.005	$&$	3.6516	$ \\
375	&$	3.657	\pm	0.006	$&$	3.6520	$ \\
400	&$	3.655	\pm	0.007	$&$	3.6524	$ \\
450	&$	3.651	\pm	0.007	$&$	3.6533	$ \\
500	&$	3.658	\pm	0.007	$&$	3.6542	$ \\
550	&$	3.653	\pm	0.008	$&$	–	$ \\
600	&$	3.66	\pm	0.01	$&$	–	$ \\
650	&$	3.66	\pm	0.01	$&$	–	$ \\
700	&$	3.66	\pm	0.01	$&$	–	$ \\
750	&$	3.66	\pm	0.01	$&$	–	$ \\
800	&$	3.68	\pm	0.02	$&$	–	$ \\
900	&$	3.66	\pm	0.02	$&$	–	$ \\
1000	&$	3.67	\pm	0.02	$&$	–	$ \\
1250	&$	3.67	\pm	0.03	$&$	–	$ \\
1500	&$	3.68	\pm	0.04	$&$	–	$ \\
1750	&$	3.67	\pm	0.04	$&$	–	$ \\
2000	&$	3.66	\pm	0.05	$&$	–	$
  \end{tabular}
  \end{ruledtabular}
\end{table*}  
}

\clearpage

\subsection{Dipolar polarizability}

\begin{figure}[h]
  \center\includegraphics[width=0.9\linewidth]{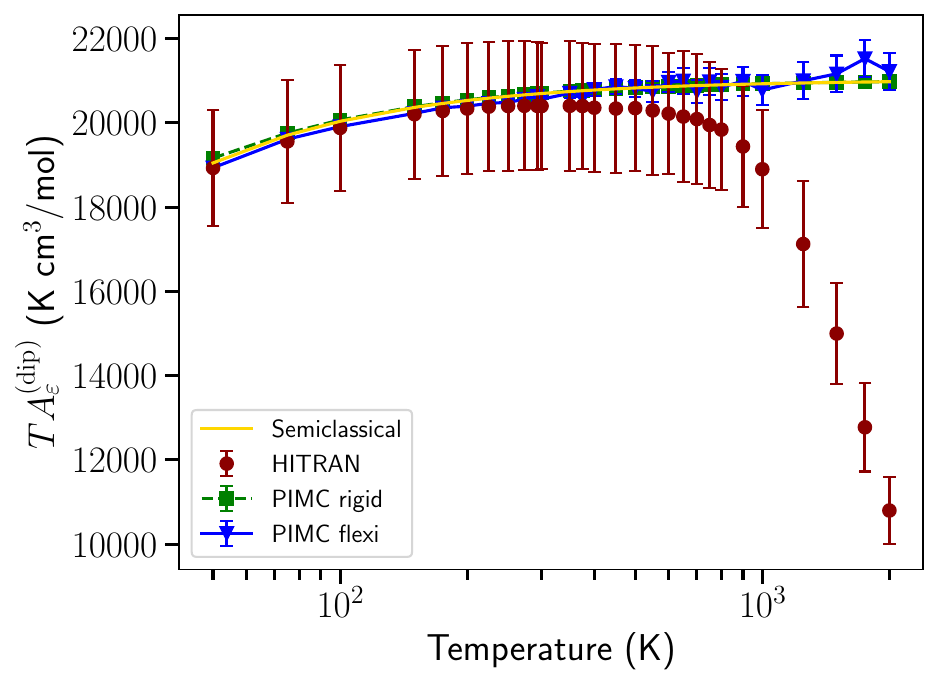}
  \caption{The dipolar contribution to $A_\varepsilon$, $A_{\varepsilon}^\mathrm{(dip)}$, of \DtO\ as
    a function of temperature. All uncertainties are reported at the $k=2$ coverage value and do not
    include the propagation of the unknown uncertainty of the water dipole-moment surface. Lines are
    a guide for the eye.}
  \label{fig:Aeps_dip_D2O}
\end{figure}

The results for the dipolar contribution to $A_\varepsilon$ for \DtO\ are reported in
Table~\ref{tab:Aeps_dip_D2O} and Fig.~\ref{fig:Aeps_dip_D2O}.
Table~\ref{tab:Aeps_dip_D2O_comp} reportes the vibrational and rotational contribution to
$A_{\varepsilon}^\mathrm{(dip)}$.

{\squeezetable
  
\begin{table*}[h]
  \caption{\label{tab:Aeps_dip_D2O} The values of $A_\varepsilon^\mathrm{(dip)}$ for
    \DtO\ using various models, and its total value (last column) from path-integral simulations.
    All of the uncertainties are reported at $k=2$ coverage and do not include the propagation of
    the unknown uncertainty of the water dipole-moment surface.}
  \begin{ruledtabular}
  \begin{tabular}{ddccc|c}
    \multicolumn{1}{c}{Temperature} &
    \multicolumn{1}{c}{$A_{\varepsilon}^\mathrm{(dip)}  $ (semiclassical)} &
    \multicolumn{1}{c}{$A_{\varepsilon}^\mathrm{(dip)}$ (HITRAN2020)} &
    $A_{\varepsilon}^\mathrm{(dip)}$ (rigid) &
    $A_{\varepsilon}^\mathrm{(dip)}$ (flexible) &
    $A_{\varepsilon}$ (flexible) \\    
    \multicolumn{1}{c}{(K)} &
    \multicolumn{1}{c}{(cm${}^3$/mol)} &
    \multicolumn{1}{c}{(cm${}^3$/mol)} &
    (cm${}^3$/mol) &
    (cm${}^3$/mol) &    
    (cm${}^3$/mol) \\
    \hline
50	&	380.995	&	379	$\pm$	14	&	383.3	$\pm$	0.2	&	379	$\pm$	2	&	382	$\pm$	2	\\
75	&	262.783	&	261	$\pm$	10	&	263.42	$\pm$	0.09	&	261.5	$\pm$	0.7	&	265.2	$\pm$	0.7	\\
100	&	200.383	&	199	$\pm$	8	&	200.68	$\pm$	0.04	&	199.1	$\pm$	0.5	&	202.8	$\pm$	0.5	\\
125	&	161.887	&	161	$\pm$	6	&	162.088	$\pm$	0.03	&	161.1	$\pm$	0.4	&	164.8	$\pm$	0.4	\\
150	&	135.785	&	135	$\pm$	5	&	135.90	$\pm$	0.02	&	134.8	$\pm$	0.4	&	138.5	$\pm$	0.4	\\
175	&	116.924	&	116	$\pm$	4	&	117.00	$\pm$	0.01	&	116.3	$\pm$	0.4	&	120.0	$\pm$	0.4	\\
200	&	102.662	&	102	$\pm$	4	&	102.737	$\pm$	0.009	&	102.0	$\pm$	0.4	&	105.7	$\pm$	0.4	\\
225	&	91.499	&	91	$\pm$	3	&	91.551	$\pm$	0.007	&	91.0	$\pm$	0.4	&	94.6	$\pm$	0.4	\\
250	&	82.525	&	82	$\pm$	3	&	82.562	$\pm$	0.006	&	82.0	$\pm$	0.4	&	85.6	$\pm$	0.4	\\
273.16	&	75.651	&	75	$\pm$	3	&	75.687	$\pm$	0.005	&	75.3	$\pm$	0.3	&	78.9	$\pm$	0.3	\\
293.15	&	70.576	&	70	$\pm$	3	&	70.604	$\pm$	0.004	&	70.4	$\pm$	0.4	&	74.0	$\pm$	0.4	\\
300	&	68.991	&	68	$\pm$	3	&	69.021	$\pm$	0.004	&	68.5	$\pm$	0.4	&	72.2	$\pm$	0.4	\\
325	&	63.762	&	63	$\pm$	2	&	63.7838	$\pm$	0.003	&	63.5	$\pm$	0.4	&	67.1	$\pm$	0.4	\\
350	&	59.269	&	58	$\pm$	2	&	59.294	$\pm$	0.003	&	59.2	$\pm$	0.4	&	62.8	$\pm$	0.4	\\
375	&	55.368	&	54	$\pm$	2	&	55.389	$\pm$	0.003	&	54.9	$\pm$	0.3	&	58.6	$\pm$	0.3	\\
400	&	51.949	&	50.9	$\pm$	1.9	&	51.962	$\pm$	0.002	&	52.0	$\pm$	0.4	&	55.7	$\pm$	0.4	\\
450	&	46.238	&	45.2	$\pm$	1.7	&	46.254	$\pm$	0.002	&	46.4	$\pm$	0.4	&	50.0	$\pm$	0.4	\\
500	&	41.658	&	40.7	$\pm$	1.5	&	41.669	$\pm$	0.001	&	41.6	$\pm$	0.4	&	45.3	$\pm$	0.4	\\
550	&	37.904	&	36.9	$\pm$	1.4	&	37.911	$\pm$	0.001	&	37.7	$\pm$	0.5	&	41.4	$\pm$	0.5	\\
600	&	34.770	&	33.7	$\pm$	1.2	&	34.7763	$\pm$	0.0010	&	34.9	$\pm$	0.4	&	38.6	$\pm$	0.4	\\
650	&	32.115	&	31.0	$\pm$	1.2	&	32.1205	$\pm$	0.0009	&	32.3	$\pm$	0.5	&	36.0	$\pm$	0.5	\\
700	&	29.836	&	28.7	$\pm$	1.1	&	29.8404	$\pm$	0.0008	&	29.6	$\pm$	0.4	&	33.3	$\pm$	0.4	\\
750	&	27.860	&	26.6	$\pm$	1.0	&	27.8639	$\pm$	0.0006	&	28.0	$\pm$	0.4	&	31.6	$\pm$	0.4	\\
800	&	26.129	&	24.8	$\pm$	0.9	&	26.1316	$\pm$	0.0006	&	26.1	$\pm$	0.4	&	29.8	$\pm$	0.4	\\
900	&	23.241	&	21.6	$\pm$	0.8	&	23.2428	$\pm$	0.0005	&	23.3	$\pm$	0.4	&	27.0	$\pm$	0.4	\\
1000	&	20.928	&	18.9	$\pm$	0.7	&	20.9286	$\pm$	0.0004	&	20.8	$\pm$	0.4	&	24.5	$\pm$	0.4	\\
1250	&	16.758	&	13.7	$\pm$	0.6	&	16.7589	$\pm$	0.0003	&	16.8	$\pm$	0.4	&	20.5	$\pm$	0.4	\\
1500	&	13.974	&	10.0	$\pm$	0.4	&	13.9756	$\pm$	0.0002	&	14.1	$\pm$	0.3	&	17.8	$\pm$	0.3	\\
1750	&	11.983	&	7.3	$\pm$	0.3	&	11.9838	$\pm$	0.0001	&	12.3	$\pm$	0.3	&	16.0	$\pm$	0.3	\\
2000	&	10.489	&	5.4	$\pm$	0.2	&	10.48890	$\pm$	0.00011	&	10.6	$\pm$	0.2	&	14.3	$\pm$	0.2
  \end{tabular}    
  \end{ruledtabular}  
\end{table*}      
}

{\squeezetable
\begin{table*}[h]
  \caption{\label{tab:Aeps_dip_D2O_comp}
    Vibrational and rotational contribution to $\Aeps^\mathrm{(dip)}$ of D${}_2{}^{16}$O from
    HITRAN2020. All of the uncertainties are reported at $k=2$ coverage.}
  \begin{ruledtabular}
  \begin{tabular}{dcc}
    \multicolumn{1}{c}{Temperature} & $A_{\varepsilon}^\mathrm{(dip,vib)}$ &
    $A_{\varepsilon}^\mathrm{(dip,rot)}$ \\
      \multicolumn{1}{c}{(K)} & (cm${}^3$/mol) & (cm${}^3$/mol) \\
      \hline
50	& $	0.103	\pm	0.004	$&$	378.5	\pm	13.7	$ \\
75	& $	0.103	\pm	0.004	$&$	260.7	\pm	9.7	$ \\
100	& $	0.104	\pm	0.004	$&$	198.7	\pm	7.5	$ \\
125	& $	0.104	\pm	0.004	$&$	160.5	\pm	6.1	$ \\
150	& $	0.105	\pm	0.004	$&$	134.6	\pm	5.1	$ \\
175	& $	0.105	\pm	0.004	$&$	115.8	\pm	4.4	$ \\
200	& $	0.105	\pm	0.004	$&$	101.6	\pm	3.9	$ \\
225	& $	0.106	\pm	0.004	$&$	90.5	\pm	3.4	$ \\
250	& $	0.106	\pm	0.004	$&$	81.5	\pm	3.1	$ \\
273.16	& $	0.106	\pm	0.004	$&$	74.6	\pm	2.8	$ \\
293.15	& $	0.106	\pm	0.004	$&$	69.5	\pm	2.6	$ \\
300	& $	0.106	\pm	0.004	$&$	67.9	\pm	2.5	$ \\
325	& $	0.107	\pm	0.004	$&$	62.7	\pm	2.3	$ \\
350	& $	0.107	\pm	0.004	$&$	58.2	\pm	2.2	$ \\
375	& $	0.107	\pm	0.004	$&$	54.3	\pm	2.0	$ \\
400	& $	0.107	\pm	0.004	$&$	50.8	\pm	1.9	$ \\
450	& $	0.108	\pm	0.004	$&$	45.1	\pm	1.7	$ \\
500	& $	0.109	\pm	0.004	$&$	40.5	\pm	1.5	$ \\
550	& $	0.110	\pm	0.004	$&$	36.8	\pm	1.4	$ \\
600	& $	0.111	\pm	0.004	$&$	33.6	\pm	1.2	$ \\
650	& $	0.112	\pm	0.004	$&$	30.9	\pm	1.2	$ \\
700	& $	0.113	\pm	0.005	$&$	28.6	\pm	1.1	$ \\
750	& $	0.114	\pm	0.005	$&$	26.5	\pm	1.0	$ \\
800	& $	0.115	\pm	0.005	$&$	24.7	\pm	0.9	$ \\
900	& $	0.115	\pm	0.005	$&$	21.5	\pm	0.8	$ \\
1000	& $	0.114	\pm	0.005	$&$	18.8	\pm	0.7	$ \\
1250	& $	0.105	\pm	0.005	$&$	13.6	\pm	0.6	$ \\
1500	& $	0.090	\pm	0.005	$&$	9.9	\pm	0.4	$ \\
1750	& $	0.074	\pm	0.004	$&$	7.2	\pm	0.3	$ \\
2000	& $	0.059	\pm	0.003	$&$	5.3	\pm	0.2	$
  \end{tabular}
  \end{ruledtabular}
\end{table*}  
}

\clearpage

\subsection{Dipole moment}

\begin{figure}
  \center\includegraphics[width=0.9\linewidth]{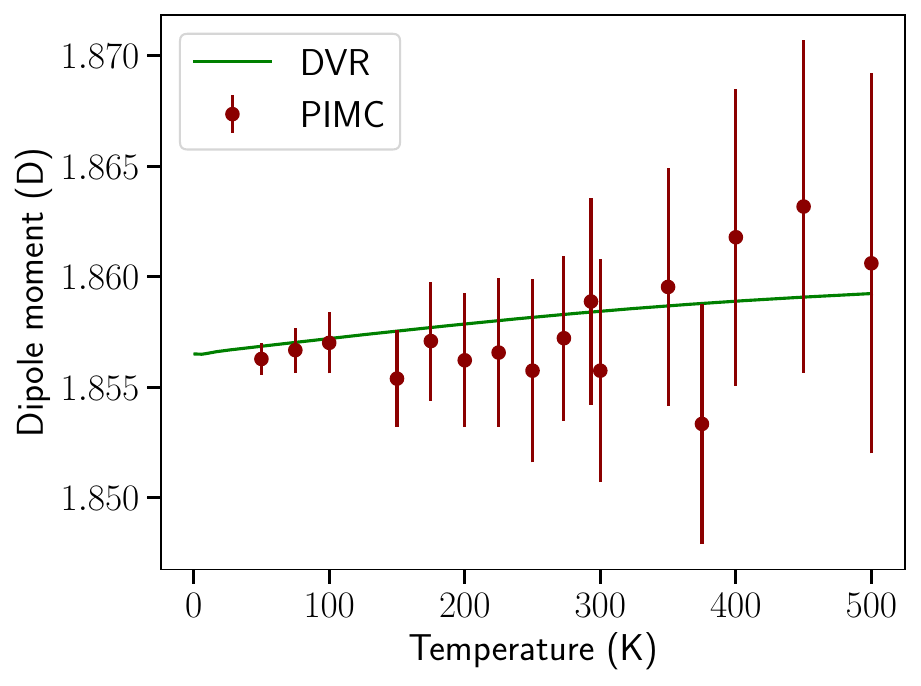}
\caption{The average dipole moment $\mu(T)$ of \DtO\ from our DVR (solid line) and PIMC (points) calculations. The
  PIMC error bars represent the expanded $(k=2)$ statistical uncertainty of the Monte Carlo
  calculation.}
  \label{fig:mu_D2O}
\end{figure}

The results for the average dipole moment of \DtO\ are reported in
Table~\ref{tab:dip_D2O} and Fig.~\ref{fig:mu_D2O}.

{\squeezetable 
\begin{table}[h]
  \caption{\label{tab:dip_D2O} The values of the average dipole moment $\mu(T)$ of \DtO\ (Debye) from our
    calculations. The PIMC uncertainties represent the expanded $(k=2)$ statistical uncertainty of
    the Monte Carlo calculation. Uncertainties do not include the propagation of the unknown
    uncertainty of the water dipole-moment surface.}
    \begin{ruledtabular}
  \begin{tabular}{dcc}
    \multicolumn{1}{c}{Temperature} &
    \multicolumn{1}{c}{$\mu(T)$ (PIMC)} &
    \multicolumn{1}{c}{$\mu(T)$ (DVR)} \\
    \multicolumn{1}{c}{(K)} &
    \multicolumn{1}{c}{(D)} &
    \multicolumn{1}{c}{(D)} \\
    \hline
1	&$		–		$&$	1.85651	$\\
10	&$		–		$&$	1.85653	$\\
25	&$		–		$&$	1.85668	$\\
50	&$	1.856	\pm	0.001	$&$	1.85685	$\\
75	&$	1.857	\pm	0.001	$&$	1.85703	$\\
100	&$	1.857	\pm	0.001	$&$	1.85721	$\\
125	&$	1.858	\pm	0.002	$&$	1.85740	$\\
150	&$	1.855	\pm	0.002	$&$	1.85754	$\\
175	&$	1.857	\pm	0.003	$&$	1.85771	$\\
200	&$	1.856	\pm	0.003	$&$	1.85786	$\\
225	&$	1.857	\pm	0.003	$&$	1.85802	$\\
250	&$	1.856	\pm	0.004	$&$	1.85817	$\\
273.16	&$	1.857	\pm	0.004	$&$	1.85830	$\\
293.15	&$	1.859	\pm	0.005	$&$	1.85840	$\\
300	&$	1.856	\pm	0.005	$&$	1.85844	$\\
325	&$	1.858	\pm	0.006	$&$	1.85870	$\\
350	&$	1.860	\pm	0.005	$&$	1.85868	$\\
375	&$	1.853	\pm	0.005	$&$	1.85879	$\\
400	&$	1.862	\pm	0.007	$&$	1.85890	$\\
450	&$	1.863	\pm	0.008	$&$	1.85908	$\\
500	&$	1.861	\pm	0.009	$&$	1.85924	$\\
550	&$	1.856	\pm	0.011	$&$	–	$\\
600	&$	1.865	\pm	0.010	$&$	–	$\\
650	&$	1.866	\pm	0.013	$&$	–	$\\
700	&$	1.854	\pm	0.012	$&$	–	$\\
750	&$	1.864	\pm	0.014	$&$	–	$\\
800	&$	1.858	\pm	0.013	$&$	–	$\\
900	&$	1.863	\pm	0.015	$&$	–	$\\
1000	&$	1.853	\pm	0.016	$&$	–	$\\
1250	&$	1.862	\pm	0.019	$&$	–	$\\
1500	&$	1.868	\pm	0.019	$&$	–	$\\
1750	&$	1.884	\pm	0.019	$&$	–	$\\
2000	&$	1.870	\pm	0.019	$&$	–	$
  \end{tabular}    
  \end{ruledtabular}  
\end{table}      
}

\clearpage

\bibliography{ms}